\documentstyle[epsfig]{aipproc}

\def\beq{\begin{equation}}
\def\eeq{\end{equation}}
\def\bea{\begin{eqnarray}}
\def\eea{\end{eqnarray}}

\def\eq#1{{Eq.~(\ref{#1})}}

\def\g{\gamma}

\def\s{\sigma}

\begin{document}

\newcounter{savefig}
\newcommand{\alphfig}{\addtocounter{figure}{1}%
\setcounter{savefig}{\value{figure}}%
\setcounter{figure}{0}%
\renewcommand{\thefigure}{\mbox{\arabic{savefig}-\alph{figure}}}}
\newcommand{\resetfig}{\setcounter{figure}{\value{savefig}}%
\renewcommand{\thefigure}{\arabic{figure}}}

\begin{flushright}
August 1998\\ 
DESY 98-120\\
TAUP 2522/98\\
\end{flushright}

\title{ \Huge \bf  An Introduction to Pomerons
 }
\author{ \bf  Eugene ~ Levin  }
\address{  School of Physics and Astronomy\\
 Raymond and Beverly Sackler Faculty of Exact Science\\
 Tel Aviv University, Tel Aviv, 69978, ISRAEL\\
and\\
DESY Theory  Group\\
22603, Hamburg, GERMANY}

\maketitle

~

\centerline{}

\centerline{ \it Talk given at  ``Workshop on diffractive
physics",}

\centerline{\it LISHEP'98, Rio de Janeiro, February 16 - 20, 1998}
\begin{abstract}
This talk is an attempt to clarify  for experimentalists   what do
we (theorists)  mean when we are saying ``Pomeron". I hope, that in
this talk they will
find answers to such questions as what is Pomeron, what we have  learned
about Pomeron both experimentally and theoretically, what is the correct   
strategy to study Pomeron  experimentally  and etc.   I also hope that
this talk could be used as a guide in the zoo of Pomerons:``soft" Pomeron,
``hard" Pomeron, the BFKL Pomeron, the Donnachie - Landschoff Pomeron and
so on. The large number of different Pomerons just reflects our poor
understanding of the high energy asymptotic in our microscopic theory -
QCD.
The motto of my talk, which gives you my opinion on the subject in short, 
is:  { \bf        
Pomeron  is  still   unknown  but
needed  for  25  years   to  describe
experimental data }. This is a status
report of our ideas, hopes, theoretical approaches and 
phenomenological successes that have been developed to achieve a
theoretical understanding of high energy behaviour of scattering
amplitude.  
\end{abstract}
\section{High energy glossary}

In this section I am going to define terminology which we use  discussing
high energy processes. Most of this terminology has deep roots in Reggeon
approach to high energy scattering and it will be more understandable
after my answer to the question:``what is the ``soft" Pomeron". 
Although  I cannot explain right now everything, I think, it will be very
instructive to have a high energy glossary in front of your eyes since it
gives you some feeling about problems and ideas which are typical for
high energy asymptotic. 

To make easier your understanding I give here the picture of a high energy
interaction in the parton model ( see Fig.1 ).

\begin{figure}
\centerline{\psfig{file=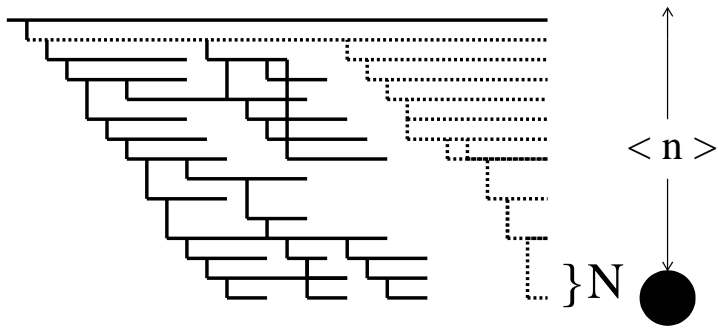,width=100mm}}
\caption{ \em The high energy interaction in the parton model.}
\label{fig1}
\end{figure}
 
In the parton approach the fast hadron decays into point-like particles (
{\it partons}) a long before ( typical time $t
\,\propto\,\frac{E}{\mu^2} $ ) the interaction with the target. However,
during time $t$ all these partons are in a coherent state which can be
described by means of a wave function.  The interaction of the slowest
parton ({\it ``wee" parton} )  with the target destroys completely the
coherence of the partonic wave function.
 The total cross section of such
an interaction is equal to
\begin{equation}
\label{1.1}
\sigma_{tot}\,\,=\,\,N\,\times\,\sigma_0
\end{equation}
where
\begin{itemize}
\item  $N$\,\,=
{ \it flux of ``wee" partons\,\,;}
\item  
$\sigma_0$\,\,= 
{ \it the cross section of the
interaction of one
``wee" parton with the target\,\,.}
\end{itemize}

One can see directly from Fig.1 that the flux of ``wee" partons is rather
large since each parton in the parton cascade can decay into its own chain
of partons. If one knows the typical multiplicity $< n >$  of a single
chain ( say,
dotted one in Fig.1 ) he can evaluate the number of ``wee" partons.
Indeed,
\begin{equation}
\label{1.2}
N\,\,=\,\,e^{< n >}
\end{equation}
where $ < n > \,\,\propto\,\,\frac{\ln s}{\Delta y}$ and $\Delta y $ is
the average distance in rapidity for partons in one chain. 
Reasonable estimate for $\Delta y $ is $\Delta
y\,\,\approx\,\,\frac{1}{\alpha_S}$. Therefore, we have finally
\begin{equation}
\label{1.3}
< n > \,\,=\,\,\omega_0 \,\ln s \,\,;\,\,\,\,\,\omega_0\,\,=\,\,Const
\,\,\alpha_s\,\,,
\end{equation}
It is obvious that \eq{1.3} leads to 
$$
N \,\,\propto\,\,s^{\omega_0}\,\,.
$$ 
Finally, \eq{1.1} can be rewritten in the form:
\begin{equation}
\label{1.4}
\sigma_{tot}\,\,=\,\,P(projectile) \,\times\,\sigma_0   ( target)
\,\times\,
(\,\frac{s}{s_0}\,)^{\omega_0}\,\,,
\end{equation}
where $P$ is a coefficient which depends only on the quantum numbers and
 variables related to  the projectile and which is  a physics meaning of 
probability to find  the parton cascade of Fig.1 in the fast hadron ({\it
projectile}). However, our picture should be relativistic invariant and
independent of the frame in which we discuss the process of scattering.
Therefore, $P$ in \eq{1.4} have to be proportional to $ P
\,\,\propto\,\,\frac{\sigma_0( projectile)}{s_0}$  and \eq{1.4} is to be
reduced to
\begin{equation}
\label{1.5}
\sigma_{tot}\,\,=\,\,\sigma_0( projectile )\,\,\times\,\,\sigma_0 (
target)\,\,\times\,\,\frac{1}{s_0}\,\,\times\,\,\left(
\,\frac{s}{s_0}\,\right)^{\omega_0}\,\,,
\end{equation}
where $s_0$ is the energy scale which says  what energy we can consider
as high one for the process of interaction. High energy asymptotic means
$s\,\,\gg\,\,s_0$ and namely the value of $s_0$ is the less known scale
in high energy physics.

It should be stressed that the value of $\omega_0$ does not depend on 
any variable related both to projectile and/or  target. It only  depends
on density of partons in the partons cascade or, in other words, on the 
parton emission in our microscopic theory.

\begin{enumerate}
{ \bf \item   Reggeon:}

The formal definition of   Reggeon is the pole in the partial wave in
$t$-channel of the  scattering process. For example, for $\pi^+\, +\, p$
elastic amplitude the Reggeon is a pole in the partial wave of the
reaction: $\pi^+\,+\,\pi^-\,\rightarrow\,p \,+\,\bar p$, namely, the
amplitude of this process can be written in the form:
\beq \label{1.6}
f_l(t)\,\,=\,\,\sum^{\infty}_{l = 0}\,\,f_l(t)\,(2 l \,+\,1)\,P_l(z)\,\,,
\eeq
where $z = cos\theta$ and $\theta$ is the scattering angle from initial
pion to final proton (antiproton).  Reggeon is the hypothesis that
$f_l(t)$ has a pole of the form
\beq \label{1.7}
f_l(t)\,\,=\,\,\frac{g_1(t)\,g_2(t)}{l\,\,-\,\,\alpha_R(t)}
\eeq
where function $\alpha_R(t)$ is the Reggeon trajectory which
experimentally has a form:
\beq \label{1.8}
\alpha_R ( t)\,\,=\,\,\alpha_R(0)\,\,+\,\,\alpha'_R(0)\,t\,\,,
\eeq
where $\alpha_R(0 $ is the intercept of the Reggeon and $\alpha'_R(0)$ is
its slope.  It turns out that such hypothesis gives the following
asymptotic at high energy ( see Refs. \cite{COLLINS} \cite{MYLEC} )
\beq \label{1.9}
A_P(s,t)\,\,=  
\,\,g_1(m_1,M_1, t)\,\times\, g_2(m_2,M_2,t) \,\times\,
\frac{s^{\alpha_R(t)}\,\,+\,\, ( -
s)^{\alpha_R(t)}}{sin\,\pi\,\alpha_R(t)}\,\,.
\eeq
This equation has two important properties: (i) at $t= m^2_R$, where $m_R$
is the mass of resonance with spin $j$ ( $ j = \alpha_R(t=m^2_R)$ ) it
describe the exchange of the resonance, namely, $$A_R(s, t\, \rightarrow
\,m^2_R)\,\rightarrow\,g_1\,g_2 \frac{s^j}{m^2_R \,-\,t}$$ and (ii)
in the scattering kinematic region ( $t\,<\,0$ )
$A(s,t)\,\propto\,s^{\alpha_R(0)}$ which gives the asymptotic for the
total cross section
$$\sigma_{tot}\,\propto\,s^{\alpha_R(0)\,-\,1}\,<\,Const\ln^2s (
Froissart\,\,bound\,)$$ for
$\alpha_R(0)\,\,<\,\,0$. 
Summarizing what I have discussed I would like to emphasize that the
Reggeon hypothesis is the only one which has solved the contradiction
between experimental observation of resonances with spin larger than 1
and a  violation of the Froissart theorem which, at first sight, the 
exchange of such resonances leads to.  

{ \bf \item  ``Soft" Pomeron:} 

Actually, ``soft" Pomeron is the only one on the market which can be
called Pomeron since the definition of Pomeron is  

{\bf Pomeron is a
Reggeon with the intercept close to 1, or in other words,
$\mathbf \alpha_P(0)
\,-\,1\,=\Delta\,\,\leq\,\,1$}.

 I want to point out three extremely
important
experimental facts, which were the reason of the Pomeron hypothesis:

1. There are no resonances on the Pomeron trajectory;

2. The measured total cross sections are approximately constant at high
energy;

3. The intercepts of all Reggeons are smaller than 1 and, therefore, an
exchange of the Reggeon leads to cross sections which fall down at high
energies in the contradiction with the experimental data.

It should be stressed that there is no deep theory  reason for the Pomeron
hypothesis and it looks only as a simplest attempt to comply with the
experimental data on total cross sections. The common belief is that the
``soft" Pomeron exchange gives the correct high energy asymptotic for the
processes which occur at long distances ( ``soft" processes ). However, 
we will discuss below what is the theory status of our understanding. 

{\bf \item  Pomeron structure:}

We say ``Pomeron structure" when we would like to understand what
inelastic processes are responsible for the Pomeron exchange in our
microscopic theory.  In some sense,  the simplest Pomeron structure is
shown
in Fig.1 in the parton model. Unfortunately, we have not reached a much
deeper insight than it  is given in the picture of Fig.1.

{ \bf \item  Donnachie - Landshoff Pomeron:}

The phenomenology based on the Pomeron hypothesis turned out to be very
successful. It survived at least two decades despite of a lack of
theoretical arguments for Pomeron exchange. Donnachie and  Landshoff
\cite{DL}  gave an elegant and economic description almost all  existing
experimental data \cite{DL} \cite{TABLE} assuming the exchange of the
Pomeron with the following parameters of its trajectory:

1. $\Delta\,=\,\alpha_P(0)\,-\,1\,\approx\,0.08$\,\,;

2.$\alpha'_P(0)\,\,\approx\,\,0.25\,GeV^{-2}$\,\,.

The more detail properties of the D-L Pomeron we will discuss below.
At the moment, we can use these parameters when we are going to make some
estimates of the Pomeron contribution.

{ \bf \item   ``Hard" Pomeron:}

 ``Hard" Pomeron is a substitute for the following sentence: the
asymptotic for the cross section at high energy for  the ``hard"
processes which occur at
small distances ($r_{\perp}$)  of the order of $1/Q$ where $Q$ is the
largest transverse momentum scale in the process.  The ``hard" Pomeron is
not universal and depends on the process. The main brick to calculate the
``hard" Pomeron contribution is the solution to the evolution equation (
DGLAP evolution \cite{DGLAP} ) in the region of high energy. The main
properties of the ``hard" Pomeron which allow us to differentiate it from
the ``soft" one are:

1. the energy behaviour is manifest itself through the variable
$ x_B\,\,=\,\,\frac{Q^2}{s}$ ;

2.  the high energy asymptotic can be parameterized as
$\sigma_{tot}\,\,\propto\,\,\frac{1}{x^{\omega_0(Q^2)}}$ but the power
$\omega_0(Q^2)$ crucially depends on $Q^2$. Recall that it is not the case
for the ``soft" Pomeron or any Reggeon;

3.  although the energy and $Q^2$ behaviour of the ``hard" Pomeron is
quite different from the ``soft" one the space - time picture for it is
very similar to that one which is shown in Fig. 1.

 The procedure of calculation of the ``hard" Pomeron is based on the
perturbation theory. Indeed, for ``hard" processes we have a natural
scale of hardness that we call $Q^2 \,\gg\,m^2$ where $m^2$ is the
typical mass scale for ``soft" interactions. The QCD coupling constant
$\alpha_S (Q^2) $ can be considered as a small parameter (
$\alpha_s(Q^2)\,\,\ll\,\,1$ ) with respect to
which we develop the perturbation theory. Generally speaking any physics
observable ( let say the gluon structure function $x_B G(x_B,Q^2)$  can be
written as a perturbative series in the form:
\beq \label{1.11} 
x_B G(x_B,Q^2)\,\,=\,\,\lim |_{Q^2\,\gg\,\,m^2}\,\, \sum^{\infty}_{n =
0}\,\,\alpha^n_S(Q^2)\,\times\,M_n(x_B, Q^2)\,\,.
\eeq
For ``hard" processes we change the order of  summation and
limit\footnote{It should be stressed that this change is not so obvious
but fortunately for this particular case we can use the powerful method of
the renormalization group approach to justify it  {\it a posteriori}.} and
get
 \beq \label{1.12}
x_B G(x_B,Q^2)\,\,=
\,\,\sum^{\infty}_{n =0}\,\,\alpha^n_S(Q^2)\,\times\,\,
\lim |_{Q^2\,\gg\,\,m^2}M_n(x_B,Q^2)\,\,.
\eeq
The analysis of the Feyman diagrams shows that for $Q^2\,\,\gg\,\,m^2$
$M_n(x,Q^2) $  has a form
\beq \label{1.13}
\lim |_{Q^2\,\gg\,\,
m^2}M_n(x_B,Q^2)\,\,\Longrightarrow 
\eeq
$$
\,\,P_n(x)\,L^n\,\,+\,\,P^{(1)}_n(x)\,\,L^{n - 1} \,\,+\,\,...\,+\,P^{n +
1}_n(x)\,\,,
$$
where $ L $ is a large log in the problem, namely, $L\,=\,\ln(Q^2/m^2)$.

Taking only the leading term with respect to $L$ ($L^n$) in \eq{1.13},
we arrive to so called leading log approximation (LLA) of perturbative
QCD:
\beq \label{1.14}
x_B G^{LLA}(x_B,Q^2)\,\,=\,\,\sum^{\infty}_{n
=0}\,\,\alpha^n_S(Q^2)\,\times\,\,P_n(x)\,L^n\\,\,.
\eeq
The practical way how we perform summation in \eq{1.14} is the DGLAP
evolution equation \cite{DGLAP} and LLA is the same as the solution of the
DGLAP evolution equation in the leading order ( LO DGLAP ). Taking a next
term in $L$ in \eq{1.13} of the order of $L^{n - 1}$ in addition to the
leading one we can arrive to the   gluon structure function in the next to
leading order (NLO DGLAP ). As I have mentioned, the DGLAP evolution
equation gives a regular procedure for calculation .

I think, that it is instructive to show here the ``hard" Pomeron
contribution to the gluon structure function in the region of large $Q^2
\,\gg\,Q^2_0$ and small $x_B$ ( see Ref.\cite{EKL} for example).
It turns out that in this case $P_n(x)$  has a leading term of the order
$\ln^n(1/x)$, reducing the problem to summation of perturbative series
with respect to parameter $\alpha_S \,\ln (1/x)\,\ln(Q^2/m^2)$:
\beq \label{1.15}
x_BG^{DLA}(x_B,Q^2)\,\,=\,\,\sum{n=0}^{\infty} c_n (\,\alpha_S \,\ln
(1/x)\,\ln(Q^2/m^2) \,)^n\,\,.
\eeq
The summation over $n$ which we can also perform  using the DGLAP
evolution equation in so called double log approximation ( DLA) gives 
\beq \label{1.10}
x_B G^{DLA}(x_B, Q^2)\,\rightarrow\,e^{2 \,\sqrt{\frac{2
N_c}{b_0}\,\ln(1/x_B)\,\ln\frac{\alpha_S(Q^2_0)}{\alpha_S(Q^2)}}}\,\,,
\eeq
where $N_c$ is the number of colours and $b_0\,=\,(11 - \frac{2}{3}N_f)/2$ 
with  $N_f$ flavours. The running QCD coupling is equal to
$\alpha_S(Q^2)\,\,=\,\,\frac{4 \pi}{b_0\,\ln (Q^2/\Lambda^2)}$.

One can see from \eq{1.10} that the real asymptotic or the ``hard"
Pomeron is quite different from the Reggeon - like behaviour and could be 
approximated by power - like function in very limited kinematic region of
$x_b$ and $Q^2$. I believe, that this fact should be known to any
experimentalist even if he  is fitting the data using the Regge - like
parameterization $x_B G(x_B, Q^2 ) = const \frac{1}{x^{\omega_0(Q^2)}}$.

{ \bf \item  The BFKL Pomeron:}

The BFKL Pomeron is what we have in perturbative QCD instead of ``soft"
Pomeron. The method of derivation for the BFKL Pomeron is designed 
specially for the processes with one but rather large scale (
$Q^2\,\approx\,Q^2_0\,\,\gg\,\,m^2$ , where $Q_0$ is the starting scale
for QCD evolution. To find the BFKL asymptotics we have to find a
different limit of the perturbative series, namely,
\beq \label{1.16} 
x_B G(x_B,Q^2)\,\,=\,\,\lim|_{x_B \,\rightarrow\,0}\,\,\,\sum^{\infty}_{n
=0}\,\,\alpha^n_S(Q^2)\,\times\,\,
M_n(x_B,Q^2)\,\,=
\eeq
$$
\sum^{\infty}_{n = 0}\,\,\alpha^n_S(Q^2)\,\times\,\,
lim|_{x_B \,\rightarrow\,0}\,\, M_n(x_B,Q^2 )\,\,.
$$  

The change of the order of two operations: sum and limit, looks even more
suspicious in this case   since we have  no arguments based on
renormalization group. For $M_n $ we have at $x\,\rightarrow\,0$
\beq \label{1.17}
lim|_{x_B \,\rightarrow\,0}\,\, M_n(x_B,Q^2 \,\Longrightarrow
\eeq
$$
\,K_n(Q^2)\,\ln^n(1/x)\,+\,K^{(1)}_n(Q^2) \,\ln^{n - 1}(1/x)\,+\,...
$$
Taking the first term in \eq{1.17} we arrive to the leading log(1/x)
approximation, namely,
\beq \label{1.18}
x_B G(x_B,Q^2)\,\,=\,\,\sum^{\infty}_{n       
=0}\,\,\,K_n(Q^2)\,(\, \alpha_s(Q^2)\ln(1/x)\,)^n\,\,.
\eeq
Actual summation has been done using the BFKL equation \cite{BFKL} and the
answer for 
the series of \eq{1.17} is the leading order solution to the BFKL equation
(LO BFKL). Taking into account term $ K^{(1)}_n(Q^2) \,\ln^{n -1}
(1/x)$ in \eq{1.17} we can calculate a next order correction to the
BFKL equation (NLO BFKL ).

The LO BFKL asymptotics has a Regge - like form:
\beq \label{1.19}
x_B
G(x_B,Q^2)\,\,\,=\,\,\,\sqrt{\,\frac{Q^2}{Q^2_0}\,}\,\,
\sqrt{\,\frac{\pi}{2\,D\,(y - y_0)}}\,\,e^{\omega_L\,(y - y_0
)\,\,-\,\,\frac{(\,r\,\,-\,\,r_0\,)^2}{4\,D\,(\,y\,\,-\,\,y_0\,)}}\,\,,
\eeq
where $r\,=\,\ln(Q^2/\Lambda^2)$ and  $r_0\,=\,\ln(Q^2_0/\Lambda^2)$,
$y\,\,=\,\,\ln(1/x_B)$ and $y_0\,\,=\,\,\ln(1/x_{B,0})$. The value of
$x_{B,0}$, which characterize the energy from which we can start to use
the
BFKL asymptotics,  cannot be evaluated in the BFKL approach. Two
parameters $\omega_L$ and $D$ are well defined by the BFKL kernel ( see
\cite{BFKL} ).

Two terms in the exponent in \eq{1.19} have two different 
meaning in physics: (i) the first one reflects the power - like energy
behaviour and,
therefore,  reproduces the Reggeon - like  exchange at high energy;
(ii) while
the second one describes so called diffusion in log of transverse momentum
and, therefore, manifest itself a new properties of our microscopic
theory - dimensionless coupling constant in QCD.

Below, we will explain in more details all characteristic features of the
BFKL asymptotics. 

{ \bf \item  Regge factorization:}

Regge factorization ( {\bf don't mix up it with ``hard"  factorization} )
is nothing more than \eq{1.9} for a Reggeon exchange. \eq{1.9} says that
the amplitude of elastic or quasi-elastic  ( like an  amplitude for the 
diffraction dissociation reaction ) process can be written as a product 
 \beq \label{1.20}
A(s,t)\,\,=\,\,g\left(\, projectile,\,\, excitation\,\, of\,\, 
projectile\,
\right)
\,\,\bigotimes
\eeq
$$
\,\,g\left(\, target,\,\, excitation\,\,
of\,\,target\,\right)
 \,\,\bigotimes\,\,P\left(\,Reggeon\,\,propagator\,\right)\,\,.
$$
In \eq{1.20} only Reggeon propagator depends on energy $s$ ( and momentum
transfer $t$ ), while all
dependence on quantum numbers and masses as well as on other
characteristics
of produced particles are factorized out in the product of two vertices.
It is easy to see that \eq{1.20} leads to a large number of different
relations between measured cross sections.

I give several examples of such relations to demonstrate how  Regge
factorization works.

1. the first relation which was suggested by Gribov and Pomeranchuk:

\beq \label{1.21}
\sigma_{tot}( \pi \pi )\,\times\,\sigma_{tot}( p p)\,\,
=\,\,\sigma^2_{tot}(\pi\,p )\,\,.
\eeq
Of course, it is impossible to check \eq{1.21} since we cannot measure 
$\sigma_{tot}( \pi \pi )$ but it can be used to estimate its value.

2. 

\beq \label{1.22}
\frac{\sigma^{SD}(\,p + p\, \rightarrow\,
M\,\,+\,\,p\,)\,B^{SD}}{\sigma^{el} ( p p )\,B^{el}}\,\,=\,\,Const (
s)\,\,,
\eeq
 where $B$ is the slope of the cross section in the parameterization
$$
\frac{\frac{d \sigma}{d t} (t)}{\frac{d \sigma}{d t}( t = 0 )}\,\,=\,\,e^{
- B\,|t|}\,\,.
$$

3.

\beq \label{1.23}
\frac{\sigma^{SD}(\,p + p\, \rightarrow\,M\,\,+\,\,p\,)\,
B^{SD}(\,p + p\, \rightarrow\,M\,\,+\,\,p\,)}{\sigma^{SD}(\,p + \pi\,
\rightarrow\,
M\,\,+\,\,\pi\,)\,B^{SD}(\,p + \pi\, \rightarrow\,M\,\,+\,\,\pi\,)}
\,\,=\,\,\frac{\sigma^{el}(pp)\,B^{el}(pp)}{\sigma^{el}(\pi p)\,B^{el}(\pi
p)}\,\,.
\eeq

I think, that these examples are enough to get the spirit what the Regge
factorization can do for you. You can easily enlarge the number of
predictions using \eq{1.20}. 

{ \bf \item  Factorization theorem ( ``hard" factorization):}

Any calculation of ``hard" processes is  based on the factorization
theorem \cite{FACT}, which allows us to separate the nonperturbative
contribution from large distances (parton densities,  $F^i_A(\mu^2)$) from
the
perturbative one (``hard" cross section, $\sigma^{hard}$). For example,
the cross section
of the high $p_t$ jet production in hadron - hadron collisions
 can be written schematically in the form:
\begin{eqnarray} \label{1.24}
\sigma( A + B \rightarrow jets(p_t) + X)\,\,=\nonumber\\
\,\,F^i_A(\mu^2)\, \bigotimes
\,F^i_A(\mu^2)\,\bigotimes\,\sigma^{hard}(partons \,\,with
\,\,p_t\geq k_t\geq \mu)\,\,.
\end{eqnarray}
$\sigma^{hard}$ in \eq{1.24} should be calculated in the framework of
perturbative QCD and not 
only in
the leading order of pQCD but also in high orders so as  to specify the
accuracy
of our calculation. Practically, we need to calculate the ``hard" cross
section  at least in the next to leading order,  to reduce the scale
dependence, which appears in the leading order calculation, as a clear
indication of the low accuracy of our calculation. Of course, we have to
adjust  the accuracy in the calculation of $F(\mu^2)$ and $\sigma^{hard}$.

{ \bf \item  Secondary Reggeons ( trajectories ):}

We call { \it secondary Reggeons or secondary trajectories} all Reggeons
with intercepts smaller than 1.    There is a plenty of different Reggeons
with a variety of the different quantum numbers. However, they have
several features in common:

1. A Reggeon describes the  family  of resonances that lies on the same
Reggeon trajectory $\alpha_R(t)$. Note, that the Pomeron does not describe
any family of resonances. In Fig. 2 you can see the experimental $\rho$
and $f$ trajectories.
\begin{figure}
\centerline{\psfig{file=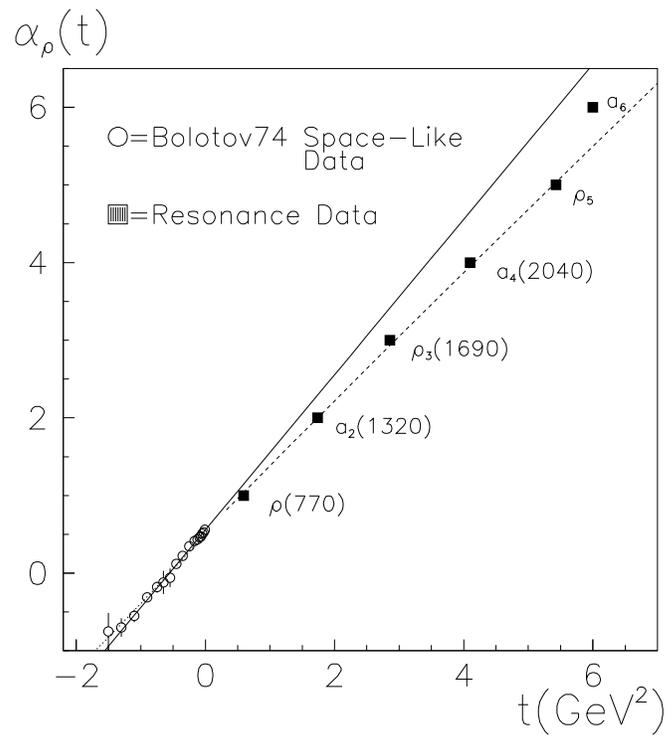,width=100mm}}
\caption{ \em The experimental `` $\rho$",
``$f$"
and ``$a$" trajectories. The picture was taken from Ref.
\protect\cite{PANCHERY}.}
\label{fig2}
\end{figure}

2. All secondary trajectories can be parameterized as 
\beq \label{1.25}
\alpha_R(t)\,\,\,=\,\,\,\alpha_R(0)\,\,\,+\,\,\,\alpha'_R(0)\,t\,\,,
\eeq
both in scattering  ( $t \,<\,0$ ) and  resonance ( $t\,>\,0 $) regions
 (see Ref.\cite{PANCHERY}  for details  and for discussion of corrections
to \eq{1.25} ). It is interesting to note that the value of the slope (
$\alpha'_R(0)$ ) is the same for all secondary Reggeons with the same
quark contents . For the Reggeons that corresponds to the resonances made
of the light quarks $ \alpha'_R(0\,\,=\,\,1\,GeV^{-2}$ ( see Fig. 2).
However, for Reggeons with heavy quark contents the slope is smaller
\cite{PANCHERY}.

3. For Reggeons we have duality between the Reggeon exchange and the
resonance contributions \cite{VEN} shown in Fig. 3.
  
\begin{figure}   
\centerline{\psfig{file=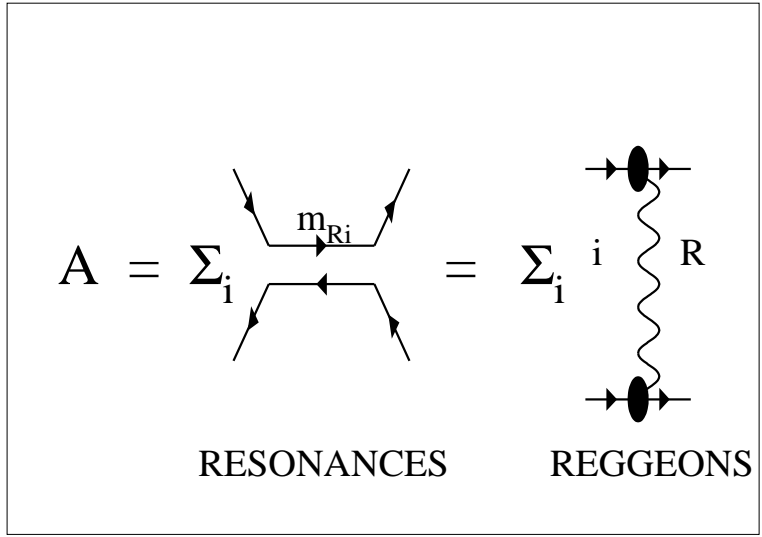,width=100mm}}
\caption{ \em  Duality between the resonances in the  s - channel and
Reggeon exchange in the t - channel.}
\label{fig3}
\end{figure}

The direct consequence of the duality approach is an idea that any Reggeon
can be viewed as the exchange of quark - antiquark pair in the $t$ -
channel ( see Fig.4 ). Therefore, the dynamics of the secondary Reggeons is
closely related to non-perturbative QCD in quark - antiquark sector while
the Pomeron is the result of the non-perturbative QCD interaction in the
gluon sector.

\begin{figure}
\centerline{\psfig{file=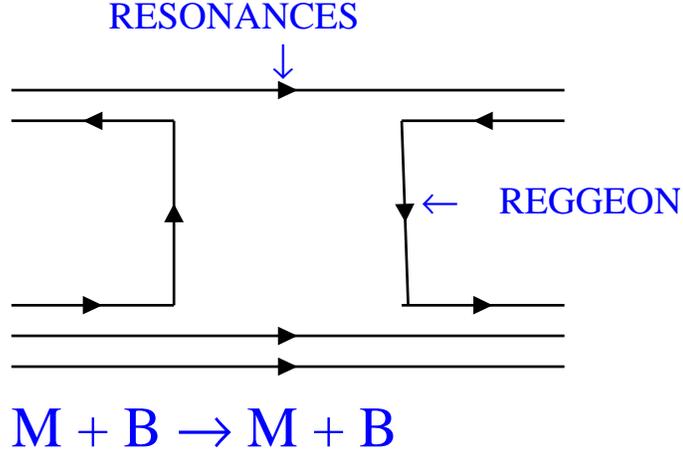,width=100mm}}
\caption{ \em  Quark diagrams for meson - meson and meson - baryon
scattering.}
\label{fig4}
\end{figure}

{ \bf \item  Shadowing Corrections ( SC ):}

To understand what is the shadowing correction ( SC ) and why we have to
deal with them let us look at Fig.1 more carefully.  Actually, the simple
formula of \eq{1.1} is correct only if the flux of ``wee" partons $N$ is
rather small. Better to say, that this flux is such that only one or less
then one of  ``wee" partons has the same longitudinal and
transverse momenta.  However, if $N\,>\,1$, several ``wee" partons can
interact with the target. For example, two ``wee" partons interact with
the target with  a cross section 
\beq \label{1.26}
\sigma^{(2)}\,\,\,=\,\,\,\sigma_0\,\times\,\kappa
\eeq
where    $\kappa$ is the probability for the second ``wee" parton to meet
and interact with the target.  All   ``wee" partons are distributed  in an
area  in the
transverse plane which is equal   $\pi\,R^2(s)$ and
$R^2(s)\,\rightarrow\,\alpha'_P(0) \,\ln (s)$. Therefore, $\sigma^{(2)}$ 
is equal to 
\beq \label{1.27}
\sigma^{(2)}\,\,\,
=\,\,\,\sigma_0\,\times\,N\,\times\,\kappa \,\,\,=
\,\,\,\frac{(\,\s^P_{tot}\,)^2}{\pi\,R^2(s)}\,\,\,.
\eeq 
Now, let us ask ourselves what should be the sign for a $\s^{(2)}$
contributions to the total cross section.   The total cross section is
just the probability that an incoming particle  has at least one
interaction with the target. However, in \eq{1.1} we overestimate the
value of the total cross section since we assumed that every parton out of
the total number of ``wee" partons $N$ is able to interact with the
target.  Actually, in the case of two parton interaction, the second
parton cannot interact with the target if it is just behind the first one.
The probability to find the second parton just behind the first one is
equal $\kappa$ which we have estimated. Therefore,  instead of  flux $N$ of
``wee" partons in  \eq{1.1} we should substitute the renormalized flux,
namely
\beq \label{1.28}
RENORMALIZED
\,\,\,FLUX\,\,\,=
\eeq
$$
\,\,\,N\,\times\,\left(\,1\,\,-\,\,\kappa\,\right)\,\,\,=\,\,\,
N\,\times\,\left(\,1\,\,-\,\,\frac{\sigma^P_{tot}}{\pi\,R^2(s)}\,\right)\,\,.
$$
The difference between {\it flux of ``wee" partons} and the {\it
renormalized  flux of ``wee" partons} if this difference is not very large
we call shadowing and/or screening corrections. In other words,

 { \bf the
shadowing corrections for the Pomeron exchange is the Glauber screening
for the flux of ``wee" partons}.
 
The analogy and terminology become more clear if you notice that
\eq{1.28} leads to Glauber formula for the total cross section, namely,
\beq \label{1.29}
\sigma_{tot}\,\,\,
=\,\,\,\s^P_{tot}\,\,\,-\,\,\,\frac{(\,\s^P_{tot}\,)^2}{\pi\,R^2(s)}\,\,\,.
\eeq
If $\kappa\,\,\ll\,\,1$ we can restrict ourselves to the calculation of
interaction of two ``wee" partons with the target. However, if
$\kappa\,\,\approx\,\,1$  we face a complicated and challenging problem of
calculating all SC. We are far away from any theoretical solution of this
problem especially in the ``soft" interaction. What we know will be
briefly reviewed in this paper.

{ \bf \item  Saturation of the parton ( gluon ) density:}

{\it Saturation of the parton (gluon ) density } \cite{GLR} is a
hypothesis about the renormalized flux of ``wee" partons in the region
where flux $N$ is much large than unity. More precisely, we assume that
the value of $\kappa$ reaches the unitarity maximum ( $ \kappa\,=\,1 $)
and ceases to increase. Fig. 5 gives a picture of the parton density in a
target which corresponds this hypothesis.
From this picture one  sees that the unitarity limit can be reached at
sufficiently large virtualities $Q^2$ ( short distances ) in the region of
applicability of pQCD. It allows us to evaluate better the flux $N$ and
parameter $\kappa$. 

  The expression for $\kappa$ can
be written in the form 
\beq \label{1.30}
\kappa\,\,=\,\,x G(x,Q^2) \frac{\sigma(GG)}{  \pi R^2}\,\,=\,\,
\frac{ 3 \,\pi\,\alpha_S}{ Q^2\,R^2}\,\,xG(x,Q^2)\,\,,
\eeq
where

1.  $ xG(x,Q^2)$ \,\,= \,\, the number of partons ( gluons) in the
parton
cascade ($N$);

2.  $R^2$ is the radius
 of the area populated by gluons in a nucleon;

3.   $\sigma (GG)$ is the
gluon
cross section inside the parton cascade  and
was evaluated in Ref.\cite{MUQI}.

Actually, at present we know $xG$ and $R^2$ well enough
\cite{LERIHC}
to estimate the kinematic region where the flux of ``wee" partons becomes
large. This is the line (see Fig.5 ) 
\beq \label{1.31}
\kappa(x,Q^2)\,\,\,=\,\,\,1\,\,.
\eeq
We will discuss below the physics of this equation.

\begin{figure}
\centerline{\psfig{file=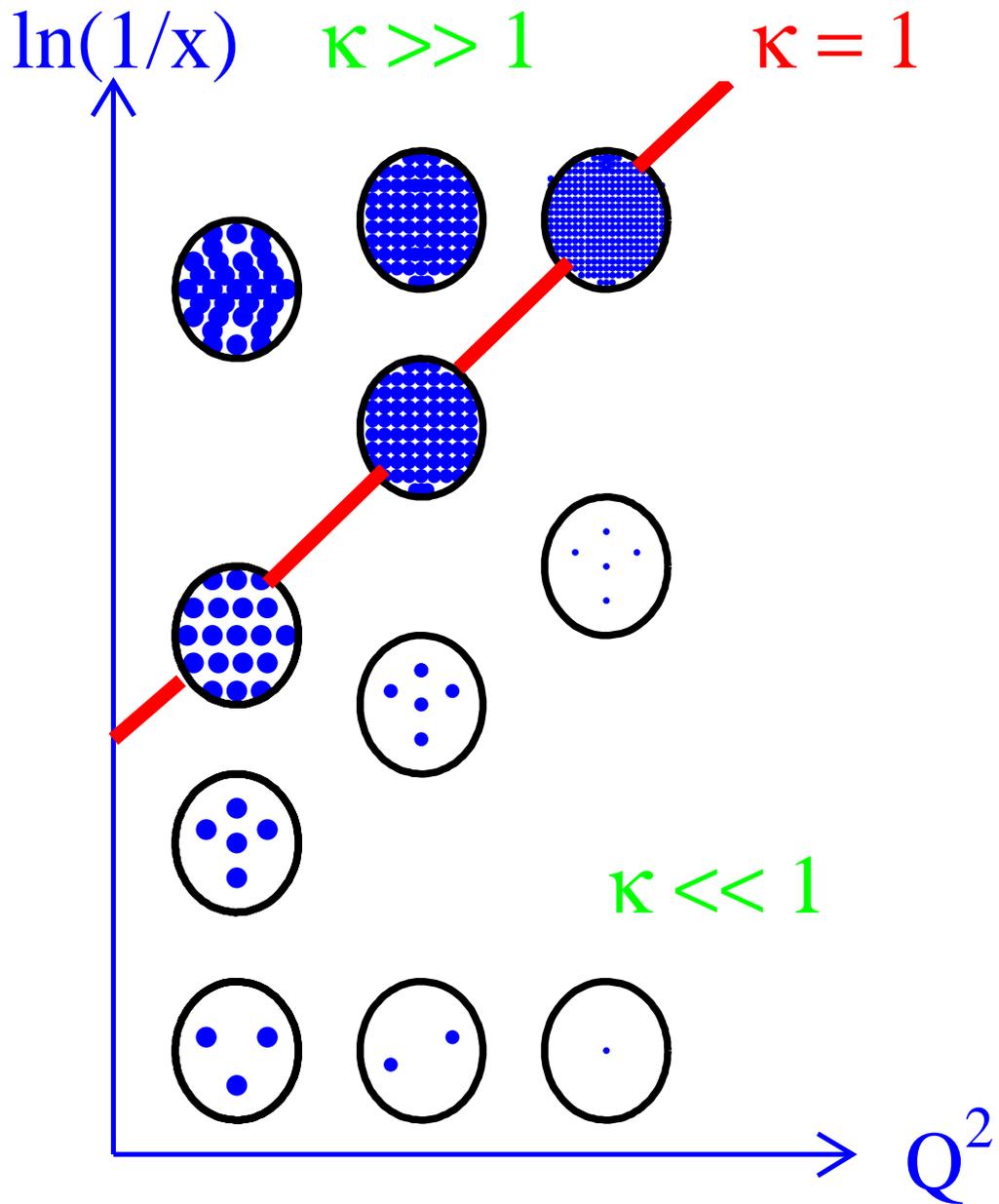,width=140mm}}
\caption{ \em  Parton distributions in the transverse plane as function
of $\ln(1/x)$ and $Q^2$.}
\label{fig5}
\end{figure}

{ \bf \item  Semiclassical gluon field approach:}

In the region where $N\,\,\gg\,\,1$ we can try to develop a different
approach \cite{MCLER}- {it semiclassical gluon field approach}. Indeed,
due to the uncertainty relation between the number of particles and the
phase of the amplitude of their production
\beq \label{1.32}
N\,\times\,\Delta \phi \,\,\geq\,\,1\,\,,
\eeq
we can consider $\Delta \phi \,\,\propto \,\frac{1}{N}\,\,\ll\,\,1$.
Therefore, we can approach such parton system semiclassically.
It means that the parton language is not more suited to discuss 
physics and we have to find an  effective Lagrangian formulated in term of
semiclassical gluonic field. A remarkable theoretical progress has been
achieved in framework of such an approach \cite{MCLER} but we mention this 
rather theoretical approach here only to illlustrate the richness
of ideas that brought to the market the experiments in the deep inelastic
processes mostly done at HERA and the Tevatron.

{ \bf \item  Impact parameter ( $\mathbf b_t$ ) representation:}

For high energy scattering it turns out to be usefull to introduce impact
parameter  $b_t$ by 
$$
l\,\,\,=\,\,k\,\times\,b_t\,\,,
$$
where $l $ is the angular momentum in $s$ - channel of our process and $k$
is the momentum of incoming particle at high energy. High energy means
that we consider the scattering process  for $ k\,\,a\,\,\gg\,\,1$ where
$a$ is the typical size of the interaction. 

The  meaning of $b_t$  as well as  useful kinematic relations
for high energy scattering is pictured in Fig.6.

\begin{figure}
\centerline{\psfig{file=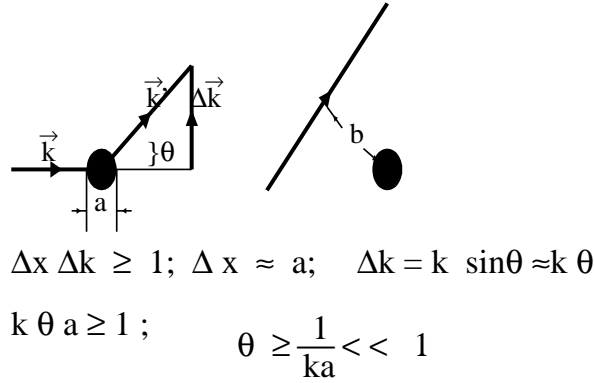,width=100mm}}
\caption{ \em  Impact parameter $b_t$ and useful kinematic relations for 
high energy scattering.}
\label{fig6}
\end{figure}

Formally,  the scattering amplitude in $ b_t $-space is defined as
 \begin{equation} \label{1.33}
 a_{el}(s,b_t) = \frac{1}{2 \pi} \int d^2 q\,\,
e^{-i \vec{q_{\perp}}\cdot\vec{ b}_t}\,\,
  A(s,t)
\end{equation}
where  $A(s,t)$ is scattering amplitude and $ t= - q^{2}_{\perp} $ is
momentum transfer squared.
The inverse transform  for $A(s,t)$ is
\beq \label{1.34}
A(s,t)\,\,=\,\,\frac{1}{2 \pi} \int a_{el}(s,b_t)\,\,d^2 b_t\,\,
e^{i\vec{q}_{\perp}\cdot\vec{b_t}}\,\,.
\eeq
 In this representation
 \begin{equation} \label{1.35}
 \sigma_{tot} = 2 \int d^2 b_t \,\, Im a_{el}(s,b_t)\,\,;
\end{equation}
 \begin{equation} \label{1.36}
 \sigma_{el} = \int d^2 b_t\,\, \vert a_{el}(s,b_t) \vert^{2}\,\,.
\end{equation}

{ \bf \item  $\mathbf s$ - channel unitarity:}

The most important property of the impact parameter representation is the
fact that the unitarity constraint looks simple in this representation,
namely, they can be written at fixed $b_t$. 

 \begin{Large}
\beq \label{UNB} 
2 \,Im \,a_{el}(s,b)\,\,=\,\,| a_{el}(s,b)|^2
\,\,+\,\,G_{in}(s,b)\,\,,
\eeq
\end{Large}   
where $G_{in}$ is the sum of all inelastic channels. This equation is our
master equation which we will use frequently   during this talk.

{ \bf \item  A general solution to the unitarity constraint:}

Our master equation ( see \eq{UNB} ) has the general solution
\beq \label{1.361}
G_{in} (s, b)\,\,=\,\,1\,\,-\,\,e^{- \,\Omega(s,b)}\,\,;
\eeq
$$
a_{el}\,\,=\,\,i\,\{\,1 \,\,-\,\,e^{ - \,\frac{\Omega(s,b)}{2}\,+\,i
\,\chi(s,b)}\,\}\,\,;
$$
where opacity $\Omega$ and phase $\chi = 2 \delta (s,b)$ are real
arbitrary but real functions.

The algebraic operations, which help  to check that \eq{1.36} gives the
solution, are:
$$
Im\, a_{el}\,\,=\,\, 1 \,\,-\,\,e^{-\,\frac{\Omega}{2}} \,cos \chi\,\,;
$$
$$
\vert a_{el} \vert^2\,\,=
\,\,\left( \, 1 \,\,-\,\,e^{-\,\frac{\Omega}{2} \,+\,i\,\chi}
\,\right) \cdot \left( \, 1 \,\,-\,\,e^{-\,\frac{\Omega}{2} \,-\,i\,\chi} 
\,\right)\,\,=\,\,1\,\,-\,\,2\,e^{- \frac{\Omega}{2}}\,cos
\chi\,\,+\,\,e^{- \Omega}\,\,.
$$

The opacity $\Omega$ has a clear meaning in physics, namely  $e^{-\Omega}$ 
is the probability to have no inelastic interactions with the
target.
  
One can check that \eq{UNB} in the limit, when $\Omega$ is  small and the
inelastic processes  can be neglected, describes the well known solution
for the elastic scattering: phase analysis. For high energies the most
reasonable assumption  just opposite, namely,  the real part of the
elastic amplitude to be  very small. It means that $\chi\,\rightarrow\,0$
and  the general solution is of   the form:
\beq \label{1.37}
G_{in} (s, b)\,\,=\,\,1\,\,-\,\,e^{- \,\Omega(s,b)}\,\,;
\eeq   
$$
a_{el}\,\,=\,\,i\,\{\,1 \,\,-\,\,e^{ - \,\frac{\Omega(s,b)}{2}}\,\}\,\,. 
$$
We will use this solution to  the end of this talk. At the moment, I do
not want to discuss the theoretical reason why the real part should be
small at high energy . I prefer to claim that this is a general feature
of all experimental data at high energy.

{ \bf \item   The great theorems:}

\centerline{ \it Optical theorem}

The optical theorem gives us the relationship between the
behaviour
of the imaginary part of the scattering amplitude at zero scattering angle
and
the total cross section that can be measured experimentally.
It follows directly from \eq{UNB}, after  integration  over $b$.
Indeed,
\beq \label{1.371}
4\pi  Im A(s,t =0)\,\,=\,\,\int 2 Im a_{el} (s,b_t) \,d^2 b_t\,\,=
\eeq
$$   
\,\,\int
\,d^2 b_t \{\,| a_{el}(s,b_t)|^2
\,+\,G_{in}(s,b_t)\,\,=\,\,\s^{el} \,+\,\s^{in}\,\,=\,\,\s_{tot}\,\,.
$$
\centerline{ \it Froissart bound}
We call the Froissart boundary the following limit of the energy growth of
the total cross section:
\beq \label{1.38}
\sigma_{tot} \,\,\,\leq\,\,\,C\,\,ln^2 s
\eeq
where $s$ is the total c.m. energy  squared of our elastic reaction:
$a(p_a ) + b (p_b)
 \rightarrow a + b$, namely $s \,=\, (p_a \,+\,p_b )^2$.
The coefficient $C$ has been calculated but we do not need to know  its
exact
value.
What is really important is the fact that $ C \,\propto \frac{1}{k^2_t}$,
where
$k_t$ is the minimum  transverse momentum for the reaction under study.
Since
 the minimum  mass in the hadron spectrum is the pion mass the
Froissart
theorem predicts  that $ C \,\propto \,\frac{1}{m^2_{\pi}}$. The exact
calculation
 gives $C \,= 60 mb$.

The proof\cite{FROISSART}  is based on \eq{1.37} and on the proven
asymptotics for
$\Omega$, namely,
\beq \label{1.39}
\Omega(s,b_t)\,|_{b_t\,\gg\,\frac{1}{\mu}}\,\,\rightarrow\,\,s^N\,e^{- \,2
\,\mu\,b_t}\,\,,
\eeq
where $\mu$ is the mass of the lightest hadron ( pion).

It consists of two steps:

1. We estimate the value of $b^{0}_t $ from the condition
\beq \label{1.40}
\Omega(s,b^{0}_t)\,\,\approx\,\,1\,\,.
\eeq
Using \eq{1.39} one contains
\beq \label{1.41}
b^0_t\,\,=\,\,\frac{N}{2 \mu }\,\ln s\,\,.
\eeq
Note, that at high energies the value of $b^{0}_t$ does  not depend on the
exact value of $\Omega$ in \eq{1.40}.

2.   \eq{1.35} for the total cross section we   integrate over $b_t$ by
dividing the integral
into  two parts $b_t >b^0_t$ and $b_t < b^0_t$. Neglecting  the second
part of the
integral where $\Omega$ is very small yields
$$
\s_{tot}\,\,=\,\,4\,\pi \int^{b^{0}_t}_0 \,b_t db_t \,
[\,1\,\,-\,\,e^{-\frac{\Omega(s,b_t)}{2}}\,]\,\,+\,\,4\,\pi
\int^{\infty}_{b^{0}_t}
\,b_t db_t \,
[\,1\,\,-\,\,e^{-\frac{\Omega(s,b)}{2}}\,]\,\,<
$$
$$
<\,\,4 \pi \int^{b_0}_0 b d b\,\,=\,\,2\,\pi \,b^2_0\,\,=\,\,\frac{2 \pi
N^2}{4 \,\mu^2}\,\ln^2 s\,\,.
$$

This is the Froissart bound.  

\centerline{ \it  Pomeranchuk theorem:}

The Pomeranchuk theorem is the manifestation of the { \bf crossing
symmetry},
 which
can be formulated in the following form: one analytic function of two
variables $s$ and $t$ describes the scattering amplitude of
 two different reactions $ a + b \,\rightarrow
a + b$  at $s > 0 $ and $t < 0 $ as well as $ \bar a + b \,\rightarrow \,
\bar a + b $ at $ s < 0$\, $  (\,u = (p_{\bar a} + p_b )^2 > 0\,) $ and $
t < 0$.

The Pomeranchuk theorem says that the total cross sections of the above
two
 reactions should be equal to each other at high energy
\beq \label{1.411}
\s_{tot} ( a\,+\,b )\,\,=\,\,\s_{tot} (\bar  a\,+\,b
)\,\,\,\,at\,\,\,\,s\,\rightarrow\,\,\infty\,\,,
\eeq
if the real part of the amplitude
is smaller than  imaginary  part.

{ \bf \item  The `` black"  disc approach:}

The rough realization of what we expect in  the Froissart limit at ultra
high energies is so called ``black disc" model, in which we 
 assume that
$$
\Omega\,\,=\,\,\infty\,\,\,\,at\,\,\,\,b\,\,<\,\,R(s)\,\,;
$$
$$
\Omega\,\,=\,\,0\,\,\,\,at\,\,\,\,b\,\,>\,\,R(s)\,\,.
$$  
The radius $R$ can be a function of energy and it can rise as
$R\,\,\approx\,\,\ln s$ due to the Froissart theorem.
  
It is easy to see that the general solution of the unitarity constraints
simplifies  to
$$  
a(s,b)_{el}\,\,=\,\,i\,\Theta( \,R\,-\,b\,)\,\,;
$$
$$
G_{in} (s,b)\,\,=\,\,\Theta( \,R\,-\,b\,)\,\,;
$$
which leads to
\begin{eqnarray} 
&
\s_{in} \,\,=\,\,\int d^2 b G_{in}\,\,=\,\,\pi\,R^2\,\,;
& \label{1.42a}\\
&
\s_{el} \,\,=\,\,\int d^2 b \,\vert a_{el} \vert^2 \,\,=\,\,\pi \,R^2\,\,;
& \label{1.42b}\\
&
\s_{tot}\,\,=\,\,\s_{el}\,\,+\,\,\s_{in}\,\,=\,\,2\,\pi\,R^2\,\,;&
\label{1.42c}
\\
&
A(s,t)\,\,=\,\,\frac{i}{2 \pi}\,\int\,\,d^2 b \Theta( \,R\,-\,b\,)
\,e^{i \vec{q_t}\cdot\vec{b}}\,\,=\,\,i \int^R_0 b d b J_0 (b
q_t)\,\,=\,\,i \frac{R}{q_t}\,J_1 (q_t R)\,\,; & \nonumber\\
&
\frac{d \s}{d t}\,\,=\,\,\pi \vert A \vert^2 \,=\,\pi \frac{J^2_1(  R
\sqrt{|t|})}{|t|}\,\,.
& \label{1.42d}
\end{eqnarray}

Comparison with the experimental data of the ``black disc" model shows
that this rather cruel model predicts all qualitative features of the
experimental data such as the value of the total cross section and the
minimum at certain value of $t$, furthermore   the errors  in the
numerical evaluation is only 200 - 300\%. This model certainly is not good
but it is not so bad as it could be. 

Actually, this model as well as the Pomeron approach is the first model
that we use to understand the new experimental data at high energy.

{ \bf \item  Generalized VDM for photon interaction:}

Gribov was the first one to  observe \cite{GRIBOVPH}  that a photon ( even
virtual one)
fluctuates into a  hadron system with  life time ( coherence length )
$\tau\,\,=\,\,l_c\,\,=\,\,\frac{1}{m x_B}$ where $x_B = \frac{Q^2}{s}$,
$Q^2$ is the 
photon virtuality and $m $ is the mass of the target. This life time is
much larger at high energy than the size of the target and therefore,
we can consider the photon - proton interaction as a processes which
proceeds  in
two stages ( see Fig.7):
\begin{enumerate}
 \item Transition  $\gamma^* \,\rightarrow \,hadrons $  which is
not affected by the target and, therefore, looks similar to  electron -
positron annihilation;

 \item   $hadron \,\,-\,\,target $ interaction,
which can be treated as standard hadron - hadron interaction, for example,
in the Pomeron ( Reggeon ) exchange approach .

\end{enumerate}

\begin{figure}[htbp]
\centerline{\epsfig{file=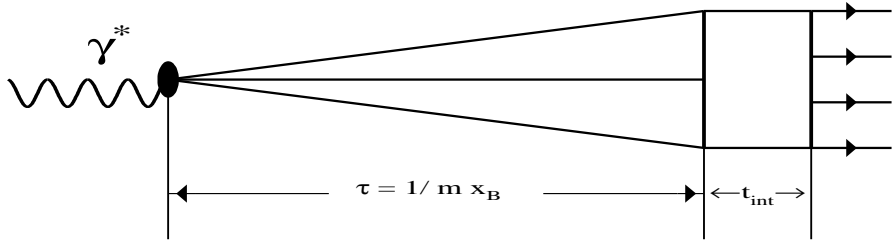,width=140mm,height= 60mm}}
\caption{{\it  Two stages of photon - hadron interaction at
high energy.}}
\label{fig7}
\end{figure}

These two separate stages of the photon - hadron interaction   allow
us to  use a
 dispersion relation with respect to the masses $M$ and $M'$
\cite{GRIBOVPH}
 to describe  the photon - hadron interaction ( see Fig.8 for notations ), 
as the correlation length $l_c =
\frac{1}{m x}\,\gg\,R_N$,  the target size. Based on this idea
we can write a  general formula for the  photon - hadron interaction,

\beq \label{GENGF}
\s(\g^* N )\,\,=\,\,
\frac{\alpha_{em}}{3\,\pi}\,\int \frac{\Gamma(M^2)\,\,d\,M^2}{\,
Q^2\,+\,M^2\,}
\,\,\s(M^2,M'^2,s)\,\,
\frac{\Gamma(M'^2)\,\,d\,M'^2}{\, Q^2\,+\,M'^2\,} \,\,.
\eeq
where $\Gamma^2(M^2) = R(M^2)$ (see \eq{RATIO} ) and $\sigma(M^2,M'^2,s)$
is
proportional to the imaginary part of the forward amplitude  for $ V\, +\,
p \,\rightarrow\,V' \,+\,p$ where $V$ and $V'$ are the vector states with
masses $M$ and $M'$. For the case of the diagonal transition ( $M = M'$)
$\sigma(M^2,s) $ is the total cross section for $V- p$ interaction.
Experimentally, it is known that a diagonal coupling of the Pomeron is
stronger than an off-diagonal coupling. Therefore, in first approximation
we can neglect the off-diagonal transition  and substitute in \eq{GENGF}
$\sigma(M^2,M'^2,s)\,=\,\sigma)M^2.s)\,M^2 \,\,\delta(\,M^2\,-\,M'^2)$.

\begin{figure}
\centerline{\psfig{file= 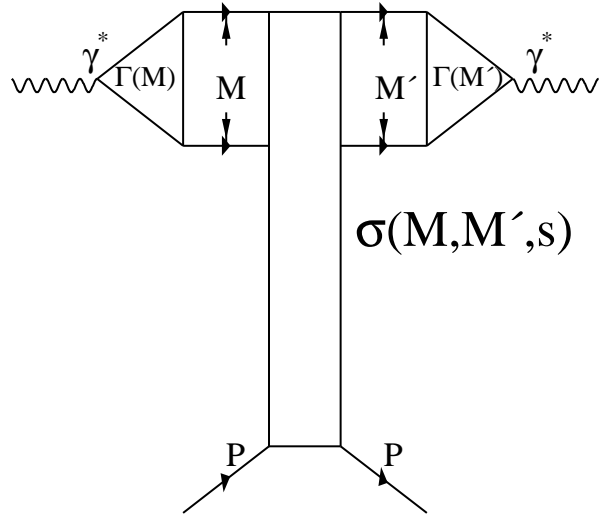,width=100mm}}
\caption{ \it The generalized Gribov's formula for DIS.}
 \end{figure}

The resulting photon - nucleon  cross section can be written written  as:
\beq \label{GRF}
\s(\g^* N )\,\,=\,\,\frac{\alpha_{em}}{3 \,\pi}\,\int\frac{R(M^2)\,M^2\,d
\,M^2}{(\, Q^2\,+\,M^2\,)^2}\,\s_{M^2 N}(s)\,\,,
\eeq
where $R(M^2)$  is defined as the ratio
\beq \label{RATIO}
R(M^2)\,\,=\,\,\frac{\s(e^{+} e^{-} \,\rightarrow\,hadrons)}{\s(e^{+}
e^{-}\,\rightarrow\,\mu^{+} \mu^{-} )}\,\,.
\eeq
The  notation is illustrated  in Fig.8 where $M^2$ is the mass squared of
the  hadronic system, $\Gamma^2(M^2)\,=\,R(M^2)$  and  $\s_{M^2 N}(s)$
 is the cross section for the hadronic system to scatter off the
nucleonic target.

 Experimentally,  $R(M^2)$ can be described in a  two component picture:
the
contribution
of resonances such as $\rho, \omega, \phi$,J/$\Psi$, $\Psi'$ and so on and
the contributions from  quarks which give a  more or  less constant term
changes abruptly with every  new open quark - antiquark channel, $R(M^2)\,
\approx\,3\,\sum_q \,e^2_q$, where $e_q$ is the charge of the quark and
the summation  is done over  all active quark team.

If we take into account only the contribution of the $\rho$ - meson,
$\omega$ - meson ,  $\phi$  and J/$\Psi$ resonances
   in $R(M^2)$  we obtain the so called vector dominance model
( VDM) \cite{JJS}  which gives for the total cross section the
following formula:
\beq \label{VDM}
\s_{tot}(\gamma^*\,+\,p\,)\,\,=\,\,\frac{\alpha_{em}}{3
\pi}\,\,\s_{tot}( \rho + p )\sum_i\,\,R(M^2=m^2_{V_i})
\{\,
\frac{m^2_{V_i}}{Q^2\,\,+\,\,m^2_{V_i}}\,\}^2\,\,
\eeq
where $m_{V_i}$ is the mass of vector meson, $Q^2$ is the value of the
photon virtuality and $ R(M^2 =m^2_{V_i})$ is the value of the $R$ in the

mass of the vector meson which can be rewritten through the ratio of the
electron - positron decay width to the total width of the resonance.
Of course, the summation in \eq{VDM} can be extended to all vector
resonances \cite{SS} or even we can return to a
general approach of \eq{GENGF} and write a model for the off-diagonal
 transition between the vector meson resonances with different masses
\cite{SH}.  We have two problems with all generalization
of the VDM: (i)  a number of unknown experimentally values
such as masses and electromagnetic width of the vector resonances with
higher masses than those that have been included in VDM  and  (ii) a lack
of   theoretical constraints on all mentioned observables. These two
reasons
 give so much
freedom in fitting of the
experimental data on photon - hadron interaction that it becomes
uninteresting and non - instructive.
One can see from \eq{VDM}, that if $Q^2\,\gg\,m^2_V$  VDM predicts a
$\frac{1}{Q^4}$
behaviour of the total cross section. Such a behaviour is in clear
contradiction with the experimental data which show an  approximate 
$\frac{1}{Q^2}$ dependence at large values of $Q^2$ ( i.e. scaling ).

The solution of this puzzle as well as the systematic description of the
photon - proton interaction can be reached on the basis of Gribov formula
but developing a general description both ``soft" and ``hard" processes.
The first and oversimplified version of such description was suggested by
Bodelek and Kwiecinski\cite{BK} and more elaborated approaches have been
recently
developed by Gotsman, Levin and Maor \cite{GLMPH} and Martin, Ryskin and
Stasto \cite{MRSPH}. 

{ \bf \item  Difftraction dissociation:}

The experimental definition of the diffractive dissociation processes:

{\bf Diffractive dissociation processes are processes of production one
( single diffraction (SD)) or
two groups of hadrons   ( double diffraction ( DD ) ) with masses (
$\mathbf M_1$
and $\mathbf M_2$ in Fig.9 ) much less than the total energy (
$ \mathbf M_1\,\ll\,s $ and
$\mathbf M_2\,\ll\,s $ ).}

In diffractive processes no hadrons are produced in the central region of
rapidity as it shown in Fig.8. This is the reason why diffractive
dissociation is the simplest process with large rapidity gap ( LRG ).

\alphfig

\begin{figure}
\centerline{\psfig{file= 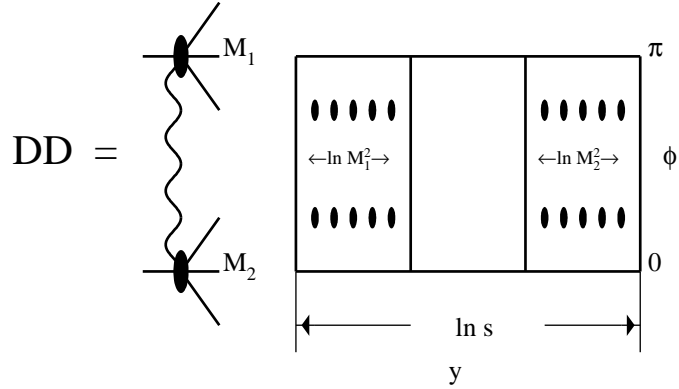,width=100mm}}
\caption{ \it Lego - plot for double diffractive dissociation.}
\label{fig9} 
\end{figure}

The above definition of  diffractive processes is the practical and/or
experimental one. From theoretical point of view as was suggested by
Feinberg \cite{FEIN} and Good and Walker \cite{GW} diffractive
dissociation is a typical quantum mechanical process which  stems from the
fact that the hadron states are not diagonal with respect to the strong
interaction. Let us consider this point in more details, denoting the wave
functions which are diagonal with respect to the strong interaction by
$\Psi_{n}$. Therefore, the amplitude  of high energy interaction is
equal to 
\beq \label{DD1}
A_n\,\,=\,\,< \Psi_n | T | \Psi_{n'}
>\,\,=\,\,A_n\,\delta_{n, n'}\,\,,
 \eeq
 where brackets denote all needed integration and $T$ is the scattering
matrix.  The wave function of a hadron is equal to
\beq \label{DD2}
 \Psi_{hadron}\,\,\,=\,\,\,\sum^{\infty}_{n=1} \,C_n \Psi_n\,\,.
\eeq
Therefore, after collision  the scattering matrix $T$ will give you 
a new wave function, namely
\beq \label{DD3}
\Psi_{final}\,\,\,=\,\,\,\sum^{\infty}_{n = 1}\,\sum^{\infty}_{n'=1}\,C_n  
< \Psi_n | T | \Psi_{n'}>  \Psi_{n'}\,\,=\,\,\sum^{\infty}_{n = 1}\,C_n
\,A_n\,\Psi_n\,\,.
\eeq    
One can see that \eq{DD3} leads to elastic amplitude
\beq \label{DD4}
a_{el}\,\,=\,\,< \Psi_{final} | \Psi_{hadron} >\,\,=\,\,\sum^{\infty}_{n =
1}\,\,C^2_n \,A_n\,\,,
\eeq
and to another process, namely, to production of other hadron state  since 
$\Psi_{final}\,\,\not\equiv \,\,\Psi_{hadron}$.
This process  we call diffractive dissociation.  The total cross section
of such diffractive process we can find from \eq{DD3} and it is equal to
\beq \label{DD5}
\s^{SD}(s,b_t)\,\,=\,\,\sum_{final}< \Psi_{final} | \Psi_{hadron}>^2
\,\,-\,\,< \Psi_{hadron} | \Psi_{hadron} >^2\,\,=
\eeq
$$\,\,
\sum^{\infty}_{n = 1}\,C^2_n \,A^2_n ( s,
b_t)\,\,-\,\,\left(\,\sum^{\infty}_{n= 1}\,\,C^2_n \,A_n (s, b_t
)\,\right)^2\,\,,
$$
where we regenerate our usual variable: energy ($s$) and impact parameter
( $b_t$ ). Using the normalization condition for the hadron wave function
( $ \sum_n \,C^2_n\,\,=\,\,1 $ ) we  can see that \eq{DD5} can be reduced
to the form \cite{GW}\cite{PUM}
\beq \label{DD6}
\s^{SD}(s,b_t)\,\,=\,\,< | \s^2(s,b_t) | >\,\,-\,\,<| \s (s, b_t ) |
>^2\,\,,
\eeq
where $<| f | >\,\,\equiv\,\,\sum_n C^2_n f_n$.

\eq{DD6} has clear optical analogy which clarify the physics of
diffraction, namely, the single- slit diffraction of ``white" light. 
Indeed, everybody knows that in the central point all rays with different
wave lengths arrive with the same phase and they give a ``white " maximum.
This maximum is our elastic scattering. However, the conditions of all
other maxima on the screen with displacement $a$, depend on the wave
lengths  ( $ a  sin \theta \,=\,k \,\lambda $ ). All of them are different
from the central one and they have different colours. Sum of all these
maxima gives the cross section of the diffractive dissociation.

{ \bf \item  Pumplin bound for SD:}

The Pumplin bound \cite{PUM} follows directly from unitarity constrain of
\eq{UNB} which we can write for each state $n$ with wave function $\Psi_n
$ separately:
\beq \label{PB1}
2\,\,Im A^{el}_n (s, b_t )\,\,=\,\,| A^{el}_n (s, b_t )|^2\,\,
+\,\,G^{in}_n (s, b_t)\,\,.
\eeq
Assuming that the amplitude at high energy is predominantly imagine, we
obtain, as has been discussed, that 
\beq \label{PB2}
| A^{el}_n (s, b_t )|^2\,\,\leq\,\,\,G^{in}_n (s,
b_t\,\,\leq\,\,\sigma^{tot}_n( s,b_t )\,\,.
\eeq
After summing over all $n$ with weight $C^2_n$ ( averaging ) one obtains
the Pumplin bound:
\beq \label{PB3}
\s_{el} (s, b_t)\,\,+\,\,s^{SD}( s, b_t)\,\,\leq\,\s_{tot} (s, b_t)\,\,,
\eeq
where 
$\s_{el} (s. b_t )\,\,=\,\,| a_{el} (s,b_t ) |^2$, $\s_{tot} (s,
b_t)\,\,=\,\,2\,Im \,a_{el} (s, b_t)$ and $ s^{SD}( s, b_t)\,\,=\,\,
\sum_{M} | a_{ M} (s,b_t ) |^2$ .

{ \bf \item  Unitarity limit of SD for hadron and photon induced
reactions:}
 
At high energy  $ A^{el}_n\,\,\Longrightarrow\,\,1$ ( see \eq{1.38} ).
It means that at high energy the scattering matrix does not depend on $n$.
Therefore from \eq{DD3} one can derive 
\beq \label{UL1}
\Psi_{final}\,\,\,=
\eeq
$$
\,\sum^{\infty}_{n = 1}\,\sum^{\infty}_{n'=1}\,C_n
< \Psi_n | T | \Psi_{n'} >\,\,  \Psi_{n'}\,\,=\,\,\sum^{\infty}_{n =
1}\,C_n    
\,\Psi_n\,\,\,\equiv\,\,\Psi_{hadron}\,\,.
$$ 
\eq{UL1} says that $\s^{SD} \,\,\Longrightarrow\,\,0$ at high energies.
Performing integration over $b_t$ it is easy to obtain that in the hadron
induced reactions
\beq \label{UL2}
\frac{\s^{SD}}{\s_{el}}\,\,\,\propto \,\,\frac{1}{\ln s}\,\,
\Longrightarrow\,\,0.
\eeq
It should be stressed, that  we should be careful with  this statement for
photon ( real or
virtual ) induced reactions. As we have seen, in Gribov formula we have an
integration over mass. For each mass we have the same property of the
diffractive dissociation as has been presented in \eq{UL2}. However,
due to integration over mass in \eq{GRF}the ratio of the total
diffractive dissociation process to the total cross section is equal to
\beq \label{UL3}
\frac{\s^{DD}_{tot} ( \g^* p \,\rightarrow\, M + p )}{\s_{tot}( \g^* p
\,\rightarrow\, \g^* + p) }\,\,\,\Longrightarrow\,\,\,\frac{1}{2}\,\,.
\eeq
 
{\bf \item Survival Probability of Large Rapidity Gaps:}

 We call any process a large rapidity gap ( LRG )process if in a
sufficiently large rapidity region no hadrons are produced.  Historically,
both Dokshitzer et al. \cite{KHOZE} and Bjorken \cite{BJLRG}, suggested
LRG as a signature for Higgs production in W-W fusion process in
hadron-hadron collisions.
 The definition of the survival probability  ( $< S^2 > $ ) is clear from
Fig. 9-b. Indeed, let us consider the typical LRG process - production of
two jets with large transverse momenta
$\vec{p}_{t1}\,\approx\,-\,\vec{p}_{t2}\,\,\gg\,\,\mu$ , where $\mu$ is
the typical mass scale of  ``soft" processes , and   with LRG between
these two jets. Therefore, we consider the reaction:
\beq \label{LRG1}
p (1)\,\,+\,\,p(2)\,\,\longrightarrow
M_1[ hadrons\,\,+\,\,jet_1(y_1,p_{t1})]
\eeq
$$
+\,\,LRG[ \Delta y = | y_1 - y_2|]
\,\,+\,\,M_2[ hadrons\,\,+\,\,jet_2(y_2,p_{t2})]\,\,,
$$
where $y_1$ and $y_2$ are rapidity of jets and   LRG $\Delta y = | y_1 -
y_2 |\,\,\gg\,\,1$.  To produce two jets with LRG between them 
we  have two possibility:
\begin{enumerate}
\item  This LRG appears as a fluctuation in the typical inelastic event.
However, the probability of such a fluctuation is proportional to 
$ e^{ - \frac{\Delta y}{L}}$ and the value of the  correlation length 
$L$ we can evaluate $L\,\approx\,\frac{1}{\frac{d n}{d y}}$, where
$\frac{d n}{d y} $ is the number of particles per unit in rapidity.
Therefore, LRG means that $\Delta y \,\gg\,L$\,\,;

\item The exchange of colourless gluonic state in QCD is responsible for
the
LRG. This exchange we denote as a Pomeron in Fig.9-b. The ratio the cross
section due to the Pomeron exchange to the typical inelastic event
generated by the gluon exchange ( see Fig. 9-b ) we denote $F_s$. Using
the simple QCD model for the Pomeron exchange, namely, the Low-Nussinov
\cite{LN} idea that the Pomeron = two gluon exchange,   Bjorken gave the
first estimate for $F_s \,\approx\,\,0.15\%$. This is an interesting
problem to obtain a more consistent theoretical estimates for $F_s$, but
$F_s$ is not the survival probability. $F_s$ is the probability of the LRG
process in a single parton shower. 
\end{enumerate}
However, each parton in the parton cascade of a  fast hadron could create
its own parton chain which could interact with one of the parton of the
target. Therefore, we expect a large contribution of inelastic processes
which can fill the LRG. To take this multi shower interaction we introduce
the survival probability $< S^2 >$ ( see Fig.9-b ).

To calculate  $< S^2 >$  we need to find the probability that all partons
with rapidity  $y_i\,>\,y_1$ in the first hadron ( see \eq{LRG1} and
partons with $y_j\,<\,y_2$ in the second hadron do not interact
inelastically and, therefore,  they cannot produce hadrons in the LRG.
As we have discuss, the  meaning of \eq{1.38}  in physics is that 
\beq \label{LRG2}
P(y_i - y_j, b^{ij}_t)\,\,=\,\,e^{- \,\,\Omega( y_i - y_j, b^{ij}_t )}
\eeq
gives you the probability that two partons with rapidities $y_i$ and $y_j$
and with $b^{ij}_t\,\,=\,\,| \vec{b}_{t, i}\,-\,\vec{b}_{t,j} |$  do not
interact inelastically.

The general formula for  $< S^2 >$ reads
\beq \label{LRG3}
< S^2 >\,\,\,=\,\,\frac{\int\, d y_i\,d y_j \,d^2 b^{ij}_t  
\,\,P_1(y_i,y_1,b_{t,i})\,P_2(y_j,y_2,b_{t,j})\,\,P(y_i - y_j,
b^{ij}_t)}{\int\, d y_i\,d
y_j \,d^2 b^{ij}_t\,\,P_1(y_i,y_1)\,P_2(y_j,y_2)}\,\,,
\eeq
where $P_1( y_i,y_1 ) ( P_2( y_j,y_2 )$ is probability to find two partons
with rapidity $y_1$ and $y_i$ ( $y_2$ and $y_j$ ) in hadron $1 $ ( $ 2$ ),
respectively. The deep inelastic structure function is $x_1
G(x_1,p^2_{t,1})\,\,=\,\,\int d y_i\,d^2 b_{t,i}\, \, P_1 (y_1 = \ln
(1/x_1 ) ,y_i,b_{t,i} ) $.
 
Unfortunately, there is no calculation using a general formula of
\eq{LRG3}. What we have on the market is the oversimplified calculation
in the Eikonal model ( see Ref. \cite{GOTSMAN} ) in which we assume that
only the fastest partons from both hadrons can interact.

\begin{figure}[h]
\centerline{\psfig{file= 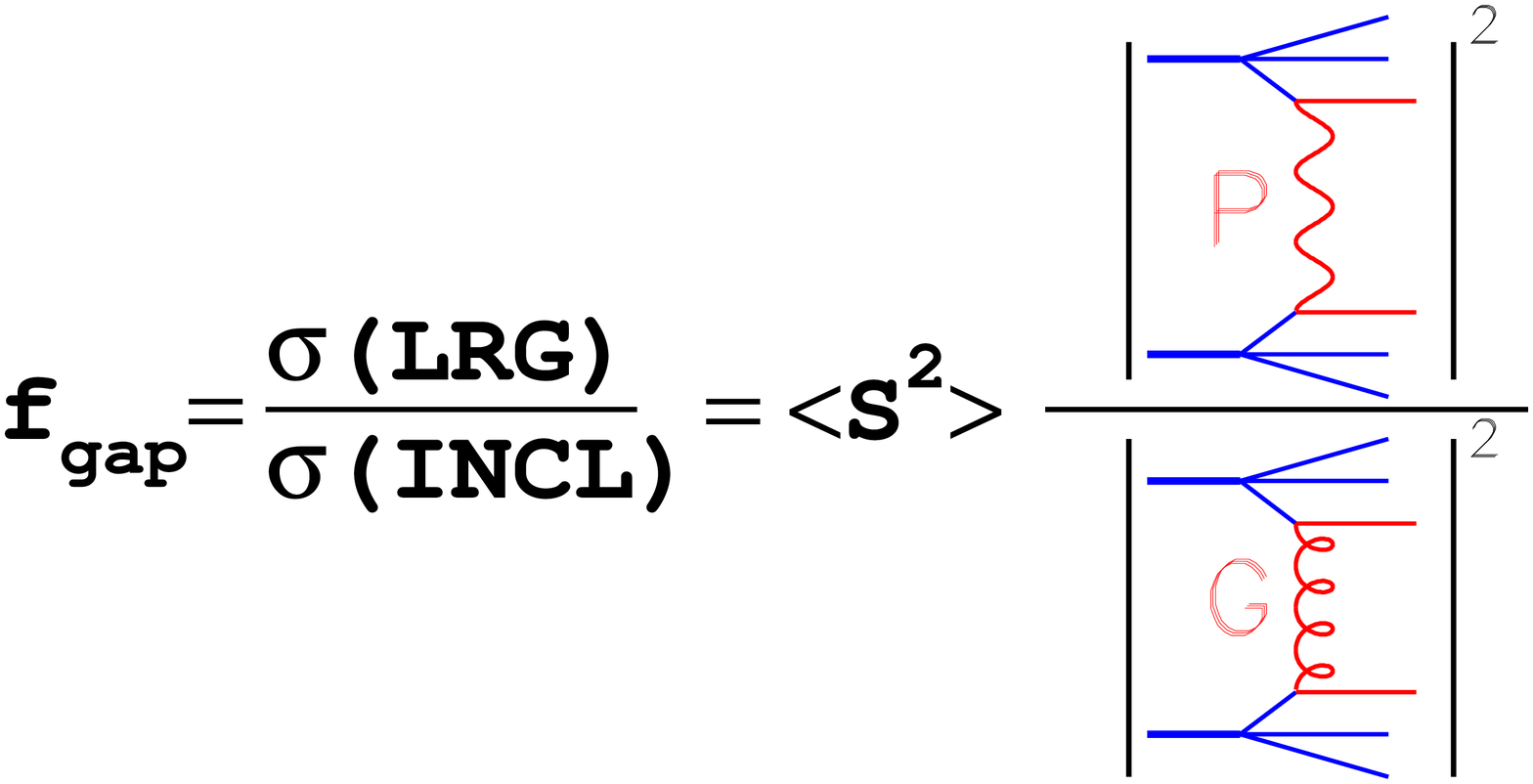,width=100mm}}
\caption{\it Typical LRG process .} 
\end{figure}

\resetfig
\end{enumerate}
\newpage   
\centerline{\Huge \bf I ``S O F T" \,\,\,P O M E R O N:}

\section{ 
``Theorems" about Pomeron}
In this section we collect main properties of the ``soft" Pomeron which we
have discussed in the previous section.
\begin{itemize}
{ \bf \item  There is only one Pomeron.}

It means that only ``soft" Pomeron has a chance to be a simple Regge pole
(Reggeon) ( see previous section for details).

{\bf \item  The  Pomeron is the Reggeon with
$\mathbf \Delta_P(0)\,=\,\alpha_P(0) - 1 \,\ll\,1$}

Being a Reggeon, the Pomeron has all typical features of a Reggeon written
in  \eq{1.9} and presented in Fig. 10. They are:

\begin{figure}
\centerline{\psfig{file= 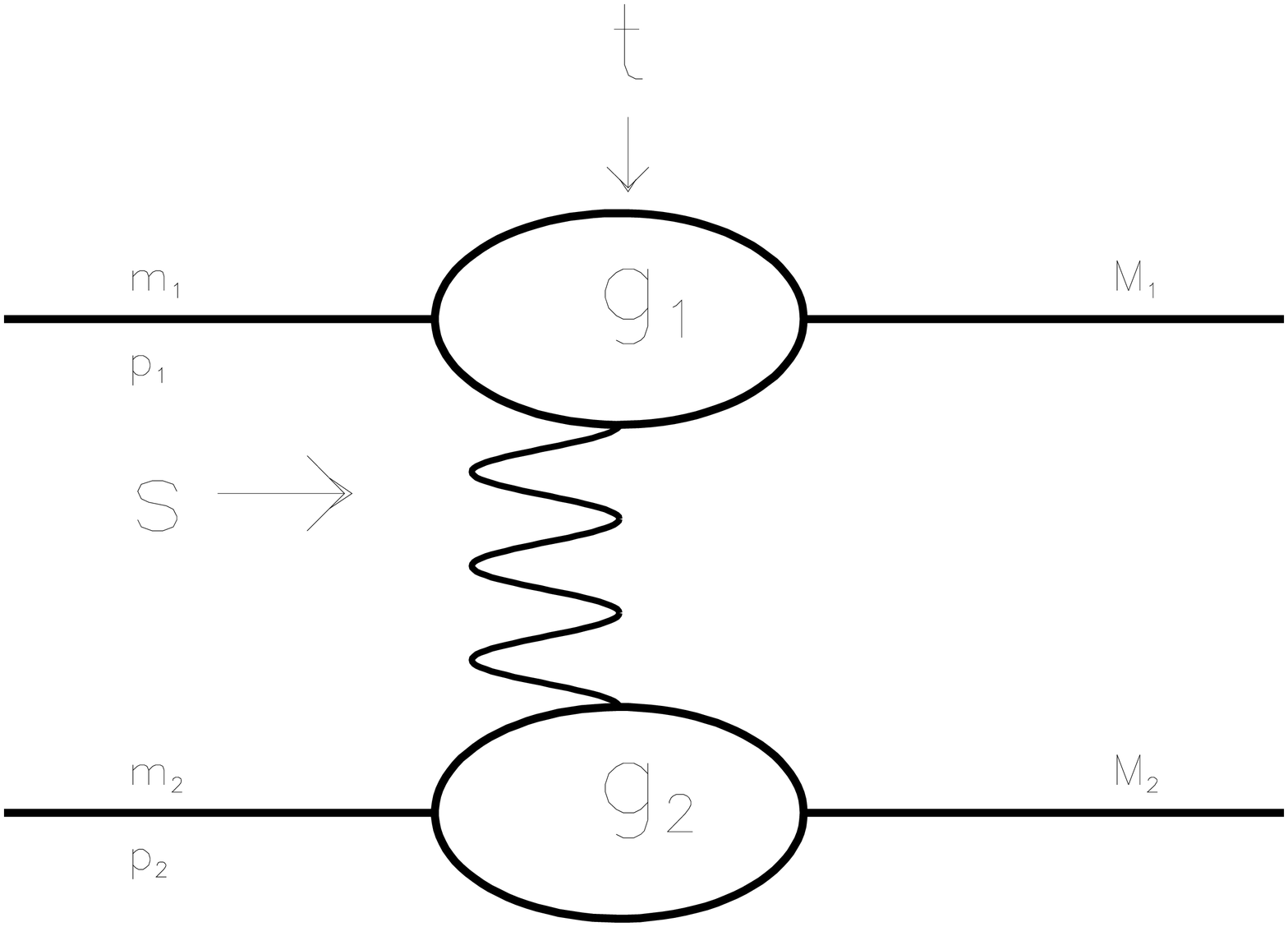,width=100mm}}
\caption{ \it The exchange of the Pomeron.}
\label{fig10}
\end{figure}

$\star$ Analiticity\,\,;

$\star$ Resonances\,\,;

$\star$ Factorization\,\,;

$\star$ Definite phase\,\,;

$\star$ Increase of the radius of interaction\,\,;

$\star$ Shrinkage of the diffraction peak\,\,;

\centerline{ \it Increase of the radius of interaction:}

\eq{1.9} can be rewritten in the impact parameter representation taking
the integral of \eq{1.33} as  follows
\beq  \label{P1}
a_{el}(s,b_t)\,=
\,g_1(0)\,g_2(0)(s/s_0)^{\Delta_P}
\frac{1}{\pi R^2(s)}e^{-
\frac{b^2_t}{R^2(s)}}
\eeq
with
$$
R^2(s)\,=\,R^2_0\,+\,4\,\alpha'_P\,\ln(s/s_0)
$$

One can see, that the typical impact parameters turns out to be of the
order of $R(s)$. This fact means that the radius of interaction increases
with energy.

 \centerline{\it Shrinkage of the diffraction peak:}

Using \eq{1.9} we can see that the $t$-dependence  is determined mostly by 
the Pomeron exchange since
\beq \label{P2} 
\frac{\frac{d \sigma}{d t}}{\frac{d \sigma}{d t}\,|_{t =
0}}\,\propto\,e^{-\,2 \alpha'_P \,\ln(s/s_0)\,|t|}\,\,.
\eeq
Therefore, the differential cross section falls down rapidly for
$|t| \,\,\geq\,\,\frac{1}{\alpha'_P(0)\,\ln(s/s_0)}$. This phenomena we
call
the
shrinkage of the diffraction peak. It should be stressed that this
shrinkage has been observed in many ``soft" reactions, but 
it seems to me that the value of $\alpha'_P(0)$ has not been determined
yet from the experimental data.

{ \bf \item  Pomeron\,\,=\,\,gluedynamics at high energy .}

There is no resonances on the Pomeron trajectory.
This is the principle difference between the secondary Reggeons and the
Pomeron. In duality approach the  Pomeron does not appear in dual
diagrams of the first order and

$$
P\,\,\propto\,\,\frac{\alpha_S}{N_c}\,(\,\frac{s}{s_0}\,)^{C
\alpha_S}\,\,.
$$
The common believe, which is based on the experience with duality approach
and QCD calculations, is that 
\centerline{ \it The Pomeron $\longrightarrow $ glueballs $\longrightarrow
$ gluon interaction at high energy.}

{ \bf \item  There is no  arguments for   a
Pomeron    but  it describes experimental data quite well}

\begin{figure}
\centerline{\psfig{file= 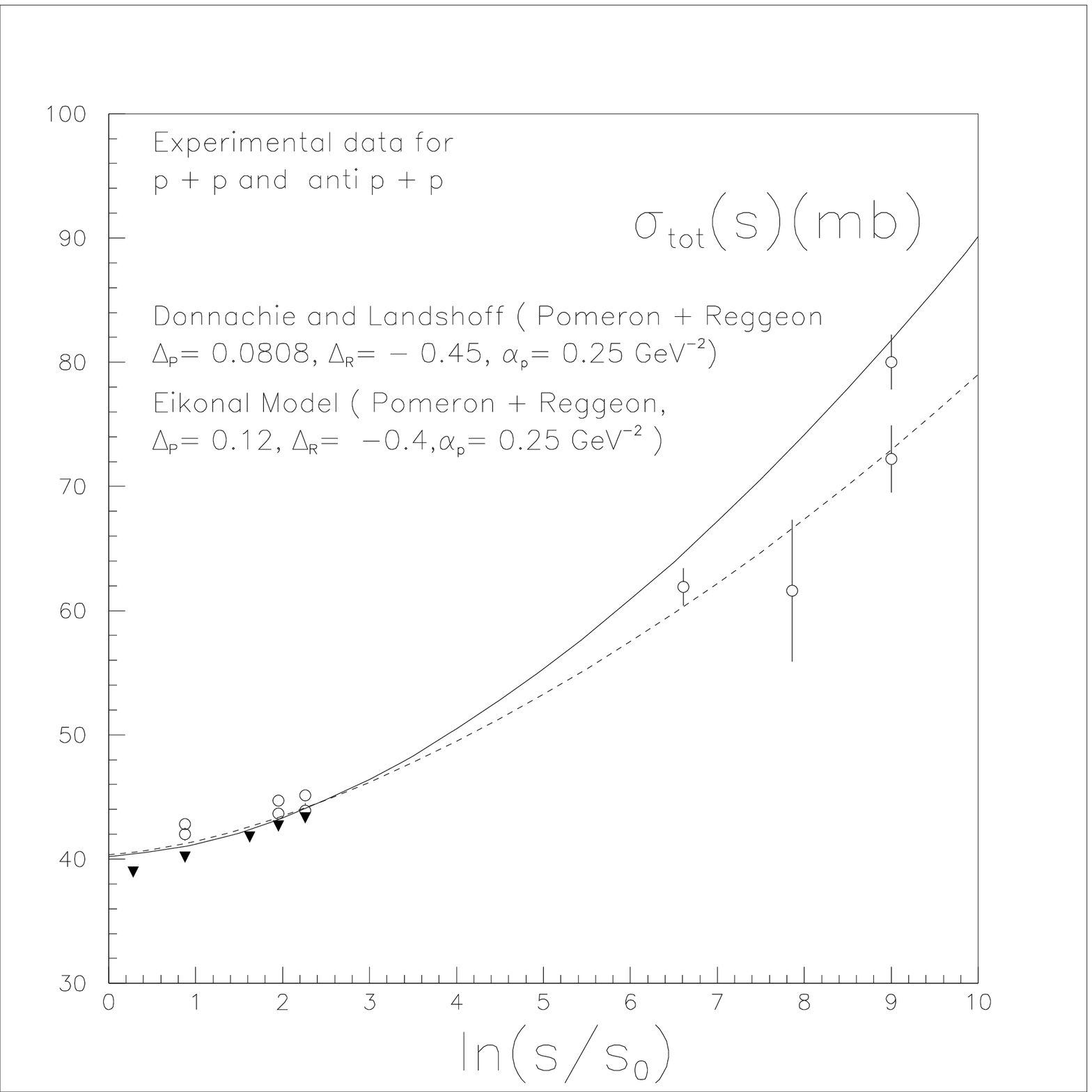,width=100mm}}
\caption{ \it The total cross section in the Pomeron approach in the
Donnachie - Landshoff approach ( full line ) and in the Eikonal model
(dotted line)}
\label{fig11}
\end{figure}
Fig.11 shows that the Pomeron exchange is able to describe the
experimental data on total proton - proton scattering at high energy.
Of course, this is an example. More systematic description of the data one
can find in Ref.\cite{TABLE} or/and in original Donnachie - Landshoff
papers \cite{DL}.

{ \bf \item Pomeron is a ``ladder" diagram for a superconvergent theory
like $\mathbf g \phi^3$.}

The charm of the Pomeron approach is in the relation between the elastic
or / and quasi - elastic reactions ( like diffractive dissociation, for
example ) and the multiparticle production at high energy. The simple
picture for the Pomeron exchange presented in Fig.12 was, is and,
unfortunately, will be our practical tool for application of the Pomeron
phenomenology to the processes of the multiparticle production.
 
\begin{figure} \centerline{\psfig{file= 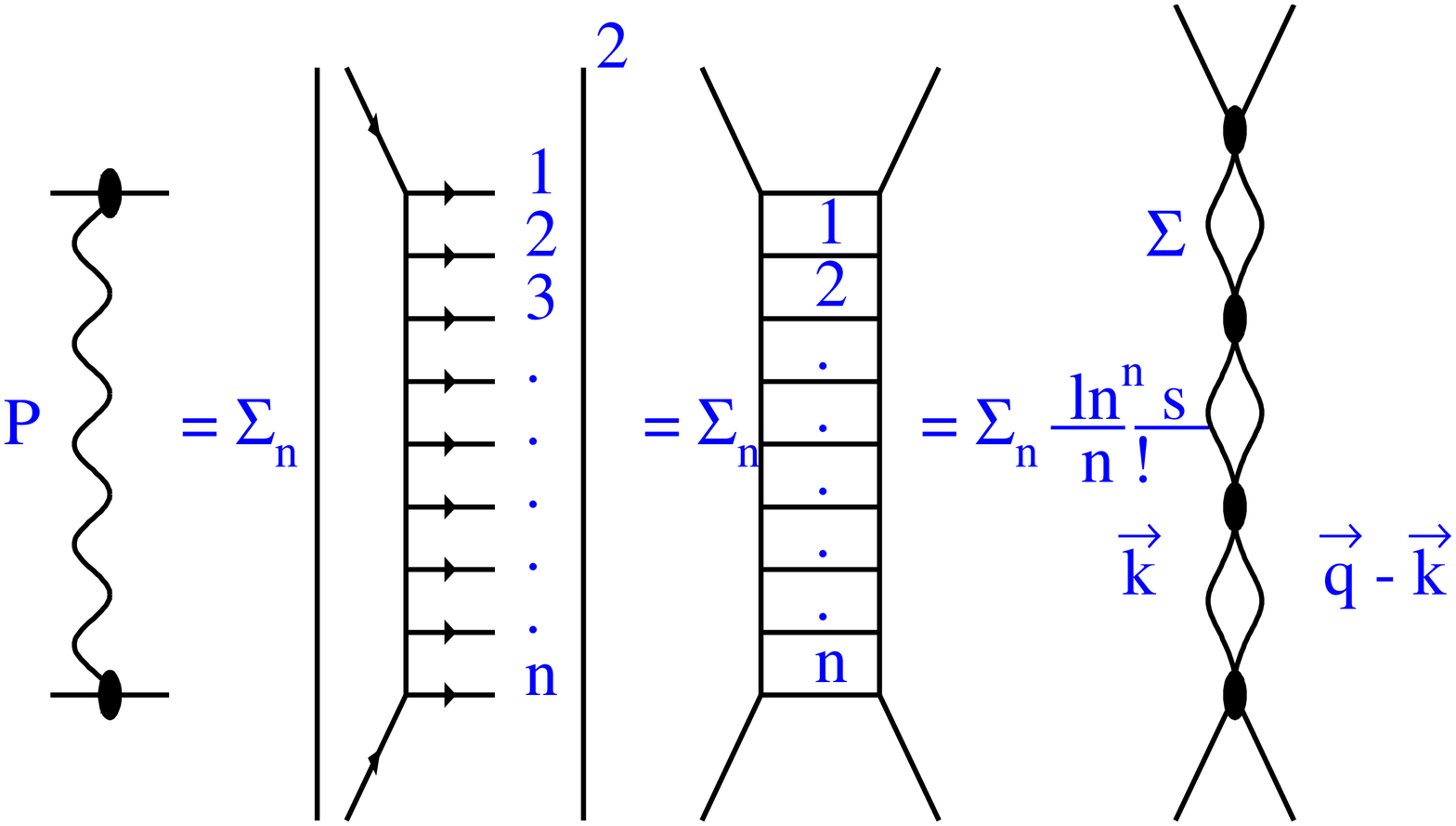,width=160mm}}
\caption{ \it The Pomeron structure in  $ g \phi^3$ - theory ( parton
model ).}
\label{fig12}
\end{figure}

In spite of the simplicity of this picture we would like to mention that it
appears in simple but strict approximation, so called leading log s
approximation. In this approximation we sum all contributions to
Feyman
diagrams of the order of $(\, \frac{g^2}{m^2} \ln s\, )^n$, neglecting
smaller
terms.

It should be stressed that this approach reproduces the main property of
observed multiparticle production at high energies.

$\star$  Power - like behaviour of the total cross section
$$ \sigma_{tot}\,\propto\,(\frac{s}{s_0})^{\Sigma(q^2)}\,\,;$$

$\star$ Pomeron  = only multiparticle production with $< N > 
\,\propto\,\ln s $\,\,;

$\star$  Uniform rapidity distribution of the produced
particles\,\,;

\begin{figure} 
\centerline{\psfig{file= 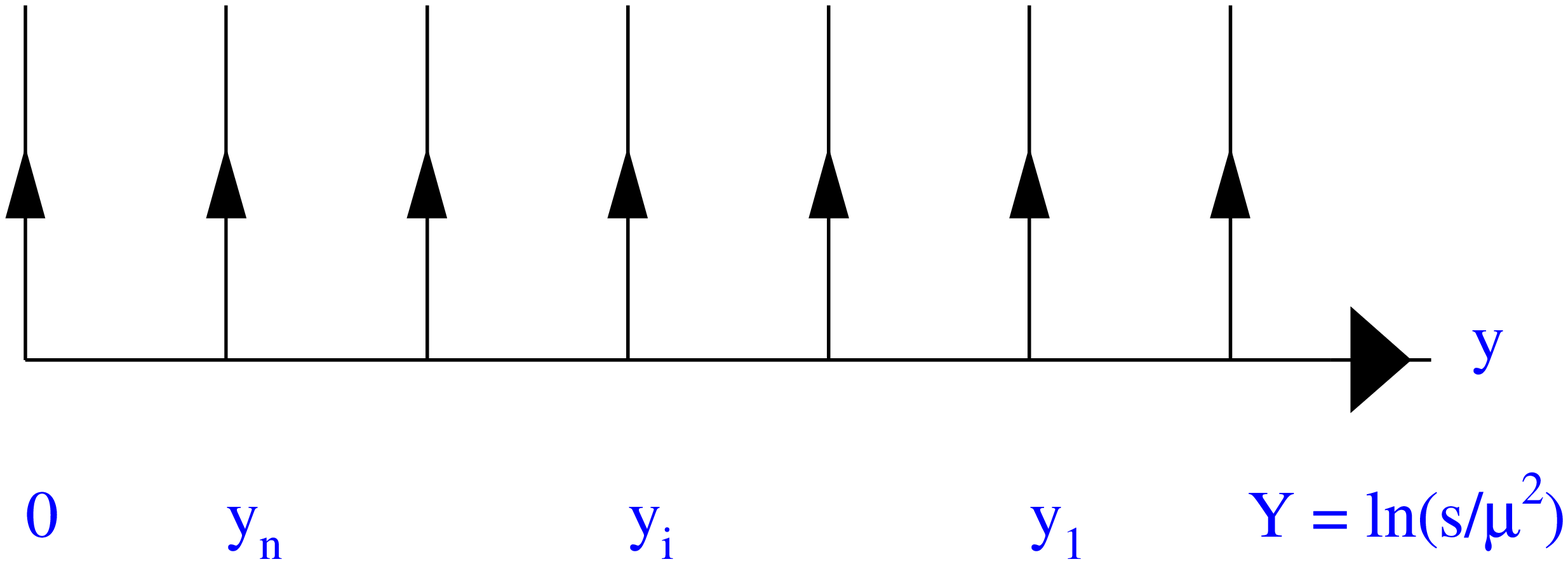,width=100mm}}
\caption{ \it Uniform distribution in rapidity in $ g \phi^3$ -
theory ( parton model ) .}   
\label{fig13}
\end{figure}

$\star$ Average transverse momentum of produced particles 
does not depend on energy\,\,;

$\star$  There are no correlation between produced
particles\,\,;

$\star$ Poisson multiplicity distribution\,\,; 
 
$\star$ Increase of the interaction radius 
due to the random walk in impact parameters\,\,;

\begin{figure}
\centerline{\psfig{file= 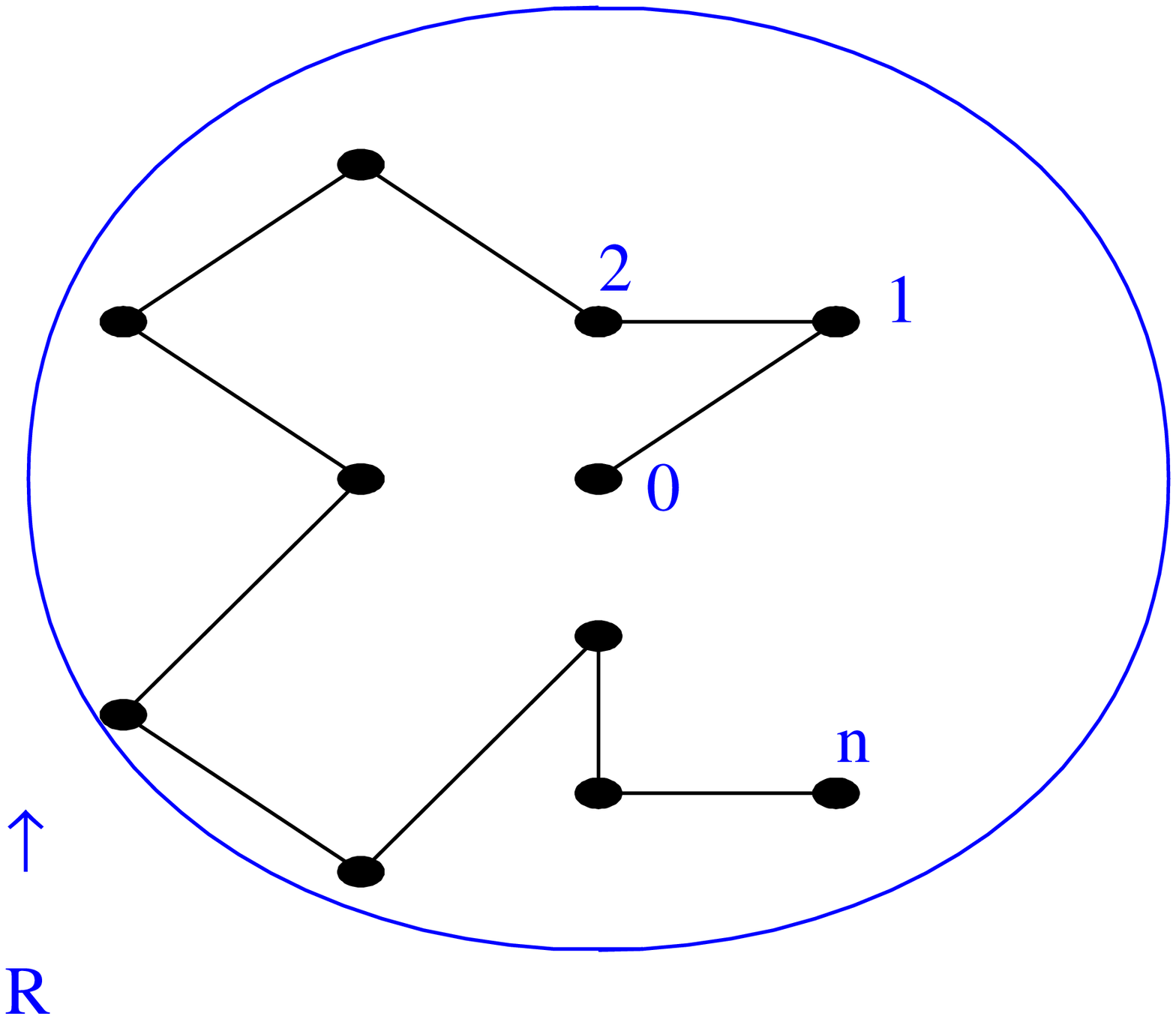,width=100mm}}
\caption{ \it Random walk in the transverse plane in  $ g \phi^3$ -
theory  ( parton model ).}
\label{fig14}
\end{figure}
\end{itemize}
The increase of the interaction radius is  seen in Fig.14 and
follows directly from uncertainty relation and from the fact that the mean
transverse momentum of particles ( partons ) does not depend on energy.
Indeed, from uncertainty principle:
\beq \label{P3}
\Delta \,b_t \,\times\,< k_t >\,\,\approx\,\,1\,\,,
\eeq 
where $< k_t >$ is the mean parton momentum. Therefore, each emission
changes the position of the parton in impact parameter space on the value
$\Delta b_t\,\,\approx\,\,\frac{1}{< k_t >}$.  After $n$ emission the
parton will be on the distance $ < b^2_{t,n} >\,\,\propto\,\frac{1}{< k_t
>}\,n$. Since the average number of emission  $N\,\,\propto\,\,\ln s $,
the radius of interaction 
\beq \label{P4}
R^2(s)\,\,=\,\, < b^2_{t,N} >\,\,\propto\,\,\frac{1}{< k_t
>}\,N\,\,=\,\,\frac{1}{< k_t >}\,\ln s\,\,.
\eeq
  
\section{ Donnachie - Landshoff Pomeron}
In practice, when we say ``soft" Pomeron, we mean so called Donnachie -
Landshoff Pomeron. Donnachie and Landshoff suggested the simplest picture
of high energy interaction - the exchange of the Pomeron which is a Regge
pole. {\bf Note, that the Donnachie - Landshoff approach to high energy
scattering is more complicated than the exchange of the Pomeron. They
also included a week shadowing corrections as we will discuss below.}
I want to stipulate   that in this section I discuss only the D-L Pomeron
but not the D-L approach. In their simple picture Donnachie and Landshoff
gave the most economic and elegant description of the experimental data
mostly related to the total and elastic cross sections. From fitting
procedure they found that the D-L Pomeron has the following features:

{ \bf 1.}\,\,\, The Pomeron intercept, $\Delta 
\,\,=\,\,\alpha_P(0)\,-\,1\,\,=\,\,0.08$\,\,;

{\bf 2.} \,\,\,The Pomeron slope
$\alpha'_P(0)\,\,=\,\,0.25\,GeV^{-2}$\,\,;

{\bf 3.}\,\,\, The linear parameterization for the Pomeron trajectory
$\alpha_P(t)\,\,=\,\,\alpha_P(0)\,\,+\,\,\alpha'_P(0)\,\,t$ can be used to
fit the experimental data\,\,;

{\bf 4.}\,\,\,The additive quark model ( AQM ) can be used to to find the
$t$ -
dependence of the Pomeron - hadron vertices. In AQM the Pomeron vertex is
proportional to the electromagnetic form factor of the hadron.

\section{Seven arguments against D - L Pomeron}
In spite of the wide use of the D-L Pomeron or, may be, because of this,
the D-L Pomeron is seriously sick. Here, we want to list seven arguments
against the D-L Pomeron which, we think, show that D-L Pomeron ( and D-L
approach, which we will discuss below) is close to its last days of
existence. In what follows we try to separate the D-L Pomeron from D-L
approach which includes not only the Pomeron but also  week SC.

\begin{enumerate}

{ \bf \item  The D-L Pomeron violates unitarity (\,\,
$ \mathbf a_{el}(s,b_t)\,\leq\,1$ 
 just around the
corner\,\, )\,\,;}

\begin{figure}
\centerline{\psfig{file= 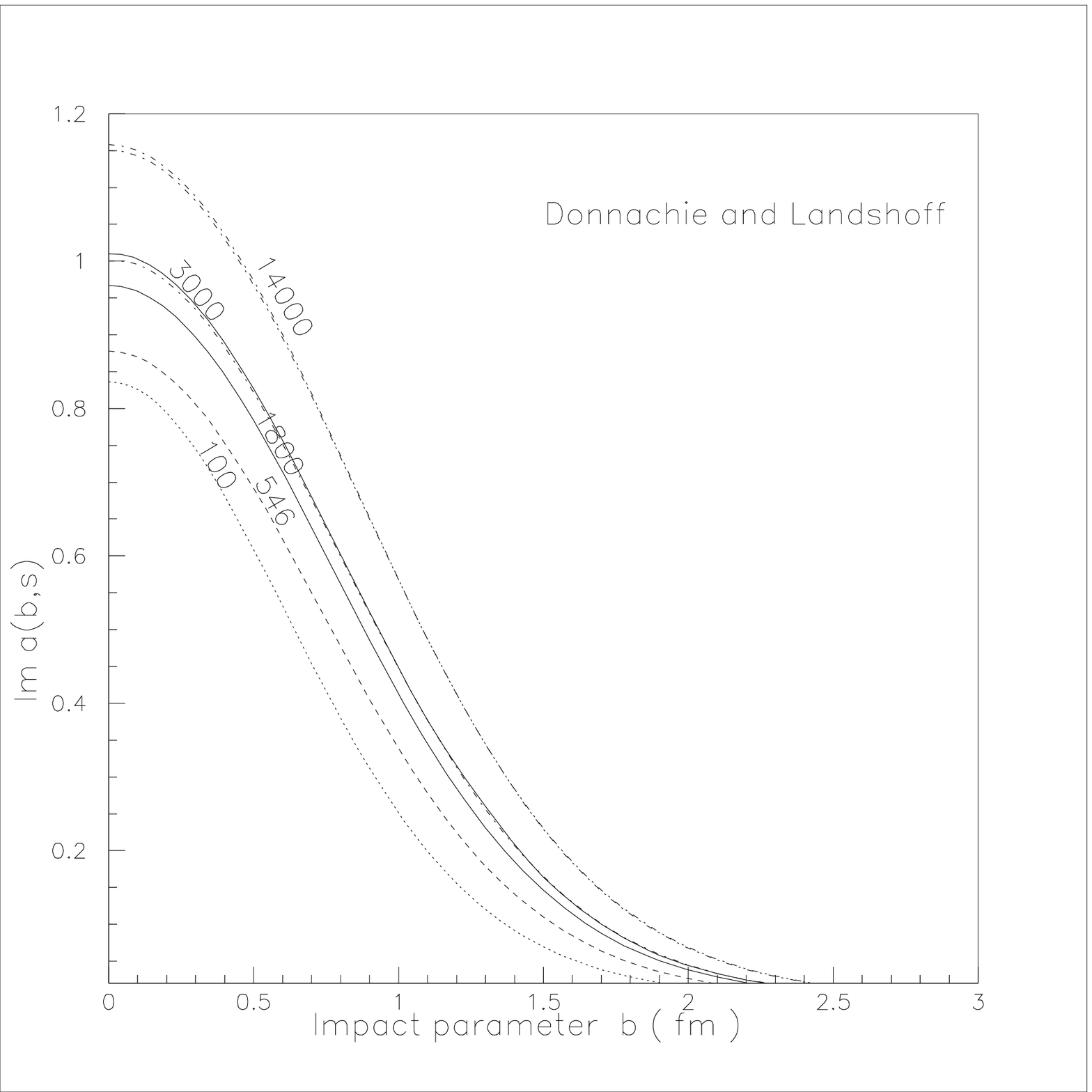,width=100mm}}
\caption{ \it  $a_{el}$ for the D-L Pomeron  .}
\label{fig15}
\end{figure}
From Fig.15 one can see that at $\sqrt{s} \,\approx\,\,3 000 GeV $ the D-L
Pomeron violates the $s$-channel unitarity constrain ( see \eq{UNB} ).
The success of the description of the experimental data on total and
elastic cross section is mostly due to the fact that the area, in which
$a_{el}(s,b_t) \,>\,1 $  in Fig. 15, is much smaller than the total area. 
However, the fact that $a_{el}$ is close to unity  should reveal itself in 
the description of other processes which are more sensitive  to the value
of $a_{el}$.    
{ \bf \item The D-L Pomeron gives $\mathbf G_{in}(s,b_t)$ which is
quite
different from the  Pomeron one \,\,;}
\begin{figure}
\centerline{\psfig{file= 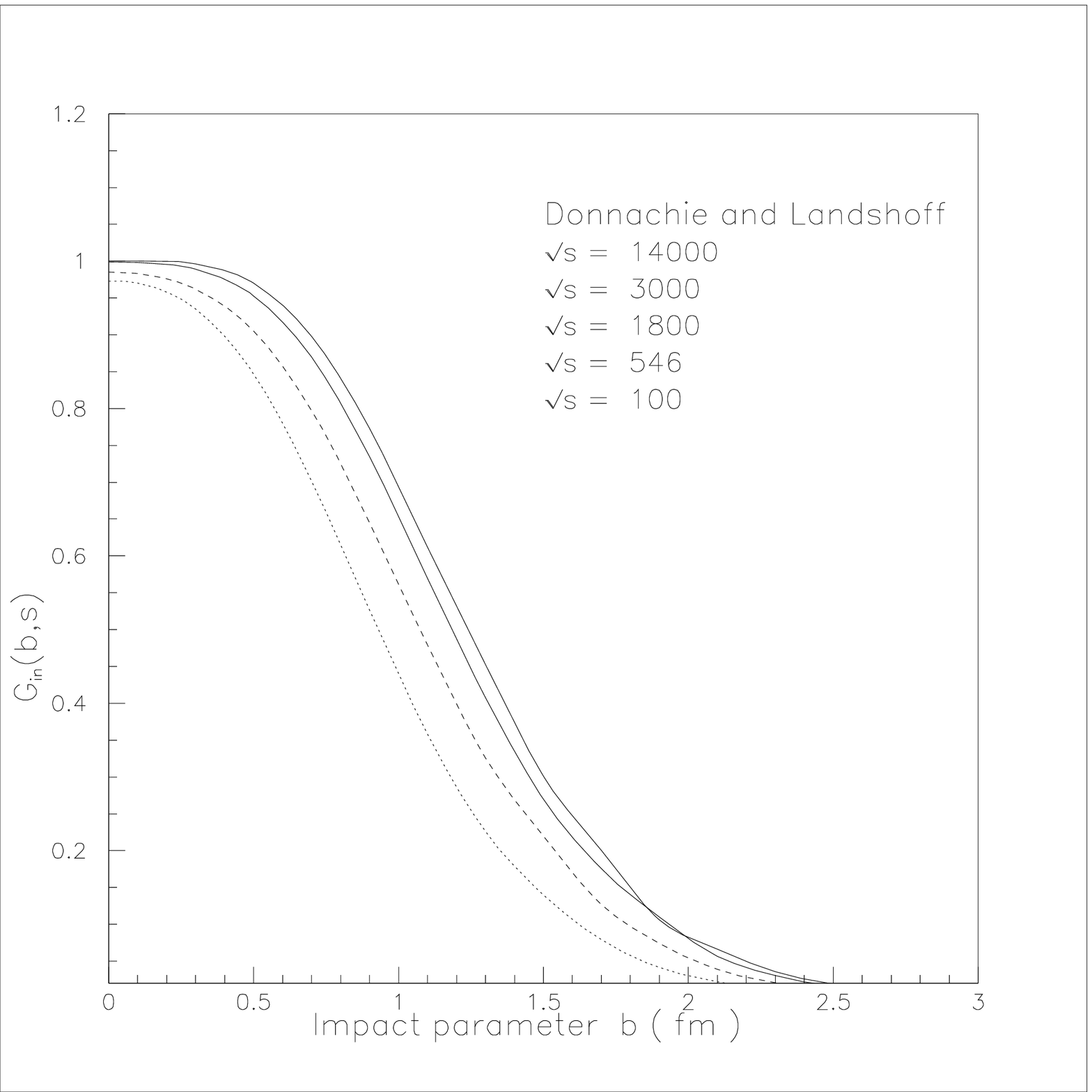,width=100mm}}
\caption{ \it  $G_{in}$ for the D-L Pomeron  .}
\label{fig16}
\end{figure}

In Fig. 16 we plot the value of $G_{in} (s, b_t )$ which we calculated
from the $s$-channel unitarity constrain of \eq{UNB}, namely,      
$$
G_{in}(s,b_t)\,\,=\,\,2\,Im \,a_{el}(s,b_t)\,\,-\,\,|a_{el}(s.b_t)|^2
\,\,,  
$$
using the D-L parameterization for $a_{el} (s, b_t ) $. One can see that 
$G_{in}$ turns out to be very close to unity in accessible energy range.
Comparing $b_t$-dependence of $a_{el} (s,b_t )$ ( see Fig.14  )
and $G_{in}(s,b_t)$ ( see Fig.15) one can see that this dependence is
quite different for them. It means that the main idea of the parton model
(see Fig.1) that the exchange of the Pomeron is closely related to the
multiparticle ( multiparton ) production is broken in the D-L approach. In
other words, the
most charming feature of the Pomeron approach as a whole, namely, the
ability of the Pomeron to describe both elastic and inelastic processes, 
turns out to be inconsistent in the D-L approach.
For me, it is too high price for the fit of the elastic and total   cross
sections even if it is a simple one.

{\bf \item  The D-L Pomeron and D-L approach cannot describe the $\mathbf
s$-dependence of the 
diffractive dissociation cross section, measured at CDF\,\,;}

\begin{figure}
\centerline{\psfig{file= 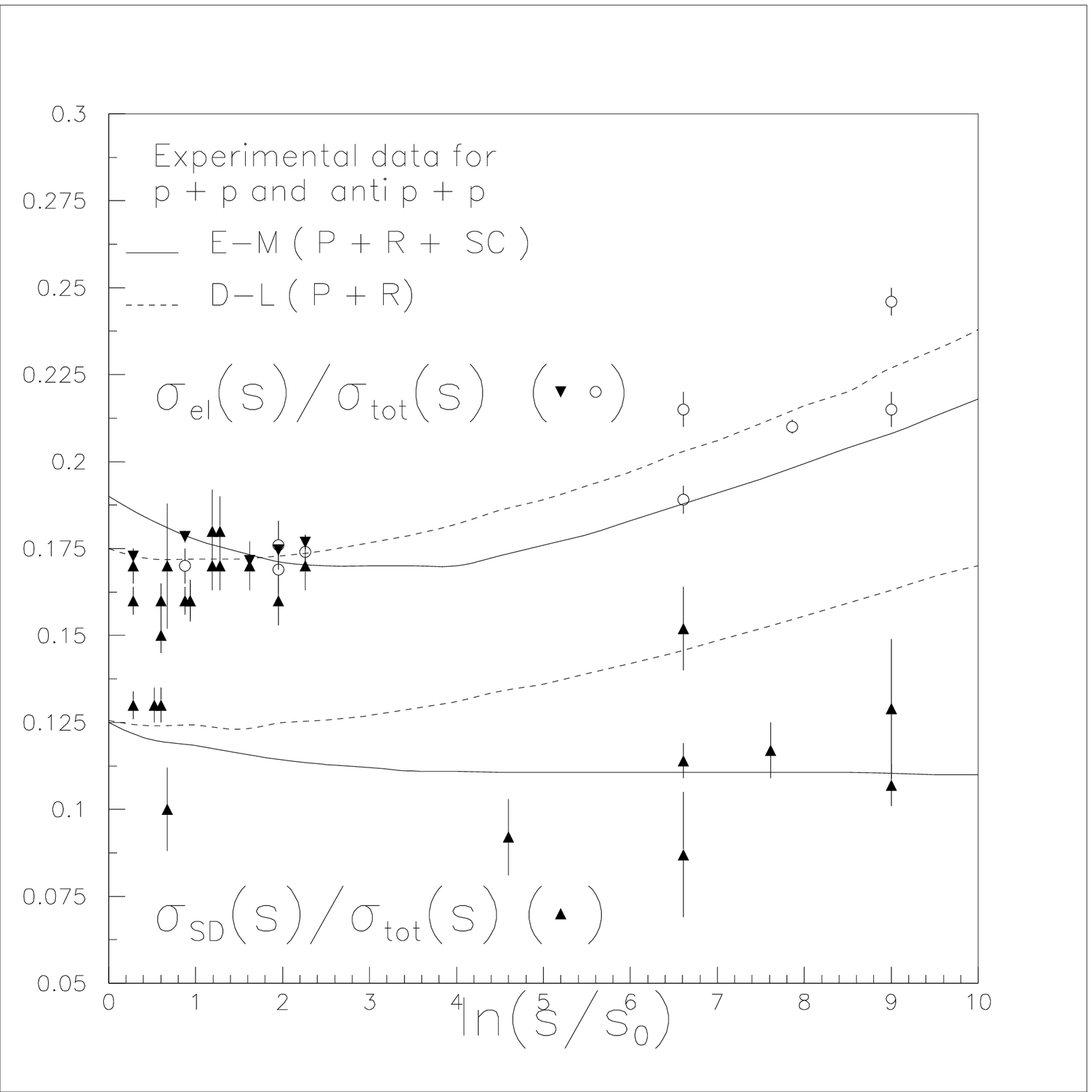,width=140mm}}
\caption{ \it $\frac{\sigma_{el}}{\sigma_{tot}}$ and
$\frac{\sigma^{SD}}{\sigma_{tot}}$ versus $\ln (s/s_0)$ with $s_0 = 400
GeV^2 $.}
\label{fig17}
\end{figure}
For D-L Pomeron $\s^{SD}\,\,\propto\,\,(\,\frac{s}{s_0}\,)^{2\Delta_P}$
while $\s_{tot}\,\,\propto\,\,(\,\frac{s}{s_0}\,)^{\Delta_P}$. Therefore,
for the D-L Pomeron
\beq \label{DL2}
\frac{s^{SD}}{\s_{tot}}\,\,\,\propto\,\,\,(\,\frac{s}{s_0}\,)^{\Delta_P}\,\,,
\eeq
while experimentally \cite{CDFSD} the energy dependence of this ratio
is quite different as Fig.17 demonstrates. Week SC of the D-L approach
cannot change this conclusion. It turns out that SC should be as strong as
in the Eikonal model or even stronger to describe the CDF data on SD.

{\bf \item The parameters of the D-L Pomeron do not fit data quite
well\,\,;}

The parameters of the D-L Pomeron were mostly fitted from the scattering
data but the D-L Pomeron trajectory predicts a resonance at $m^2
\,\approx\,\,4\,GeV^2$ with spin 2 which was not found experimentally.
On the other hand, the experimental data. especially new HERA  data on
photon - proton scattering show that the shrinkage of the diffraction peak
( $\alpha'_P(0)$ ) is different for different particles \cite{HERASHR}.
     
{\bf \item  The  D-L Pomeron and /or D-L Approach cannot describe
the CDF data on double parton cross section\,\,; }

We will comment on this below after discussion of the shadowing
corrections. We will show that the double parton cross section is closely
related to the collision of the two parton showers which are absent in the
D-L Pomeron  and too small in the D-L approach.
 
{\bf \item  The D-L Pomeron and /or D-L Approach predicts
Survival Probability for LRG processes $\rightarrow$ 1 and, therefore,
cannot describe the measured LRG processes at the Tevatron \,\,;}
       
As we have discussed in section 1 the origin of the survival probability
is the fact that there are many parton shower interactions with the
target. If we have only  the exchange of the Pomeron or, in other words,
only;y one parton shower interaction the survival probability is equal to
unity. Experimentally \cite{D0LRG}, this survival probability is, at least,
0.1 and  it shows a substantial $s$ - dependence ( see Ref \cite{GOTSMAN}
for details ). In D-L approach this survival probability is close to unity 
since the value of the SC that they suggested is too small.

{ \bf \item The D-L Pomeron and /or D-L Approach cannot
reproduce the Glauber Shadowing for hadron - nucleus interactions\,\,; }

It is well known that the Glauber Shadowing can be described in the Reggeon
approach as the multiPomeron  exchange of the incoming particle
 with nucleons in a nucleus.  In D-L approach such multiPomeron exchange
is suppressed. On the contrary for nucleus - nucleus interaction 
D-L  approach gives the Glauber formula since the exchange of the Pomerons
between different nucleons from different nuclei are not suppressed.
Therefore, in the D-L approach the Glauber shadowing for hadron - nucleus
and nucleus - nucleus interactions look quite different. In the extreme
case of the D-L Pomeron approach they predict that 
$$\s_{tot}( hadron + 
nucleus )\,\,\propto\,\, A$$ 
while 
$$
\s_{tot}( nucleus + nucleus
)\,\,\propto\,\,A^{\frac{2}{3}}\,\,.$$

\end{enumerate}

\section{Shadowing corrections ( SC )}
In this section I discuss the shadowing ( screening, absorption )
corrections to the Pomeron exchange. Formally speaking,in the Reggeon
approach such corrections can be pictured as the exchange of many
Pomerons (see Figs.18a - 18b for examples). The only question, that we
should answer, is how can we sum all these complicated diagrams?
\alphfig
\begin{figure}
\centerline{\psfig{file= 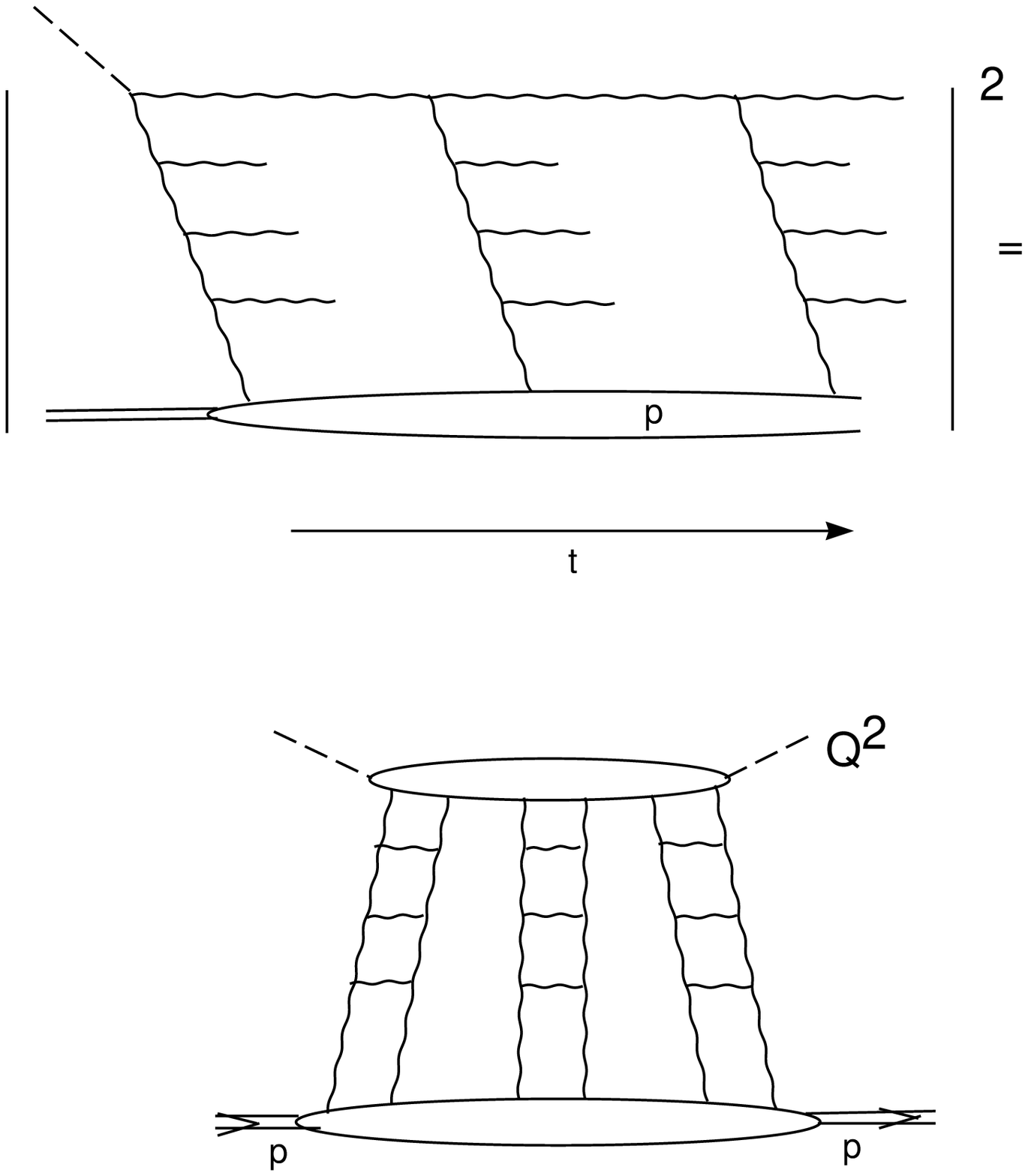,width=120mm}}
\caption{ \it The space - time picture in the parton approach for the SC
induced by three Pomeron exchange .}
\label{fig18a}
\end{figure}

\begin{figure}
\centerline{\psfig{file= 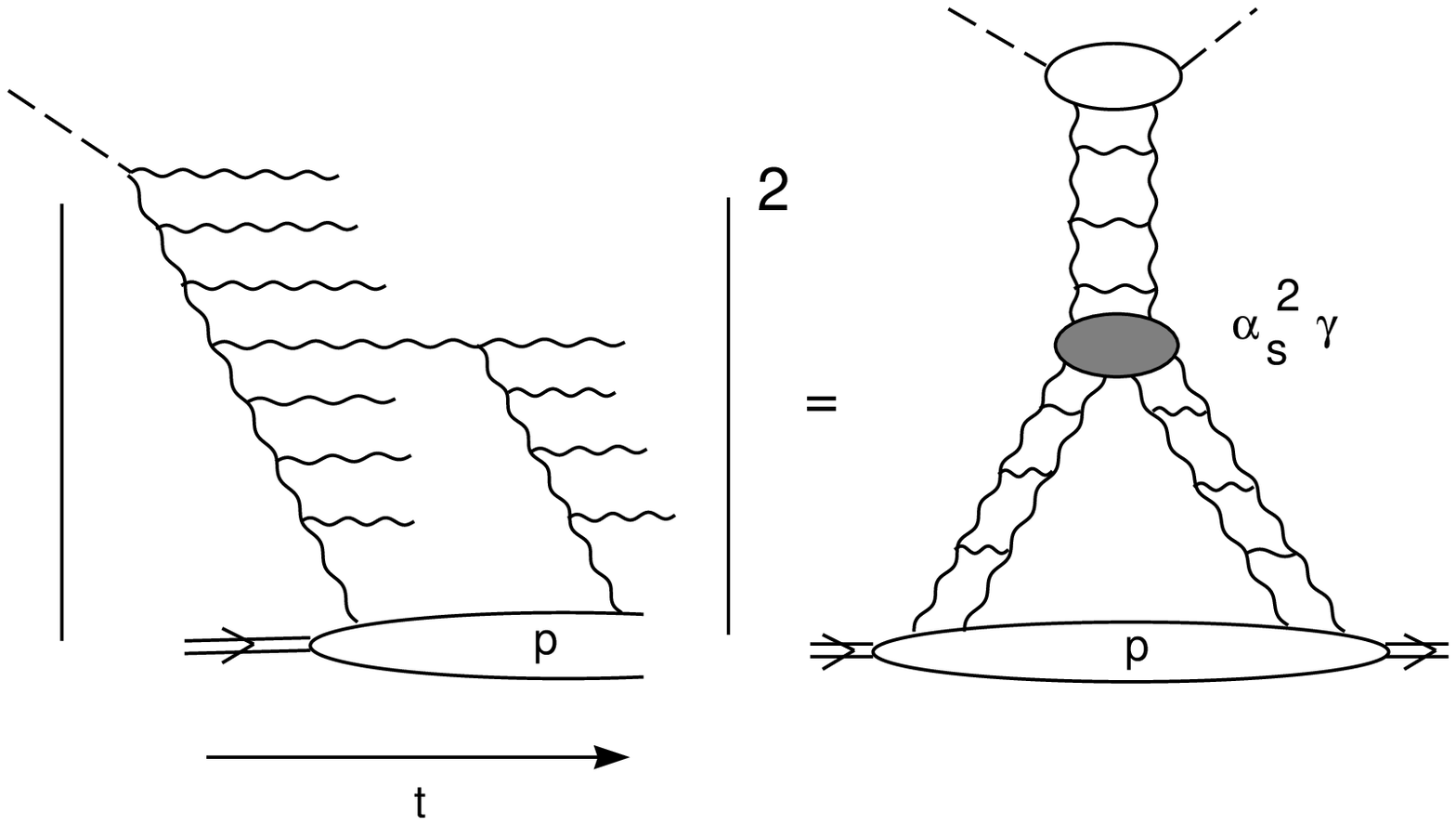,width=120mm}}
\caption{ \it  The space - time picture in the parton model for the first
``fan"  Pomeron diagram .}
\label{fig18b}
\end{figure}
\resetfig

\subsection{Several general remarks:} 
\begin{itemize}

 { \bf \item There is no Pomeron without SC\,\,;}

It means that there is no any  theoretical idea why many Pomeron exchange
can be equal to zero. Therefore, our main strategy in operating with the
SC is to find the kinematic region in which the SC are small, to develop a
technique how to calculate the SC when they are small and to built a
model ( better theory of course but practically this is a problem for
future) to approach SC in the kinematic region when they are large.

{ \bf \item SC follow from the s-channel unitarity\,\,;}

To understand why SC follow from the $s$-channel unitarity let us
generalize the solution to the unitarity constrains of \eq{UNB} using the
hadronic states which are diagonal with respect to strong interaction and
which have wave functions $\Psi_n$, as we did discussing the diffraction
dissociation processes in the previous section ( see \eq{DD1} - \eq{DD6}
).  For each of this state we have a unitarity constrain of \eq{PS1} and
we can find a solution to \eq{PB1} assuming that $Im\,
A_n(s,b_t)\,\,\gg\,\,Re \, A_n(s,b_t)$:
\begin{eqnarray}
&
A^{el_n}\,\,\,=\,\,\,i\,\frac{1}{2}\,
\{\,1\,\,\,-\,\,\,e^{-\frac{\Omega_n(s,b_t)}{2}}
\,\}\,\,;\label{SC1}
&\\
&
G^{in}_n\,\,\,\,=\,\,\,\,1\,\,\,-\,\,e^{- \Omega_n(s,b_t)}\,\,.
\label{SC2}
&
\end{eqnarray}

Let us assume that $\Omega_n\,\,\ll\,\,1$ and expand \eq{SC1} and \eq{SC2}
with respect to $\Omega_n$:
\begin{eqnarray}
&
A^{el}_n(s,b_t)\,\,=\,\,\frac{\Omega_n(s,b_t)}{2}\,\,\,-
\,\,\frac{\Omega^2_n(s,b_t)}{8}\,\,\,+\,\,\,O (\Omega^3_n)\,\,;
& \label{SC3}\\
&
G^{in}_n(s,b_t)\,\,\,=\,\,\,\Omega_n(s,b_t)\,\,\,-\,\,\,
\frac{\Omega^2_n(s,b_t)}{2}\,\,\,+\,\,\,O (\Omega^3_n)\,\,.& \label{SC4}
\end{eqnarray}

Using \eq{DD2} - \eq{DD5} we obtain for the  observables:
\begin{eqnarray}
&  
\s_{el}(s,b_t)\,\,\,=\,\,\,\frac{1}{4}\,\,
\left(\,\sum^{\infty}_{n=1}\,C^2_n\,\Omega_n(s,b_t)\,\right)^2
\,\,;& \label{SC5}\\
&
\s_{tot}\,\,\,=\,\,\,\sum^{\infty}_{n=1}\,
C^2_n\,\Omega_n(s,b_t)
\,\,\,-\,\,\,\frac{1}{4}\,\left(\,\sum^{\infty}_{n=1}
\,C^2_n\,\Omega^2_n(s,b_t)\,\right)
\,\,; & \label{SC6}\\
&
\s_{in}\,\,\,=\,\,\,\sum^{\infty}_{n=1}\,C^2_n\,
\Omega_n(s,b_t)
\,\,\,-\,\,\,\frac{1}{2}\,\left(\,\sum^{\infty}_{n=1}\,C^2_n\,
\Omega^2_n (s,b_t)\,\right)\,\,;& \label{SC7}\\
&
\s^{SD}\,\,\,=\,\,\,\frac{1}{4}\,\{\,\sum^{\infty}_{n=1}\,C^2_n 
\,\Omega^2_n(s,b_t)\,\,\,-\,\,\,\left(\,\sum^{\infty}_{n=1}\,C^2_n\,
\Omega_n(s,b_t)\,\right)^2\,\}\,\,. &\label{SC8}
\end{eqnarray}
As we have discussed, in the framework of converged theories or in the
parton model, the one Pomeron exchange corresponds to the typical
inelastic event with production of the large number of particles.
Therefore, we can associate this exchange with $\Omega$ since one can see
that  $\s_{tot}(s,b_t )\,\,=\,\,\s_{in}( s, b_t )\,\,\propto \,\,\Omega$.
All terms which are proportional to $\Omega^2$ describe the two Pomeron
exchange and  they induce the SC.

{\bf \item  The scale of SC  from the experimental
data on $\mathbf \s_{tot},\s_{el}$ and $\mathbf \s^{SD}$;}

We can evaluate the scale of the SC using experimental data on $
\s_{tot},\s_{el}$ and $ \s^{SD}$.
Indeed, we can write the expression of the total cross section in the
form:
\beq \label{SC9}
\s_{tot}\,\,\,=\,\,\s^P_{tot}\,\,\,-\,\,\,\Delta \s^{SC}_{tot}\,\,,
\eeq
where $\s^P_{tot} $ is the contribution of the Pomeron exchange to the
total cross section.  Summing \eq{SC5} and \eq{SC8} we derive  that
\beq \label{SC10}
\Delta \s^{SC}_{tot}\,\,\,=\,\,\,\s_{el}\,\,+\,\,\s^{SD}\,\,
\eeq
or 
\beq \label{SC11}
\kappa\,\,=\,\,\frac{\Delta \s_{tot}}{\s_{tot}}\,\,=\,\,\frac{\s_{el}
\,+\,\s^{SD}}{\s_{tot}}\,\,\propto\,\,\frac{\int d^2 b_t \,\Omega^2
(s,b_t)}{\int d^2 b_t \,\Omega(s,b_t)}\,\,.
\eeq
 Fig.17 shows that in the wide range of energy $\kappa
\,\,\approx\,\,0.34$ and, therefore, we can consider the single Pomeron
exchange as a good approximation only with the errors of about 34\%.
I do not think,  we need to argue that such an approximation cannot be
considered as a good one.
 
{\bf \item  The scale of SC  from the inclusive correlation function\,\,;} 
 
It is well known ( see for example Ref.\cite{MYLEC} ) that the Reggeon
approach gives the two particle rapidity correlation function, which
defined as:
\beq \label{SC12}
R \,\,=\,\,\frac{\frac{d^2 \sigma( y_1, y_2 )}{\sigma_{tot} d y_1 d y_2}}{
\frac{d \sigma(y_1)}{\s_{tot}\, d y_1} \frac{d \sigma (y_2)}{\s_{tot} d
y_2}}\,\,-1
\eeq
where  $ \frac{d^2 \sigma}{d y_1 dy_2}$ is the double inclusive
cross section for  the reaction:
$$
a\,\,+\,\,b\,\,\rightarrow \,\,1 (y_1)\,\,+\,\,2 (y_2)\,\,+
\,\,anything\,\,.
$$

The correlation function can be written in the form:
\beq \label{SC13}
R(y_1,y_2)\,\,=\,\,SR\,\times\,e^{- \frac{\Delta
y}{L_{cor}}}\,\,\,+\,\,LR\,\,.
\eeq
These two terms in \eq{SC13} correspond to two Reggeon diagrams in Fig.19.
The first diagram gives the short range correlations which falls at
$\Delta y \,=\,| y_1 - y_2 |\,>\,L_{cor}$ with
$L_{cor}\,=\,\frac{1}{\alpha_R(0)}$, where $\alpha_R(0)$ is the intercept
of the
secondary Reggeon trajectory ( see Fig.2 ) and $\alpha_R(0)\,\approx\,
0.5$.  The second term is closely related to the shadowing correction 
to the total cross section and it is equal:
\beq \label{SC14}
LR\,\,\,=\,\,2\,\frac{\Delta \s_{tot}}{\s_{tot}}\,\,.
\eeq
Therefore, if we can separate the long range rapidity correlation from the
short range one we will measure the value of the SC.

\begin{figure} 
\centerline{\psfig{file= 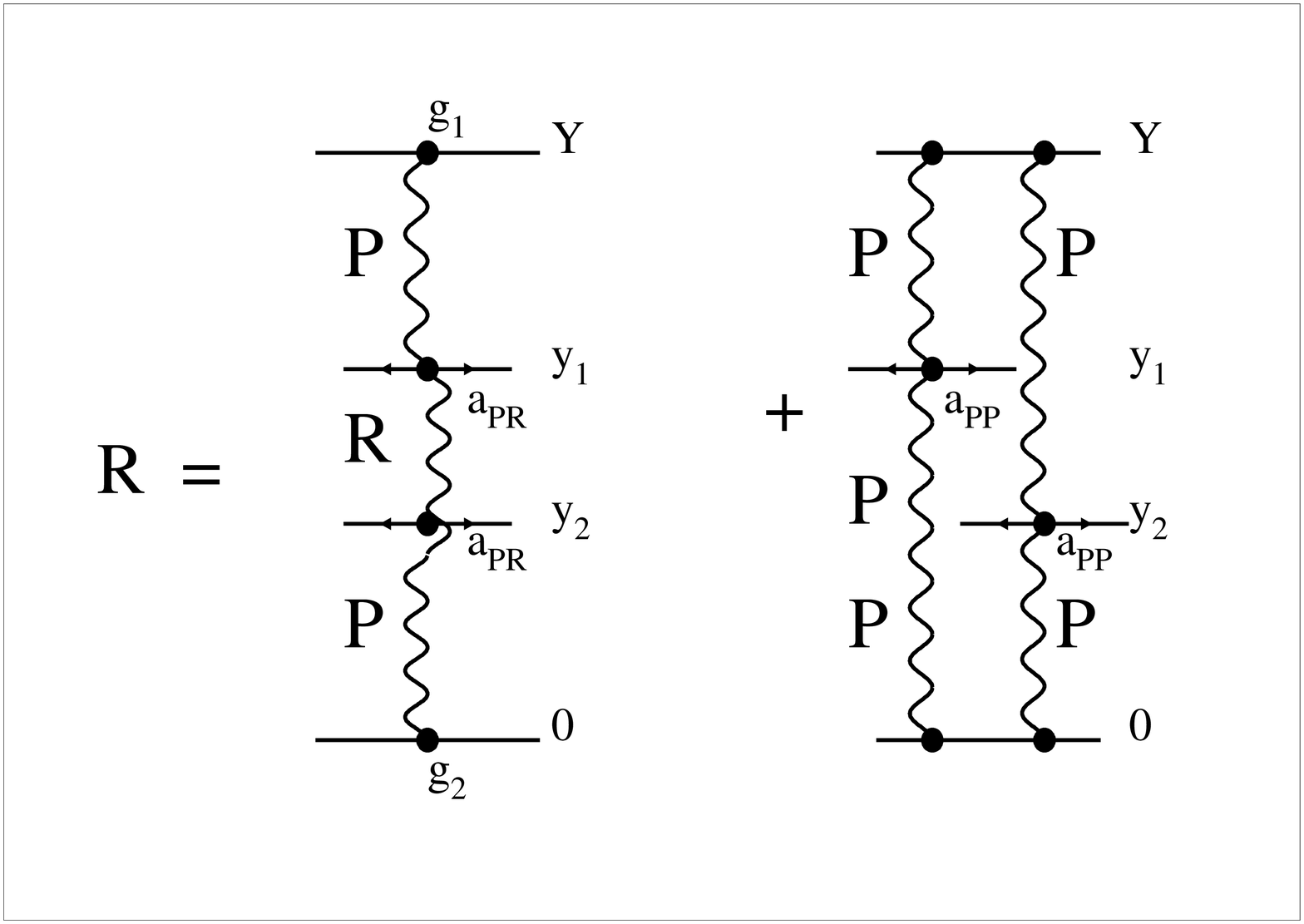,width=120mm}}
\caption{ \it  The rapidity correlation function $R$ in the Reggeon
approach .}
\label{fig19}
\end{figure}

The CDF collaboration at the Tevatron did this. The CDF measured the
process of inclusive production of two ``hard" jets with large but almost
compensating  transverse momenta in each pair( $\vec{p}_{t1}\,\approx\,-
\vec{p}_{t2}\,\gg\,\mu$, where $\mu$ is the scale of ``soft"
interactions)  and with values of
rapidity that are very similar. Such pairs cannot be produced by one
Pomeron exchange or in other words in one parton shower collision if the
difference in rapidity of these pairs is small that $1/\alpha_S(p^2_t)$.
They can only be produced in double parton shower interaction and their
cross section can be calculated using the Mueller diagrams \cite{MUD}
given in Fig.20. It can be written in the form \cite{CDFDP}
\beq \label{SC15}
\s_{DP}\,\,=\,\,m\,\frac{\s_{incl}(\,2 \,\,jets\,)\,\,\s_{incl}(\,2
\,\,jets\, )}{2\,\,\s_{eff}}\,\,,
\eeq
where factor $m$ is equal 2 for different pairs of jets and to 1 for
identical pairs. The experimental value of
$\s_{eff}\,\,=\,\,14.5\,\pm\,1.7\,\pm\,2.3\,mb$ \cite{CDFDP}.

Comparing \eq{SC15} with \eq{SC14} one can obtain the estimates for
\beq \label{SC16}
\frac{\Delta \s_{tot}}{\s_{tot}}\,\,=\,\,\frac{\s_{in}}{4
\,\s_{eff}}\,\,\approx\,\,40 -  50\,\%\,\,.
\eeq

\begin{figure}
\centerline{\psfig{file= 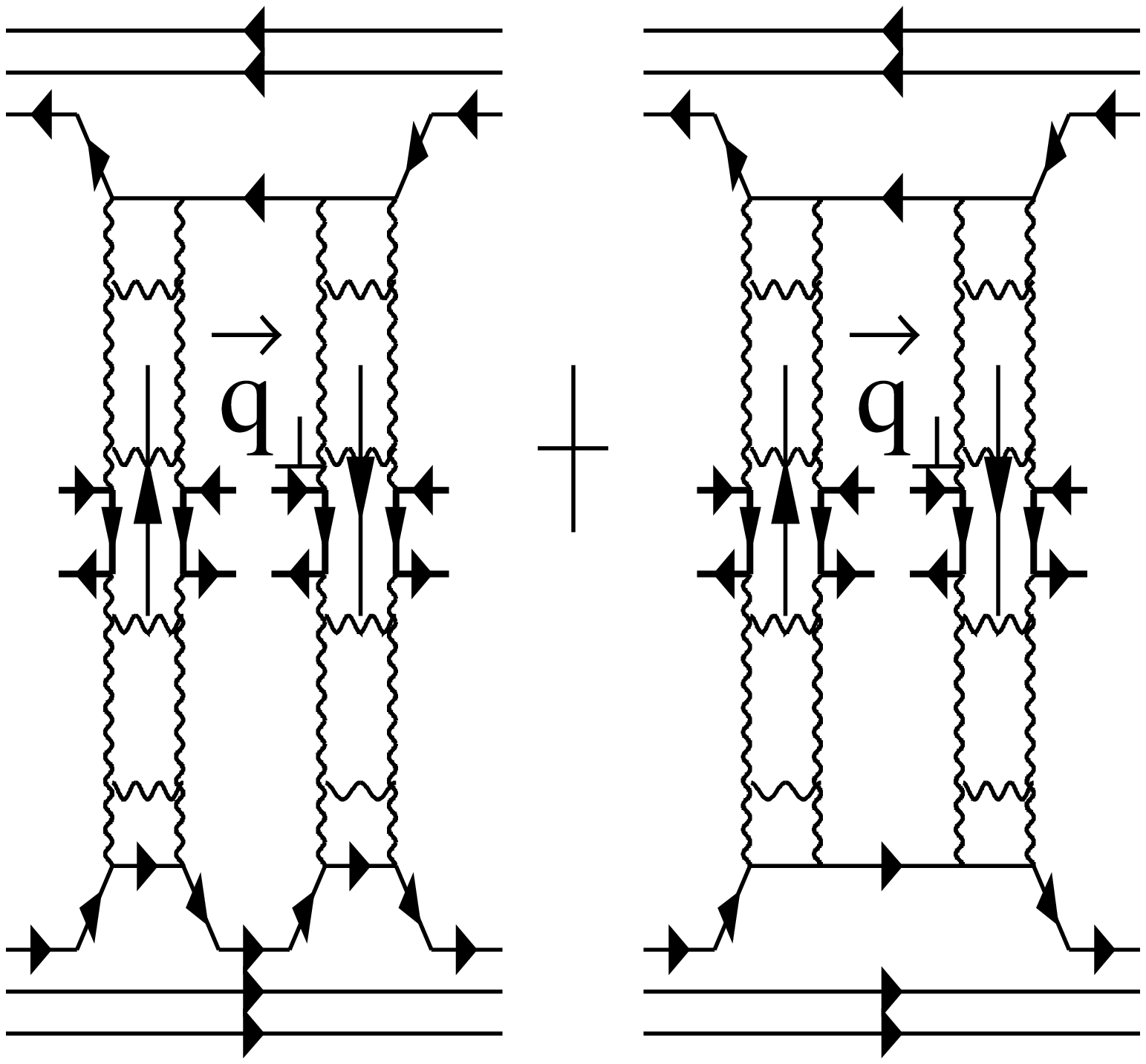,width=120mm}}
\caption{ \it  The   Mueller ( Reggeon ) diagram for double parton
interaction .}
\label{fig20 }
\end{figure}

{\bf \item  The scale of SC  from diffractive dissociation at HERA\,\,;}

As we have discussed the value of the SC crusially depends on the size of
the target ( see \eq{1.27} and \eq{1.28} ) and
$\kappa\,\,\propto\,\,\frac{\s^P_{tot}}{\pi \,R^2(s)}$. Actually, $R^2$
reflects the integration over $q_t$ ( $b_t $ ) in the first diagrams for
the SC ( see Figs. 21 and 22 ). In some sense
$$
\int \,\,d^2 q_t\,\,\,\Longrightarrow\,\,\,\frac{1}{\pi\,R^2}\,\,.
$$

\begin{figure}
\centerline{\psfig{file= 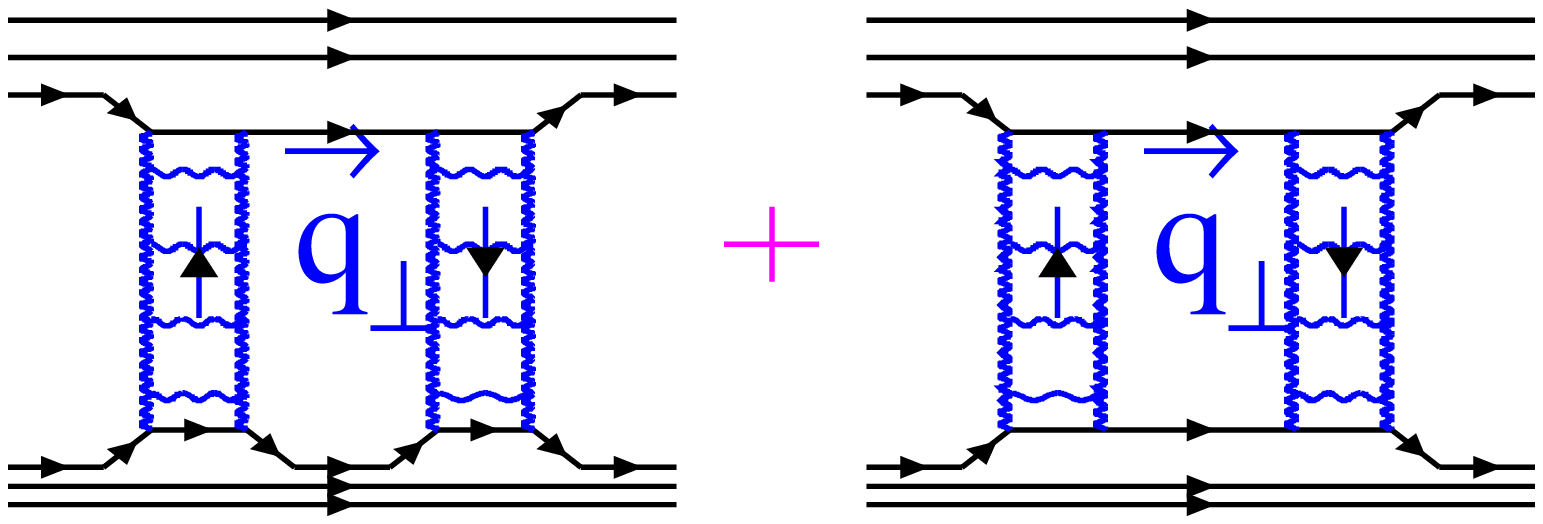,width=120mm}}
\caption{ \it  The first order SC $\propto\,\,\int\, d^2 b_t
\Omega^2(s,b_t)$ in the additive quark model for the  proton .}
\label{fig21 }
\end{figure}
\begin{figure}
\centerline{\psfig{file= 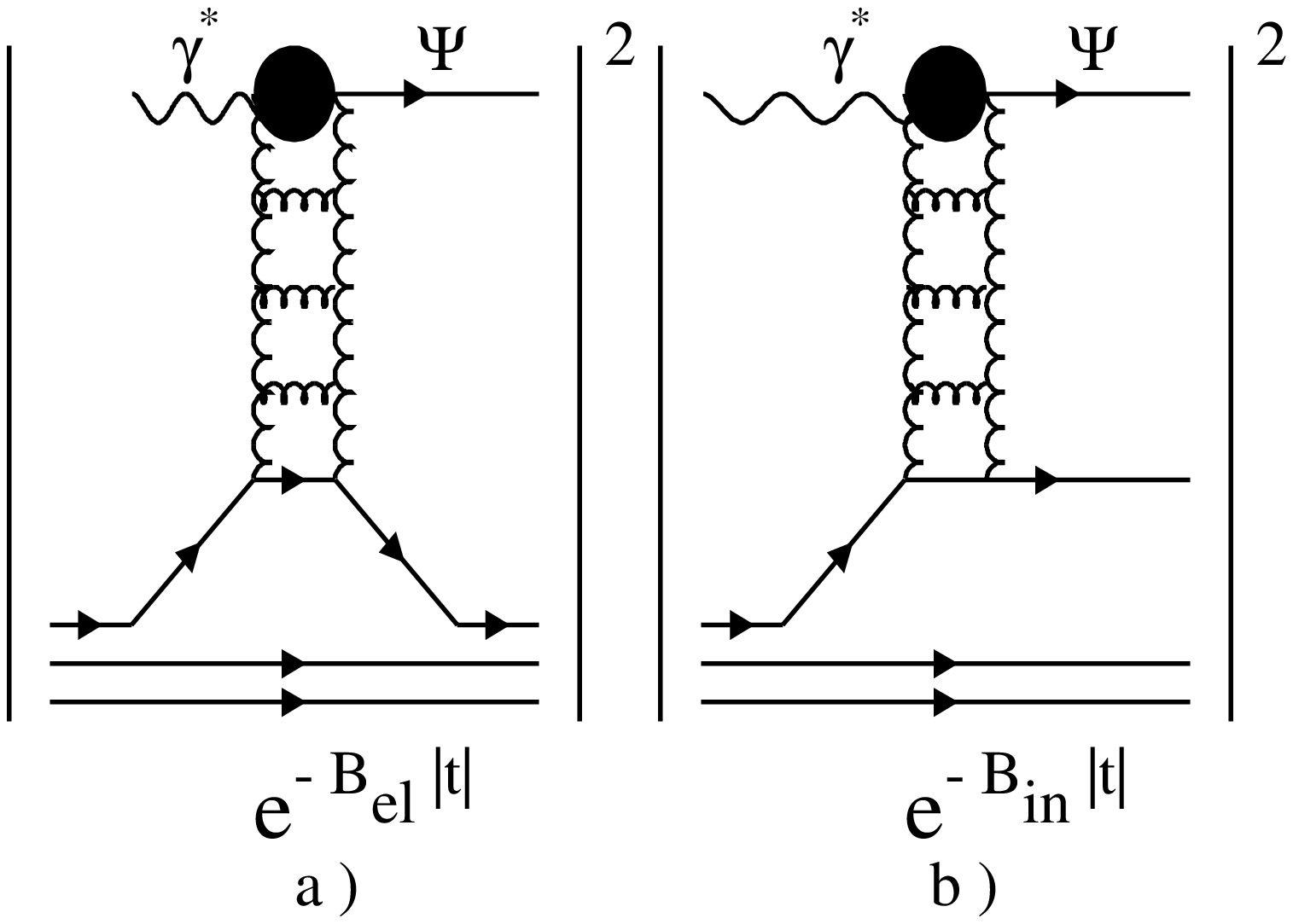,width=120mm}}
\caption{ \it  The first order SC $\propto\,\,\int\, d^2 b_t 
\Omega^2(s,b_t)$ DIS with the proton  in the additive quark
model for the proton .}
\label{fig22 }
\end{figure}

The HERA data on diffractive photoproduction of $J/\Psi $ meson
\cite{HERAPSI} give a unique possibility to find $R^2$. 
Indeed, (i) the experimental values  for the slope ( see Fig.23 ) are
$B_{el}\,=\,4\,GeV^{-2}$  and  $B_{in}\,=\,1.66\,GeV^{-2}$ and (ii) the
cross section for $J/\Psi $ diffractive production   with and without
proton dissociation are equal. Neglecting the $t$ dependence of the upper
vertex in Fig.22, we can estimate the value of $R^2$. It turns out that
$R^2\,\approx\,5\,GeV^{-2}$  or in other word it approximately in 2 times
smaller than the    radius of the proton. Using  \eq{1.27} and this value
for $R^2$ we obtain that $\Delta \s/\s \,\approx \,\,40\%$.

\end{itemize}
\subsection{ Weak SC ( Donnachie - Landshoff approach ):}
Let us ignore everything that I have discussed in the previous section
about the size of the SC. Let us pretend that we have not learned 
\eq{SC5} - \eq{SC8} and let us try to estimate the value of the SC using
only  experimental data. In my opinion this was the logic of the Donnachie
- Landshoff approach. Indeed, they achieved a good description of the
experimental data on $\s_{tot} $ and $\s_{el}$ and, therefore, they could
expect that the value of SC would  be small.  However, they had to
introduce the SC to describe the $t$- dependence of the elastic cross
section.  Indeed, the experiment shows a clear structure in $t$ dependence
-  a minimum at $t\,\approx 1.3 - 1.4 \,GeV^2$ ( see Fig.23 ).
Everybody knows that such a structure is very
 typical for interaction in optics as well as for interaction of any waves
with the target with definite size. However, the Pomeron exchange does not
reproduces this typical diffractive structure. The reason for this is
clear since the Pomeron corresponds to  predominantly inelastic production
and strictly speaking the elastic cross section should be considered as
small in pure Pomeron approach.   The $t$ structure appears in the Pomeron
approach only due to the SC. To understand this let us consider a model
with one and two Pomeron exchanges. Namely, this model corresponds
\eq{SC5} - \eq{SC8}.  For purpose of simplicity we assume that the
asymptotic $b_t$ - dependence of the Pomeron exchange give by  Eq.(69)
holds also for medium  energies.

\eq{SC6} can be rewritten in the form:
\beq \label{SC17}
a_{el} (s,b_t)\,\,=
\eeq
$$
\s_0\,\left(\,\, 
(\,\frac{s}{s_0}\,)^{\alpha_P(0)}\,e^{- \frac{b^2_t}{R^2(s)}}
\,\,\,-\,\,\lambda\,\frac{\s_0}{4\,\pi 
R^2(s)}\,(\,\frac{s}{s_0}\,)^{ 
2\,\alpha_P(0) - 1}\,\,e^{- \frac{2\,b^2_t}{R^2(s)}}\,\right)\,\,.
$$
Returning to $t$ representation we have
\beq \label{SC18}
A(s,t)\,\,=
\eeq
$$
\s_0 \left(\,(\,\frac{s}{s_0}\,)^{\alpha_P(0)}\,e^{ -
\frac{B_{el}}{2}\,|t| }
\,\,\,-\,\,\lambda\,\frac{\s_0}{16 \pi B_{el}}\,(\,\frac{s}{s_0}\,)^{
2\,\alpha_P(0) - 1}\,\,e^{- \frac{B_{el}}{4} \,|t|}\,\right)\,\,,
$$
where $B_{el}$ is the $t$ slope of the elastic cross section $B_{el}\,=\,
\frac{R^2(s)}{2}$. One can see that at sufficiently large |t| the second
term in \eq{SC18} becomes more important and at $|t| = |t_0|$ the
amplitude has a zero which, actually, manifests itself as a minimum in the
differential cross section. Taking  $|t_0|\,\approx\,1.3 GeV^2$
one can find $\lambda \,\approx\,0.04$. Donnachie and Landshoff fitted the
data using  a formula of \eq{SC18} - type \cite{DLPPT}.

\begin{figure}
\centerline{\psfig{file= 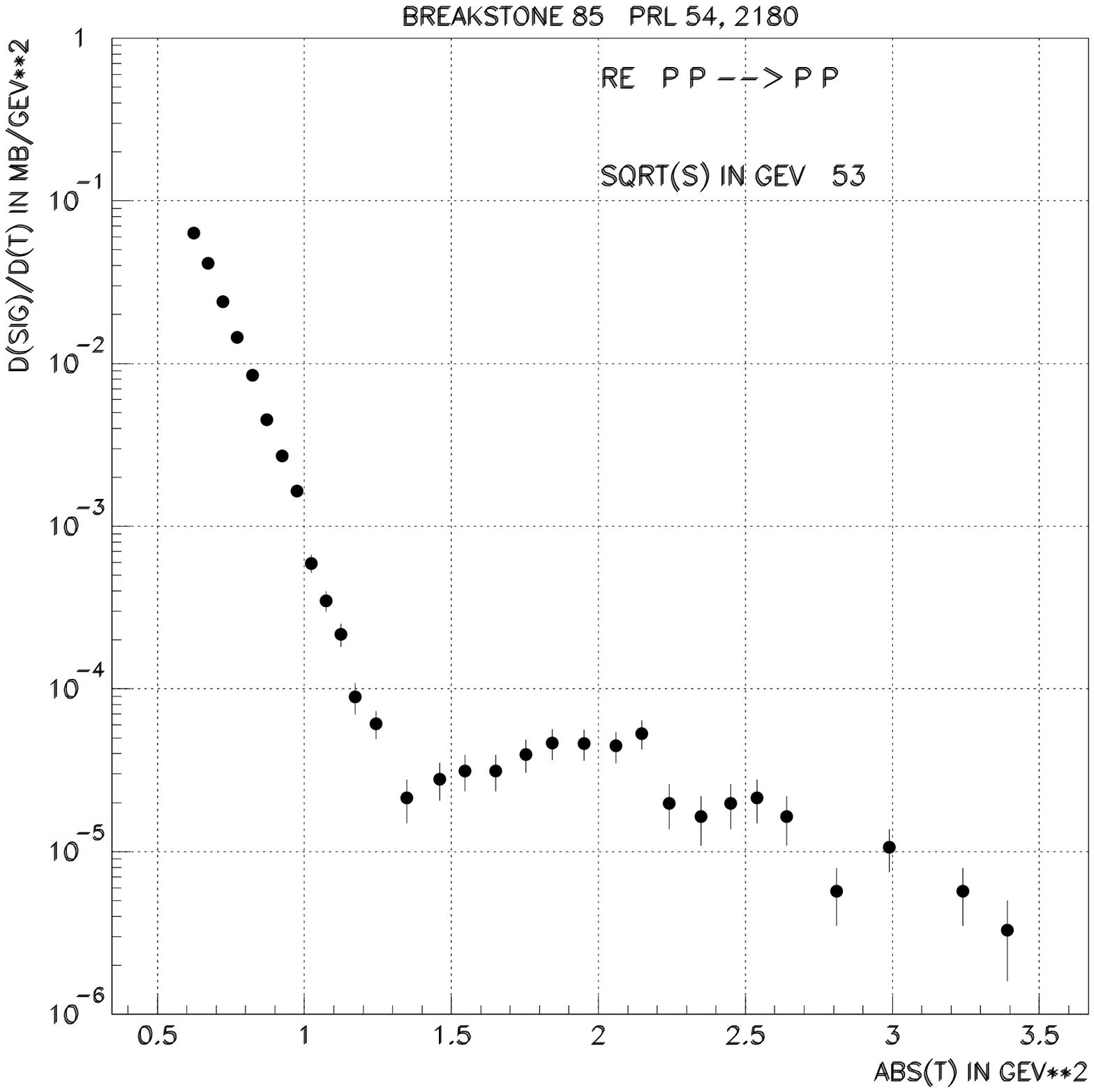,width=120mm}}
\caption{ \it  Differential elastic cross section for proton - proton
collisions at $\sqrt{s} \,=\,53\,GeV$ versus $|t|$.}
\label{fig23 }
\end{figure}

It should be stressed that this small amount of SC  gives about 3 - 5 \%
contribution to the total cross section and, therefore, can be neglected.
It is interesting also to see how SC helps to preserve unitarity. Fig. 24
shows the $b_t$-dependence of $a_{el}$ in the D-L approach with the  SC.
One can see that the unitarity will be violated only at
$\sqrt{s}\,\approx\, 5\,TeV$ (see Fig.24 ).

\begin{figure}
\centerline{\psfig{file= 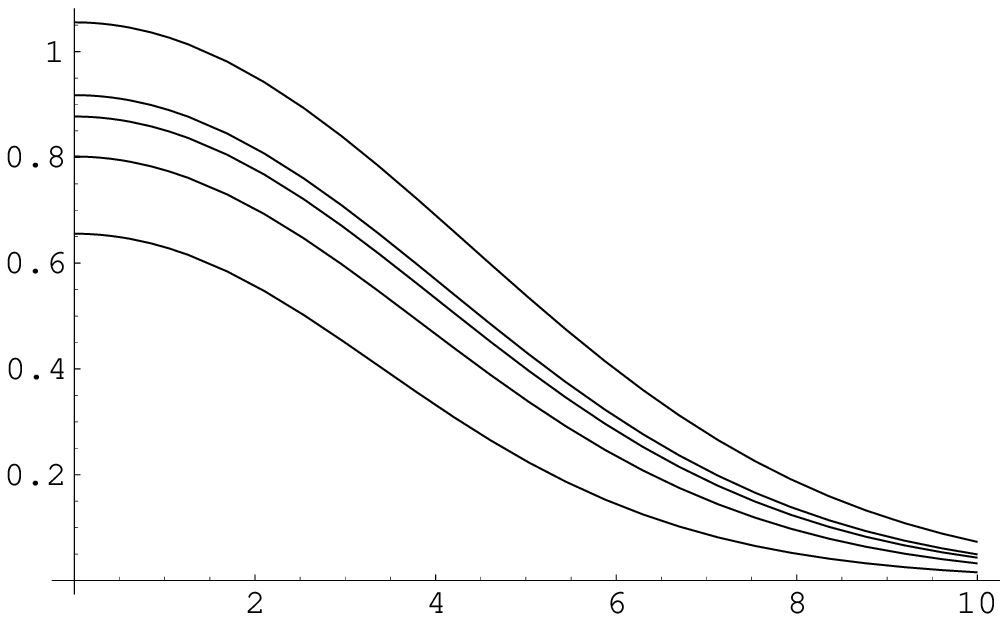,width=120mm}}
\caption{ \it  $a_{el} (s,b_t )$ versus $b_t$ in Donnachie - Landshoff
approach \protect\cite{DLPPT} The values of energies are the same as in
Fig.15.}
\label{fig24 }
\end{figure}

\subsection{Minimal SC ( Eikonal approach) :}

We assume in the Eikonal approximation, that the hadron states are
diagonal with the strong interaction. As we have discussed ( see \eq{DD1}
- \eq{DD6} ), in this case $  \sigma^{SD}$ = 0. Therefore, the Eikonal
approach has a well defined precision, namely, we cannot trust the
Eikonal approach within the accuracy better than
$\frac{\sigma^{SD}}{\sigma_{tot}}$. Fig.17 shows that this estimates give 
 errors of the order of 17 - 12 \%. This is an improvement in comparison
with the single Pomeron exchange or with the D-L approach with their weak
SC,  but it is not a perfect approach. 
 
Let me list here all pluses and minuses of this approach. I hope, that the
honest discussion of them  will lead us to a better understanding of the
main theoretical problems in high energy scattering.

{\Huge \bf +\,'\,\,s\,\,:}
\begin{enumerate}
\item The Eikonal approach is the first thing that we have to do to comply
the unitarity constraints  ( see \eq{UNB} ).  Indeed, this approach is the
consistent way to take into account two main processes: (i) the inelastic
 interaction with the typical Pomeron - like event with large multiplicity
of produced particles; and (ii) the elastic cross section. The structure
of the typical structure of the final state is pictured in Fig.25.

\begin{figure}
\centerline{\psfig{file= 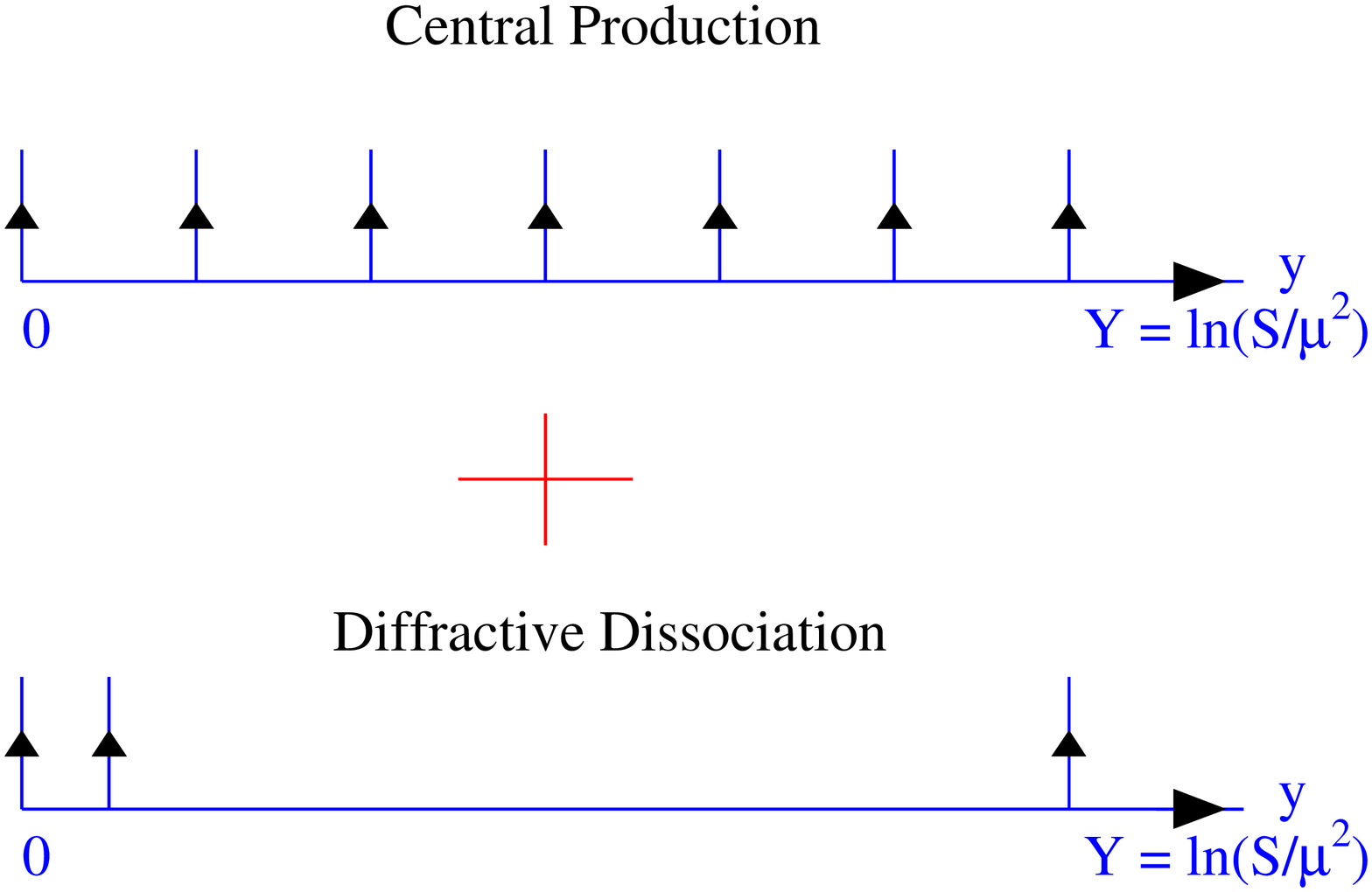,width=120mm}}
\caption{ \it  Structure of the final state in the Eikonal approach.}
\label{fig25 }
\end{figure}  

\item The Eikonal approach gives what we expect from the wave picture of
the interaction. Everybody knows that light ( or any other wave process )
interaction with the target gives always a white spot in the middle of the
shade.  Therefore, we cannot even dream to be consistent with the quantum
mechanique without the Eikonal approach. Of course, you can calculate the
elastic cross section in the D-L approach considering the elastic cross
section as a small perturbation. However, we have to check how small the
calculated elastic cross section using the unitarity constraints of
\eq{UNB} and reconstructing from it the total and inelastic cross
sections. Fig. 16 shows that the correction are large and 
changes  the behaviour of $G_{in}(s,b_t)$.

\item  The Eikonal approach is certainly the simplest way of calculating
of SC and can be used for the first estimates of how essential they could
be. The simplest is not always the best and we will show below that the
Eikonal approach has a serious pathology. 

\item The Eikonal approach describes  well the experimental data on total
and elastic cross sections ( see Ref.\cite{GLMSOFT} and Figs.11 and 17 ).

\end{enumerate}
{\Huge \bf --\,'\,\,s\,\,:}
\begin{enumerate}
\item  In the parton model the Eikonal approach means that only the
fastest parton interacts with the target. This is a very unnatural
assumption and, in my opinion, it is a great defect of this approach.

\item In the Eikonal approach the errors are of the order of
$\frac{\sigma^{SD}}{\sigma_{tot}}\,\,\approx\,\,10 - 17 \%$  and they are
not small.

\item The Eikonal approach cannot explain the CDF data on double parton
interaction and it give $\sigma_{eff}\,\approx\,30 \,mb$ ( see \eq{SC15} ) 
which is in about
two times large that experimental value ( see \eq{SC12} - \eq{SC16} ).

\item The intristic problem of the Eikonal model is the energy behaviour
of $\sigma^{SD}$ at large masses $M^2\,\,\propto\,\,s$. We will show below
that in this region of mass the diffractive cross section $\sigma^{SD}
\,\,\propto\,\,s^{\Delta_P}$ and it exceeds the value of the total cross
section $\sigma_{tot}\,\,\propto\,\,\ln s$. This is our payment for
simplicity and unjustified ( $ ad hoc $ ) assumption that only the fastest
parton interacts with the target.

\end{enumerate}

The main assumption of the Eikonal approach is the identification the
opacity $\Omega(s,b_t)$ with the single Pomeron exchange, namely
\begin{eqnarray}
&
\Omega(s,b_t)\,=\,POM(s,b_t)\,\,; &\label{EA1}\\
&
POM(s,b_t)\,\,=\,\, \frac{\sigma_0}{\pi R^2(s)}\,( \frac{s}{s_0}
)^{\Delta_P}\,e^{-
\frac{b^2_t}{R^2(s)}}\,\,;&\label{EA2}\\
 &
R^2(s)\,=\,R^2_0\,\,+\,\,2\alpha'_P\,\ln(s/s_0)\,\,; &\label{EA3}\\
&
\nu(s)\,\,=\,\,\frac{\sigma_0}{\pi R^2(s)}\,( \frac{s}{s_0}
)^{\Delta_P}\,\,=\,\,\Omega(s,b_t = 0 )\,\,. & \label{EA4}
\end{eqnarray}

\eq{EA1} is certainly correct in the kinematic region  where $\Omega$ is
small or, in other words, at rather small energies or at high energies but
al large values of $b_t$. Therefore, the Eikonal approach is the natural
generalization according to $s$-channel  unitarity of the single Pomeron
exchange.  Using \eq{UNB} and \eq{EA1} - \eq{EA3}, one can obtain ( see
Ref.\cite{GLMSOFT} for details ):
\begin{eqnarray}
&
\sigma_{tot}\,\,=\,\,2\,\pi
\,R^2(s)\,\{\,\ln(\nu(s)/2)\,\,+\,\,C\,\,-\,Ei( - \nu/2)
\,\}\,\,; & \label{EA5}\\
&
\sigma_{in}\,\,=\,\,\,\pi
\,R^2(s)\,\{\,\ln(\nu(s))\,\,+\,\,C\,\,-\,Ei( - \nu)\,\}\,\,; &
\label{EA6}\\
&
\sigma_{el}(s)\,\,=\,\,\sigma_{tot} (s)
\,\,=\,\,\sigma_{in}(s)\,\,; & \label{EA7}\\
&
B\,\,=\,\,\frac{d \ln \frac{d \sigma}{d t}}{d t}\,|_{t - 0}\,\,=\,\,
\frac{R^2(s)}{2}\,\frac{hypergeom([1,1,1,],[2,2,2],-
\nu/2)}{hypergeom([1,1],[2,2],-\nu/2)}\,\,. & \label{EA8}
\end{eqnarray}
Here, the notations of Maple were used for the generalized hyper
geometrical functions. Figs. 11 and 17 show that the Eikonal approach is
able to describe the experimental data.

The Eikonal approach gives a very simple procedure how to calculate the
survival probability of LRG processes ( see our high energy glossary ).
Indeed, in this approach the general formula of \eq{LRG3} can be
simplified and it has a form:
\beq \label{EA9}
< S^2 >\,\,=\,\,\frac{\int \,d^2 b_t S_H(b_t) \,\,P(s,b_t)}{\int \,d^2 b_t
S_H(b_t)}\,\,=\,\,\frac{\int \,d^2 b_t S_H(b_t)
\,\,e^{\Omega(s,b-t)}}{\int\,d^2\,b_t\,\,S_H(b_t)}\,\,,
\eeq
where $ S_H(b_t)$ is the profile function for the ``hard" processes for 
two jets production.  For all numerical estimates we took a Gaussian
parameterization for $ S_H(b_t)\,=\,\frac{1}{\pi \,R^2_H}e^{-
\frac{b^2_t}{R^2_H}}$. In the framework of this simple parameterization
all integrals can be taken analytically and we have
\beq \label{EA10}
< S^2 >\,\,=\,\,\frac{a(s)\,\gamma[a(s),\nu(s)]}{[\,\nu(s)\,]^{a(s)}}\,\,
,
\eeq
with
\beq \label{EA11}
a(s)\,\,=\,\,\frac{R^2(s)}{R^2_H}\,\,=\,\,\frac{interaction
\,\,area\,\,for\,\,``soft"\,\,collisions}{interaction
\,\,area\,\,for\,\,``hard"\,\,collisions}\,\,>\,\,1\,\,.
\eeq
$\gamma[a,\nu]$ is incomplete gamma function.

Fig.26 shows the Eikonal model prediction for $< S^2 > $ which are in a
good agreement with the D0 data and reproduce the energy behaviour of the
LRG rate( see Fig.9-b ).

\begin{figure}
\centerline{\psfig{file= 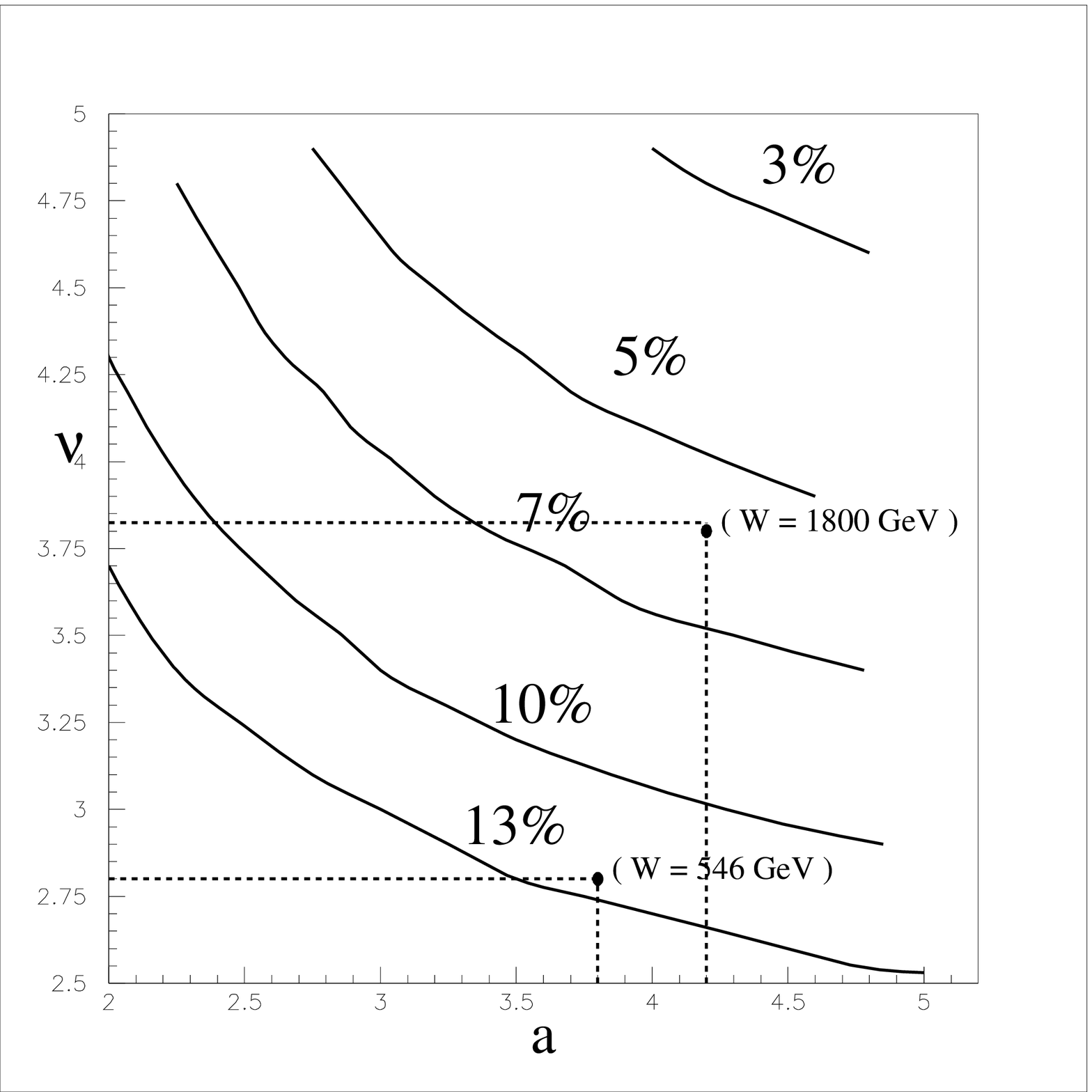,width=120mm}}
\caption{ \it  $< S^2 > $ in the Eikonal approach.}
\label{fig26 }
\end{figure}

The key problem in the Eikonal approach is  diffraction production which
has not been included in our unitarization procedure. This is a more
complicated way to express that DD is equal to zero in the Eikonal
approach. However, we can treat the diffraction dissociation as a
perturbation considering that  its cross section is much smaller than the
elastic one. Fig.17 shows that such an approach cannot not be very good
since $\frac{\sigma^{DD}}{\sigma_{el}}\,\,\approx\,\,50\%$. The first step
is to consider the triple Reggeon diagram of Fig.18-b type $A_{PPP}$. The
second one
is to take into account  the rescatterings of the fast partons  which
suppress the value of $A_{PPP}$. Actually,  the resulting cross section
has the same form of the survival probability for LRG processes since the
DD is the simplest LRG process:
\beq \label{EA12}
\frac{ M^2\,d \sigma^{SD}}{d M^2 }\,|_{t = 0}\,\,=\,\,\int \,d^2\,b_t
A_{PPP}(s,M^2)\,\,P(s,b_t)\,\,=
\eeq
$$
\frac{\sigma^2_0}{2
\,\pi\,R^2_1(\frac{s}{M^2})}\,G_{PPP}\,\times\,(\,\frac{M^2}{s}\,)^{2\,\Delta_P}
\,(\,\frac{M^2}{s_0}\,)^{\Delta_P}\,\frac{
a(s)\,\gamma[a(s),\nu{s}]}{[\nu]^{a(s)}}\,\,,
$$
where
\beq \label{EA13}
a(s)\,\,=\,\,\frac{2
R^2(s)}{R^2_1( \frac{s}{M^2})\,\,+\,\,2\,R^2_1( \frac{M^2}{s_0})}
\eeq
with $R^2_1(\frac{s}{M^2})\,=\,2 \,R^2_{01}
\,+\,r^2_{01}\,+\,4\,\alpha'_P(0)\,\ln(\frac{s}{M^2})$.
$R^2_{01}$ denotes the radius of the proton - Pomeron vertex while
$r^2_{01} $ is the radius of the triple  Pomeron vertex \cite{GLMDD}.
$R^2(s) \,=\, 4 \,R^2_{01}\,\,+\,\,4\,\alpha'_P \ln(s)$.

\eq{EA12} describes experimental data much better than the single Pomeron
exchange  after adding a secondary Reggeon contributions. It shows that we
are on the 
 correct  way. The most important feature of Eikonal approach description
is the flattering of the energy dependence. However, in Fig.17 we plot the
integrated  cross section over $M^2$ defining the region of integration $
4\,<\,M^2\,0.3\,\sqrt{s\,1\,GeV^2}$.  the experimental $\sigma^{SD}$ was
integrated in the region  $4\,<\,M^2\,0.05\,s$. We cannot trust our
calculations for $M^2\,\propto\,s$ since 

\beq \label{EA14}
a(s)\,|_{s\,\,\gg\,\,1}\,\longrightarrow\,\,\,2\,\{\,\,1\,\,\,-\,\,\,\frac{\ln 
M^2}{\ln s}\,\,\}\,\,.
\eeq
Using the asymptotics for the incomplete gamma function we obtain that
\beq \label{EA15}
\frac{ M^2\,d \sigma^{SD}}{d M^2 }\,\,\Longrightarrow
\,\,(\,M^2\,)^{\Delta_P}\,\,.
\eeq
Integrating \eq{1.15} over $M^2$ up to $M^2\,=\, k\,s$ one obtain
\beq \label{EA16}
\sigma^{SD}\,\propto \,\int^{k s} \frac{d M^2}{M^2}\,(\,M^2\,)^{\Delta_P}
\,\,=\,\,\frac{1}{\Delta_P}(\,k\,s\,)^{\Delta_P}\,\,
\gg\,\,\sigma_{tot}\,\,\propto\,\,\ln s\,\,.
\eeq
\eq{1.16} shows that there is no way to incorporate the
diffractive
production in the Eikonal model since the ``small" parameter of such an
approach cannot be small.

This example is the most transparent illustration of the difficult
problem in taking into account the SC - the description of high mass
diffraction.

\subsection{ Correct SC ( ? )}
Certainly, we do not know an answer to the question what is a correct
 procedure to take into account SC. Here, I  discuss two models for SC
which both are better than Eikonal approach.

\centerline{\it Quasi-Eikonal Model.}

The idea of this model is very simple: to take into account the
diffractive dissociation in the same Eikonal way as elastic rescatterings.
Doing so, we assume that we have two wave functions which are diagonal
with respect to the strong interactions: $\Psi_1$ and $\Psi_2$. The general
\eq{DD2}  has the form
\beq \label{QE1}
\Psi_{hadron}\,\,\,=\,\,\,\beta_1\,\Psi_1\,\,\,+\,\,\,\beta_2
\,\Psi_2\,\,\,,
\eeq
with condition: $\beta_1^2\,+\,\beta_2^2\,\,=\,\,1$, which follows from
the
normalization of the wave function.
The wave function of the diffractively produced bunch of hadrons should be
orthogonal to $\Psi_{hadron}$ and  looks as
follows:
\beq \label{QE2}
\Psi_D\,\,\,=\,\,\,-\,\beta_2\,\Psi_1\,\,\,+\,\,\,\beta_1\,\Psi_2\,\,.
\eeq
 The elastic and single diffraction  amplitude in this model can be
rewritten, using \eq{DD3}, \eq{QE1} and \eq{QE2}  in the form:

\begin{eqnarray}
&
a_{el}(s,b_t)\,\,=\,\,< \Psi_{final} \ \Psi_{hadron} >\,\,
=\,\,\beta^2_1\,A_1\,\,+\,\,\beta^2_2\,A_2\,\,; &\label{QE4}\\
&
 a_{SD}(s,b_t)\,\,=\,\,< \Psi_{final} \ \Psi_{D} >\,\,=\,\,
\beta_1\,\beta_2\,\{\,A_2\,\,-\,\,A_1\,\}\,\,.&\label{QE5}
 \end{eqnarray}
For amplitude $A_n(s,b_t)$ we have a unitarity constraint of \eq{PB1} and
therefore:
\begin{eqnarray}
&
A_n(s,b_t)\,\,=\,\,i\,\{\,1\,\,-\,\,e^{-
\frac{\Omega_n(s,b_t)}{2}}\,\}\,\,;&\label{QE6}\\
&
G_n(s,b_t)\,\,=\,\,1\,\,\,-\,\,\,\,e^{-
\Omega_n(s,b_t)}\,\,.&\label{QE7}
 \end{eqnarray}
 Assuming the simplest eikonal form for $\Omega_n$ ( see \eq{EA1} -
\eq{EA4} ), namely
\beq \label{QE8}
\Omega_n(s,b_t)\,= \, \frac{\sigma_{0n}}{\pi R^2_n(s)}\,( \frac{s}{s_0}
)^{\Delta_P}\,e^{-
\frac{b^2_t}{R^2_n(s)}}\,\,,
\eeq
we have the final answer:
\begin{eqnarray}
&
\sigma_{tot}\,=\,2 \int\, d^2 b_t \{ \beta^2_1 \,[\,1 \,\,-\,\,e^{-
\frac{\Omega_1}{2}}\,]\,\,+\,\,(\,1\,-\,\beta^2_1\,)\,[\,1\,-\,e^{-
\frac{\Omega_2}{2}}\,]\,\}\,\,;&\label{QE9}\\
&\sigma_{el}\,=\,\int\,d^2 b_t\,[\,\beta^2_1\,[\,1 \,\,-\,\,e^{-
\frac{\Omega_1}{2}}\,]\,\,+\,\,(\,1\,-\,\beta^2_1\,)\,[\,1\,-\,e^{-
\frac{\Omega_2}{2}}\,]\,]^2\,\,;&\label{QE10}\\
&
\sigma^{SD}\,\,=\,\,\int\,d^2b_t\,\beta^2_1\,(\,1\,-\,\beta^2_1)\,[
 \,e^{-\frac{\Omega_2}{2}}\,\,-\,\,e^{-
\frac{\Omega_1}{2}}\,]^2\,\,.&\label{QE11}
\end{eqnarray}
It is obvious from \eq{QE9} - \eq{QE11} that we introduce three new
parameters to describe  SC including the processes of diffractive
dissociation: $R^2_2$ $\sigma_{20}$ and $\beta_1$. They are correlated
with the total cross section of the diffractive dissociation and its  $t$-
dependence. At first sight it is strange that we have to introduce three
parameters not two. The third parameter $\sigma_{02}$ is closely related
to the cross section of ``elastic" rescattering of the hadrons produced in
the diffractive dissociation. The meaning of all parameters is clear from 
Fig. 27.
\begin{figure}
\centerline{\psfig{file= 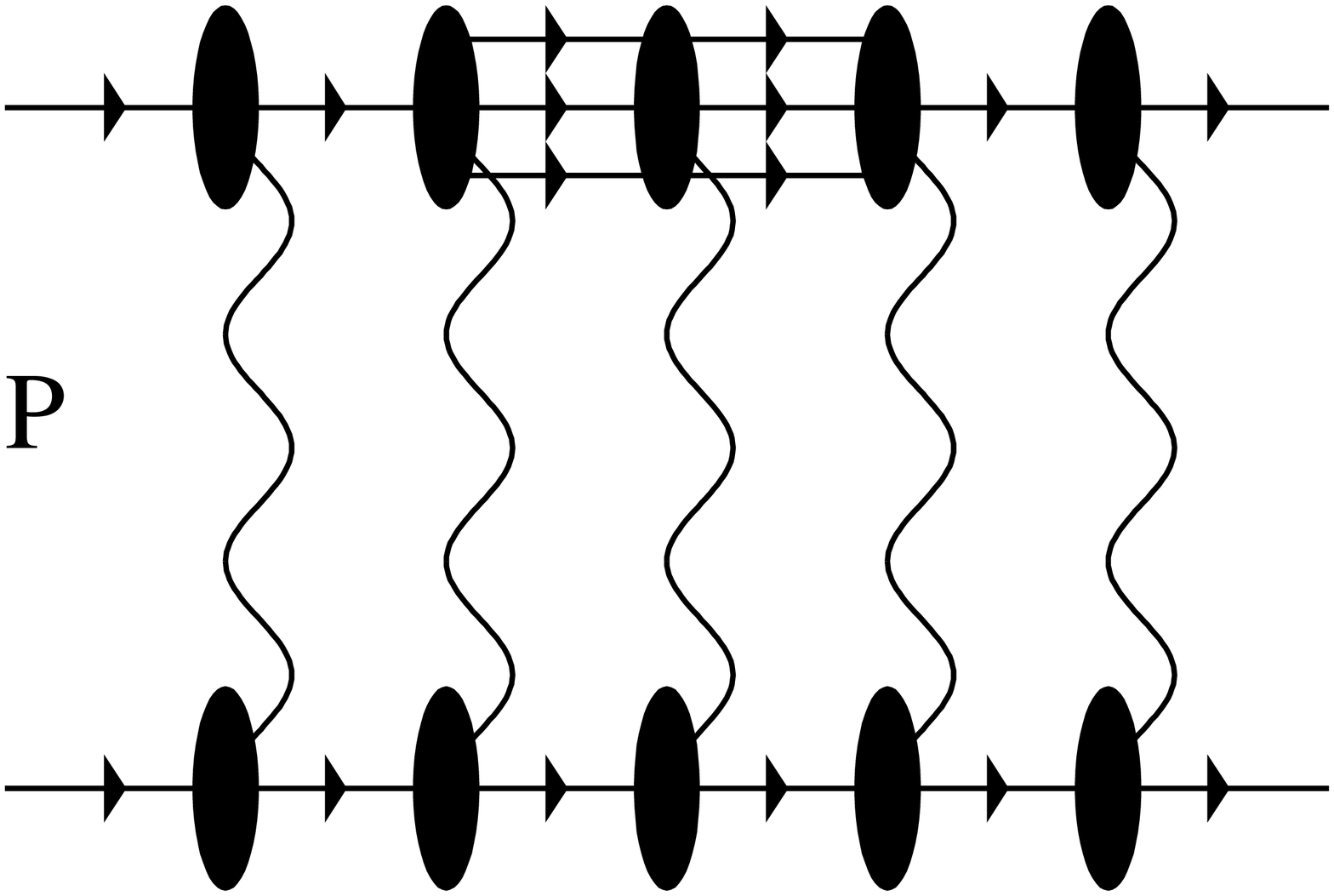,width=120mm,height=25mm}}
\caption{ \it  The Pomeron exchanges in the Quasi-Eikonal model.}
\label{fig27 }
\end{figure}

We can rewrite \eq{QE9} - \eq{QE11} in more convenient form, using the
experimental fact that the exponential parameterization works quite well
for differential cross section of DD. Indeed, we can assume that
\beq \label{QE12}
\Omega_2 \,\,-\,\,\Omega_1\,\,=\,\,\Omega_D(s,b_t)\,= \,
\frac{\sigma_{D0}}{\pi R^2_D(s)}\,( \frac{s}{s_0})^{\Delta_P}\,e^{-
\frac{b^2_t}{R^2_D(s)}}\,\,,
\eeq
 while for $\Omega_1$ we use the parameterization of \eq{QE8}. 
In such a parameterization we have a very simple form for $a_{el}$ ands
$a_D$, namely
\begin{eqnarray}
&
a_{el}(s,b_t)\,\,=\,\,1\,\,-\,\,e^{-\frac{\Omega_1}{2}}\,\,+\,\,( 1 -
\beta^2_1)\,e^{-\frac{\Omega_1}{2}}\,\{
\,1\,-\,e^{-\frac{\Omega_D}{2}}\,\}\,\,; &\label{QE13}\\
&
a_D(s,b_t)\,\,=\,\,\beta^2_1\,(1
- \beta^2_1)\, \{\,1\,-\,e^{-\frac{\Omega_D}{2}}\,\}\,\,
e^{-\frac{\Omega_1}{2}}\,\,.&\label{QE14}
\end{eqnarray}

 Fig. 28 shows the result of fitting of energy behaviour of different
``soft" observables. The parameters, that were used, are

1. $\Delta_P$\,\,=\,\,0.1\,\,;

2.  $\frac{\sigma_{10}}{R^2_1}\,\,=\,\,0.06
\,\,\frac{\sigma_{D}}{R^2_D}\,\,=\,\,\frac{50}{\pi R^2_1}$\,\,;

3.$\beta_2$= 0.5

4. $R^2_1\,\,=\,\,26\,\,+\,\,4\,\alpha'_P(0)\,\ln(s/s_0)$ with
$\alpha'_P(0) = 0.25\,\,GeV^{-2}$.

5.  $R^2_D\,\,=\,\,13\,\,+\,\,4\,\alpha'_P(0)\,\ln(s/s_0)$ with
$\alpha'_P(0) = 0.25\,\,GeV^{-2}$.
 
The only parameter which we have to comment is a sufficiently big
difference in $\frac{\sigma{0i}}{R^2_i} $ for elastic and diffractive
scattering. We think,  that it happened because of the fact that we took
effectively into account the integration over produced mass in
our parameterization.

\begin{figure}
\begin{tabular}{c c}
\psfig{file= 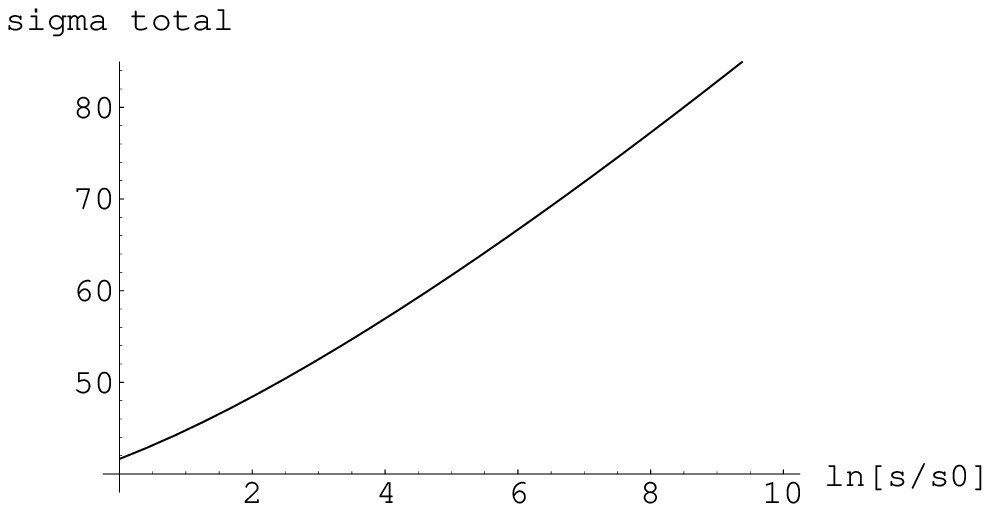, width=70mm,height=70mm} &\psfig{file= 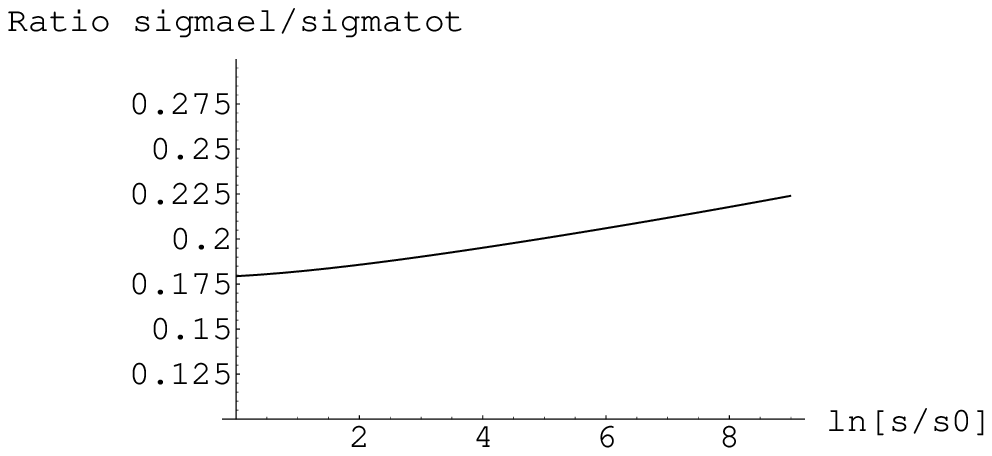,
width=70mm,height=70mm}\\
Fig. 28-a & Fig. 28-b\\
\psfig{file= 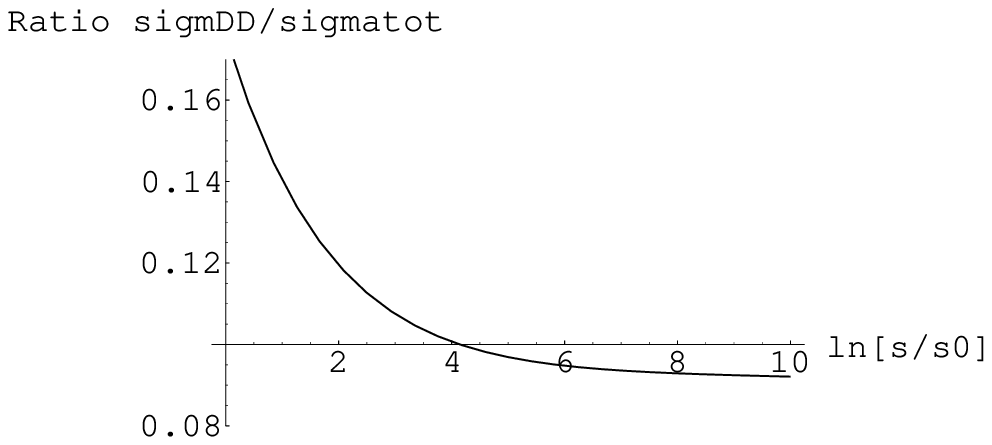,width=70mm,height=70mm} &\psfig{file=
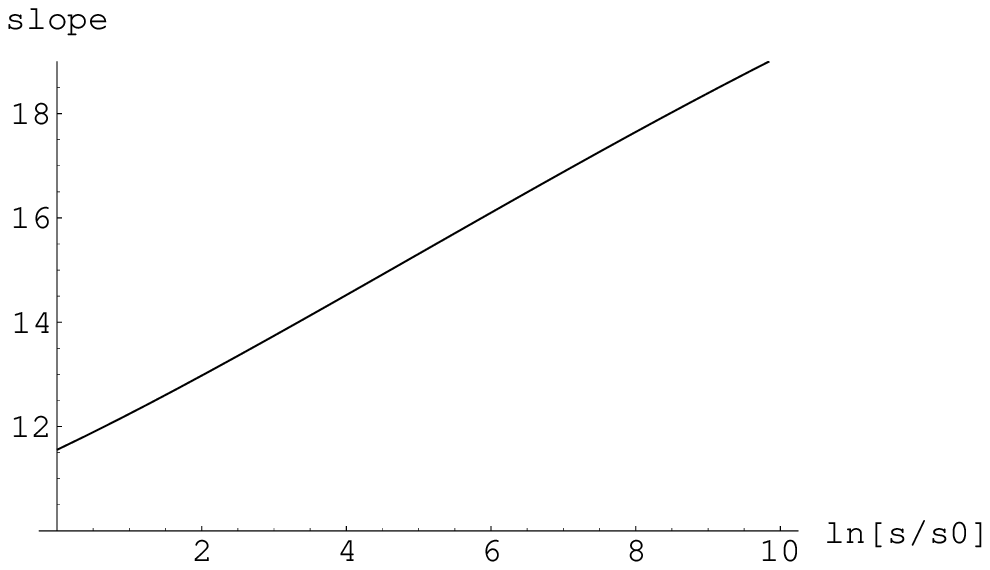,width=70mm,height=70mm}\\
Fig. 28-c & Fig. 28-d\\
\end{tabular}
\vspace{2cm}
\caption{ \it  The ``soft" observables in the Quasi - Eikonal model.}
\label{fig29 }  
\end{figure}
\centerline{\it ``Fan" diagrams (Schwimmer resummation ).}
The technical problem that we are going to solve here is the resummation
of all ``fan" diagrams (see for example Fig. 18-b and Fig. 29-a). it is
interesting to mention that there is a problem which can be solved by
such a resummation, namely, the interaction of a fast hadron with heavy
nuclei ( see Refs. \cite{SCHWIM} \cite{LR81} ). Indeed, the
vertex for
Pomeron - nucleus interaction is proportional to  $g_N\,A$, where $g_N$ is
the Pomeron-nucleon vertex and $A$ is the atomic number. Therefore, the
``fan" diagrams give you the maximal contribution ( $\propto\,( g_N G_{3P} 
A)^n$  where $G_{3P}$ is the vertex for triple Pomeron interaction ).
Neglecting all momentum transferred dependence  in nucleon - nucleon
scattering since $R^2_N\,=\,R^2_0\,+\,\alpha_P(0) \ln(s/s_0)
\,\,\ll\,\,R^2_A$ we can write a simple equation which sums all ``fan"
diagrams. The graphic form of this equation is presented in Fig.29-b and
it has the following analytical form for p + A interaction at
$Y\,=\,\ln(s/s_0)$:
\beq \label{FD1}
G_P(Y,b_t)\,\,=\,\, g^2_N A F_A(b_t)\,e^{\Delta_P\,\,Y}\,\,-\,\,
g_N\,G_{3P}\,\,\int^{Y}_0\,dy\,\,e^{Y\,-\,y}\,\,G^2_P(y,b_t)\,\,.
\eeq
 
\begin{figure}
\begin{tabular}{c}
\psfig{file= 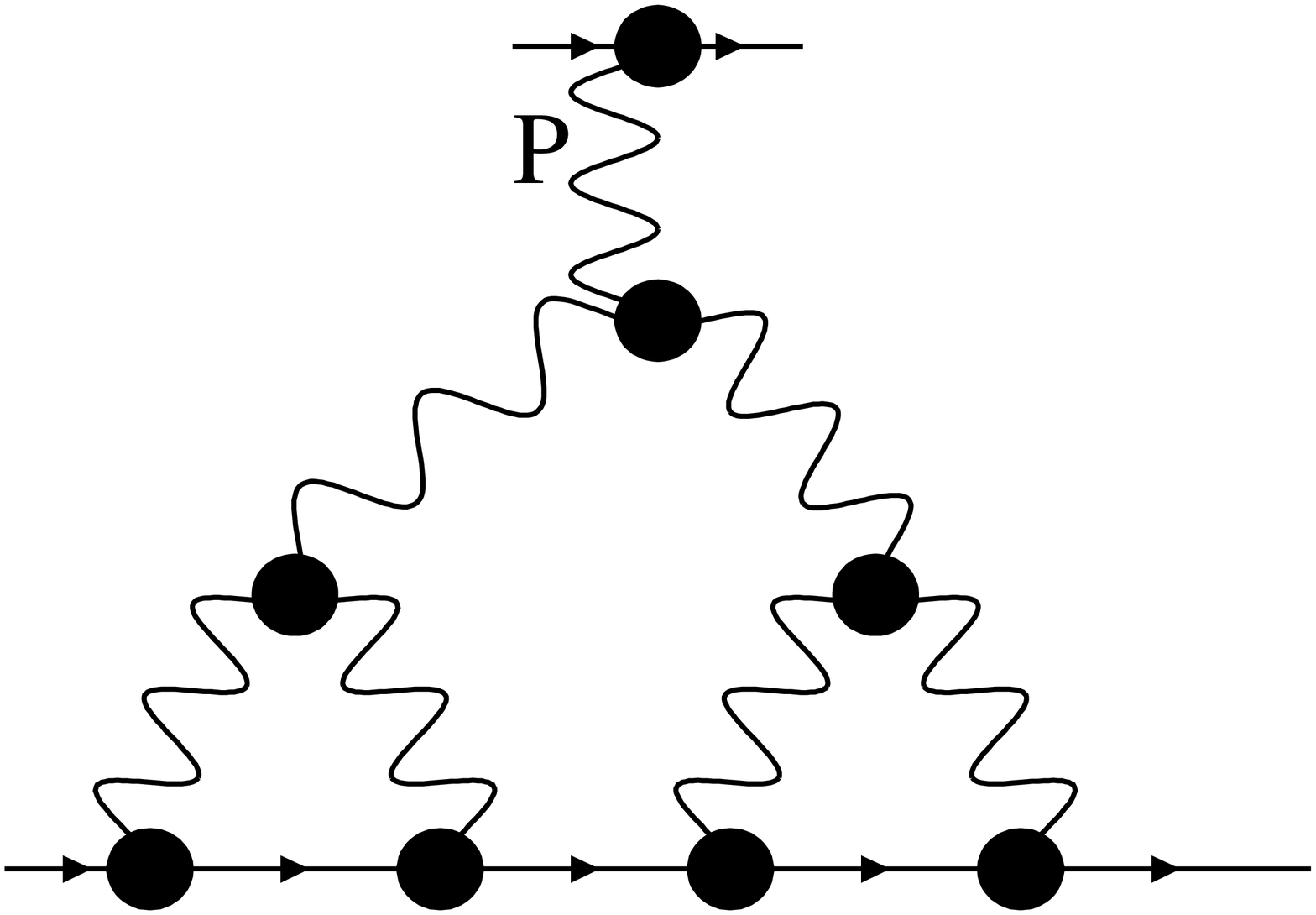, width=100mm,height=70mm}\\
Fig. 29-a \\
\psfig{file= 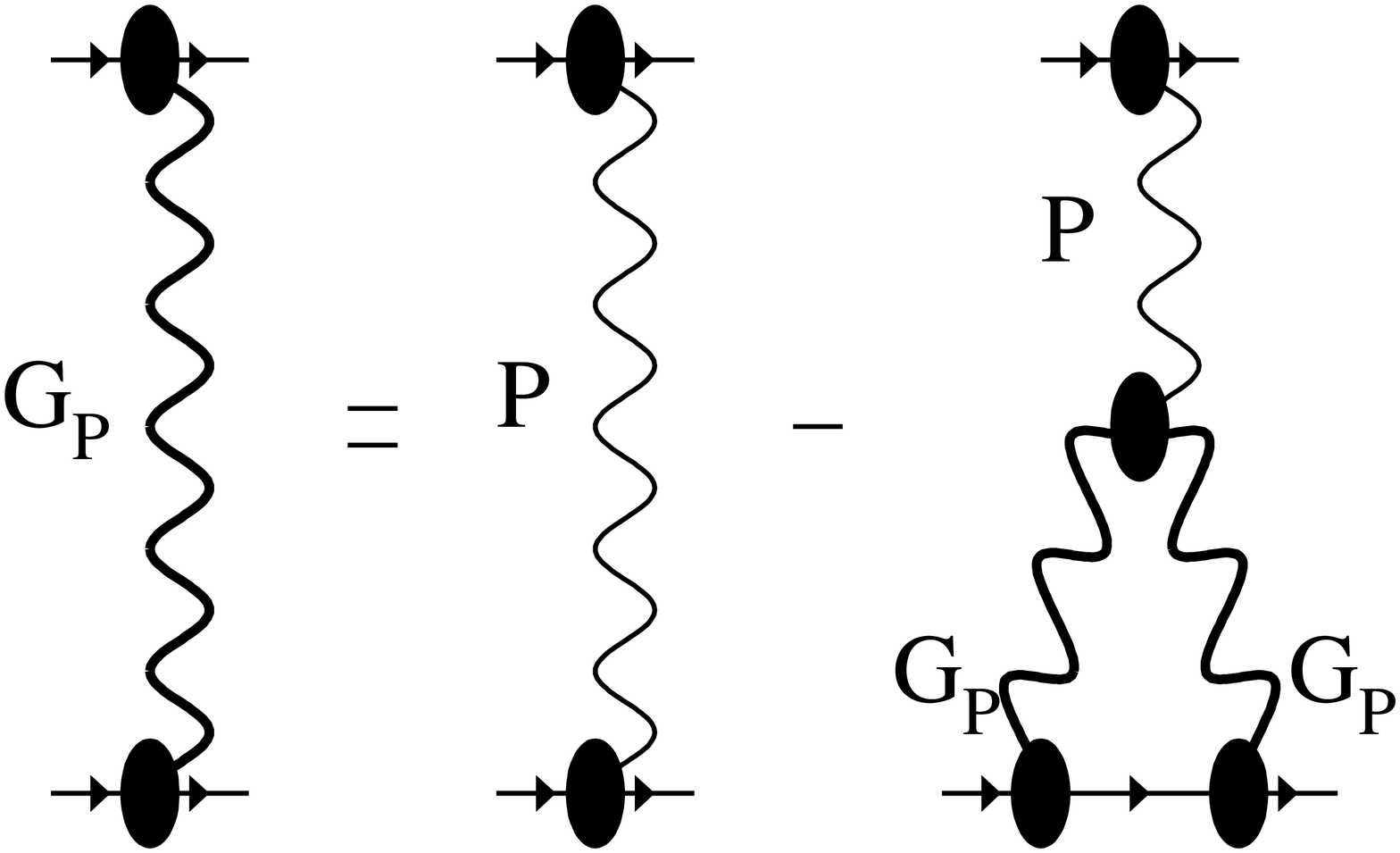,width=140mm,height=50mm}\\
Fig. 29-b\\
\end{tabular}  
\vspace{2cm}  
\caption{ \it ``Fan" diagrams for nucleon - nucleusa interaction: (a) an 
example and (b) the graphic form of equation for summation of all ``fan" 
diagrams.} 
\label{fig30 }
\end{figure}  

Taking derivative with respect to $Y$  and using \eq{FD1} we can reduce
this equation to simple differential equation:
\beq \label{FD2}
\frac{d \,G_P(Y.b_t)}{d Y}\,\,=\,\,\Delta_P
G_P(y.b_t)\,\,-\,\gamma\,G^2_P(Y,b_t)\,\,,
\eeq
where $gamma = g_N \,G_{3P}\,$.  The solution should satisfy the
following initial condition which is obvious directly from \eq{FD1}:
\beq \label{FD3}
G_P(Y,b_t)|_{Y = 0}\,\,=\,\,g^2_N A F_A(b_t) \,\,,
\eeq
where $F_A(b_t)$ directly related to form factor of  a nucleus
$F(t)\,\,=\,\,\int\,\, d^2b_t\,e^{i\,\vec{q} \cdot \vec{b}_t}\,\,F_A(b_t)$
with $q^2 = -t$.

One can obtain the solution: 
\beq \label{FD4}
G_P(Y,b_t)\,\,\,=\,\,\frac{g^2_N\,A\,F_A(b_t)\,\,e^{\Delta_P\,Y}}{1\,\,+\,\,
g^2_N\,A\,\,F_A(b_t)\frac{\gamma}{\Delta_P}\,\,
\{\,e^{\Delta_P\,Y}\,\,-\,\,1\,\}}\,\,.
\eeq

One can see that this solution has a nice properties:
\begin{enumerate}
\item If $A\,F_A(b_t)\,e^{\Delta_P \,Y}\,\,\gg\,\,1$
\,\,$G_P\,\,\Longrightarrow\,\,\frac{\Delta_P}{\gamma}$\,\,;

\item If $A\,F_A(b_t)\,e^{\Delta_P\,Y}
\,\,\ll\,\,1$\,\,\,
$G_P\,\,\Longrightarrow\,\,g^2_N\,A\,F_A(b_t)\,\,e^{\Delta_P\,Y}$\,\,;

\item the equation 
\beq \label{FD5}
g^2_N\,A\,F_A(b_0)\,\,\frac{\gamma}{\Delta_P}\,\,e^{\Delta_P
\,Y}\,\,=\,\,1
\eeq
defines the value of $b_t = b_0$, which is the border between the above
two conditions. Namely, for $b_t\,\,<\,\,b_0$ the first one holds, while
for $b_t\,\,>\,\, b_0$ the form  factor is so small that we are in the
region of the second constraint.

\item Talking for simplicity the exponential form for form factor
$F_A(b_t)\,\,= \,\,\frac{1}{\pi R^2_A}\,e^{-\,\frac{b^2_t}{R^2_A}}$ 
we obtain the solution to \eq{FD5}:
\beq \label{FD6}
b^2_0\,\,=\,\,R^2_A\,\,\Delta_P\,Y\,
\,\,\ln(g^2_N\,A\,F_A(b_0)\,\,)\frac{\gamma}{\Delta_P})\,\,.
\eeq

\item From \eq{FD6} we can obtain the asymptotic formula for the total
cross section:
\beq \label{FD7}
\sigma_{tot}(nA)\,\,\Longrightarrow\,\,2 \pi
b^2_0\,\,\propto\,A^{\frac{2}{3}}\Delta+p\,Y\,\ln(A Y)
\eeq

\item The cross section of the diffractive dissociation turns out to be
constant at large $M^2 \,\sim\,s$. Therefore, this approach has no such
difficulties as Eikonal and/or Quasi-Eikonal Model. The reason for this
is quite simple: In the Eikonal model we took into account only the
rescatterings of the fastest parton (see Fig.18-a ) in the parton cascade.
The ``fan" diagrams describes the interaction of all partons with the
target as it can be seen from Fig.18-b.

\item \eq{FD7} gives a nice illustration that we can have an increase in
the radius of interaction without assuming any slope for Pomeron
trajectory.
\end{enumerate}
The ``fan" diagrams gives a rather selfconsistent approach which heal a
lot of difficulties but the main problem is to justify this approach. I
gave one example - hadron - nucleus collision-  for which the problem can
be reduced to summation of the ``fan" diagrams. You will see more in this
lecture. 

To finish with hadron - nucleus interaction we have to treat $G_P$ as
opacity in the eikonal formula $\Omega = G_P$ (see \eq{EA1} - \eq{EA4} ).

\centerline{}

\centerline{\Huge \bf II\,\,\,``H A R D"\,\,\, P O M E R O N}.

\section{EVOLUTION EQUATIONS AT $\mathbf x\,\longrightarrow\,\,0$}

\subsection{ Parton cascade and evolution equations:}
As it has been mentioned ( see \eq{1.10} - \eq{1.15} ) ``hard" Pomeron is
nothing more than the high energy behaviour of the scattering amplitude at
short distances that follow from the pQCD. In spite of the fact that
solution of the evolution equations gives nothing like the Regge pole -
Pomeron, the parton cascade picture  ( see Fig.1 ) is still correct for
interpretation.  
Let me derive for you \eq{1.15} and show you that that simple equations
(\eq{1.1} - \eq{1.5} ) work quite well for the ``hard" Pomeron too.

The probability to emit one extra gluon in QCD is equal to
\beq \label{HP1}
P_i\,\,=\,\,\frac{N_c\alpha_S(k^2_{ti})}{\pi}\,\frac{d
x_i}{x_i}\,\frac{d k^2_{ti}}{k^2_{ti}}\,\,,
\eeq
where $x_i$ is a fraction of energy carried by gluon $``i"$ and $k_{ti}$
is its  transverse momentum. Factor $\frac{d x_i}{x_i} d k^2_{ti}$ is
nothing more than the invariant volume for emission ( $d^3 k_i/E_i $ ) and
the extra $k^2_t$ comes from the gluon propagator. You can check that
\eq{HP1} gives the only possible dimensionless combination. Recall, QCD is
dimensionless theory.

To calculate the probability to emit $n$ ( $W_n$ ) gluons we need to find
the
product of probabilities, namely
\beq \label{HP2}
W_n\,\,=\,\,\prod^{n}_{i = 0} \,\,\,P_i\,\,,
\eeq
and to find the probability to emit arbitrary number of gluons we have to
sum \eq{HP3} over $n$:
\beq \label{HP3}
W\,\,=\,\,\sum^{\infty}_{n = 0}\,\,W_n\,\,=\,\,\sum^{\infty}_{n =0}
\,\,\prod^{n}_{i = 0} \,\,\,P_i
\eeq
$$
=\,\,\sum^{\infty}_{n =0}\,\,\prod^{n}_{i = 0} 
\,\,\,\frac{N_c\alpha_S(k^2_{ti})}{\pi}\,\frac{d
x_i}{x_i}\,\frac{d k^2_{ti}}{k^2_{ti}}\,\,.
$$
To obtain the deep inelastic structure function we have to integrate over
all possible emission with the conditions:
$ x_i\,\,\geq\,\,x_B$ and $k^2_i\,\,\leq\,\,Q^2$ where
$x_B\,\,=\,\,\frac{Q^2}{s}$ and $q^2$ is the virtuality of the probe.
The largest contributions we can obtain from the phase space in which both
integrals with respect $xi$ and $k^3{ti}$ give  logs ( $ ln (1/x_B)$ and $ln
Q^2$, respectively). This phase space is the following:
\begin{eqnarray}
&
x_0\,\gg\,x_1\,\gg\,x_2\,\,\gg\, ...\,\gg\,x_i\,\gg \,x_{i +
1}\,\gg\,...\,\gg\,x_n\,\gg\,x_B \,\,;&\label{HP4}\\
&
Q^2\,\gg\,k^2_{tn}\,\gg\,k^2_{t,n-1}\,\gg \,...\,\gg\,k^2_{t
1}\,\gg\,Q^2_0\,\,.&\label{HP5}
\end{eqnarray}

Therefore, we have a strong ordering in transverse momenta and fractions 
of energy. Taking into account \eq{HP4} and \eq{HP5}, we can easily
rewrite ( in the simplest case, neglecting $k^2$ dependence of
QCD coupling constant )
\beq \label{HP6}
\int
\prod^{n}_{i=0}\,\frac{N_c\alpha_S(k^2_{ti})}{\pi}\,\frac{d
x_i}{x_i}\,\frac{d k^2_{ti}}{k^2_{ti}}\,\,=\,\,\frac{1}{n!^2}\,
\{\,\frac{N_c\alpha_S(Q^2_0)}{\pi}\,\ln (Q^2/Q^2_0) \,\ln
(1/x_B)\,\}^n\,\,\,.
\eeq
Therefore, the structure function can be found just summing all
contributions of \eq{HP6}:
\beq \label{HP7}
xG(x,Q^2) \,\,=\,\,\sum^{\infty}_{n=0}\,\,\frac{L^n}{n!^2}\,\,
\eeq 
where scale $L$ is equal to
\beq \label{HP8}
L\,\,\,=\,\,\frac{N_c\alpha_S(Q^2_0)}{\pi}\,\ln (Q^2/Q^2_0) \,\ln
(1/x_B)\,\,=\,\,\xi \,\,y\,\,\,
\eeq
where $\xi \,\,=\,\,\frac{N_c\alpha_S(Q^2_0)}{\pi}\,\ln (Q^2/Q^2_0)$ and
$y =
\ln(1/x_B)$. For large $Q^2$ and small $x_B$ parameter $L\,\,\gg\,\,1$
and, therefore, only sufficiently large $n$ will be dominant in \eq{HP8}.
Since 
\beq \label{HP9}
2n!\,\,=\,\,\Gamma(2n + 1)\,\,=\,\,\frac{2^{2n}}{\sqrt{\pi}}\,\Gamma(n +
1) \,\,\Gamma(n +
\frac{1}{2})\,\,\Longrightarrow\,\,\frac{2^{2n}}{\sqrt{\pi}} (n!)^2\,\,
\eeq
we can rewrite \eq{HP7} in the form
\beq \label{HP10}
xG(x,Q^2) \,\,\Longrightarrow\,\,\sum^{\infty}_{n=0}\,\,\frac{(\,2
\,\sqrt{L}\,)^{2n}}{2n!}\,\,=\,\,e^{2\sqrt{L}}
\eeq
On the other hand, using Stirling formula, we can evaluate 
$$ 2n!\,\,\,\rightarrow\,\,\,(2n)^{2n} \,e^{-2n}\,\,,$$  
which, being substituted in \eq{HP10} gives the estimate for the average
$n$, namely,
\beq \label{HP11} 
< n >\,\,= \,\,2 \sqrt{L}\,\,.
\eeq
One can see that \eq{HP10} can be written as
\beq \label{HP12}
xG(x,Q^2)|_{L\,\gg\,\,1} \,\,\Longrightarrow\,\,e^{< n >}    \,\,.
\eeq

For running coupling constant the only difference is the redefinition of
the variable $\xi = \ln \frac{\alpha_S (Q^2)}{\alpha_S(Q^2_0)}$ in
\eq{HP8}.

\subsection{Why do  we use  evolution equations?} 
The first question is what is the meaning of $< n >$. To answer this
question we need to look in the picture of our cascade more careful.
Let us assume that the number of ``wee" partons $N$ is not extremely
large or in other words there is only one ``wee"  parton with definite
fraction of energy ($x_i$ ) and transverse momentum ($k_{ti}$ ). 
In this case only one ``wee" parton interacts with the target and all
others gather together without loosing their coherency. In the Feynman
diagrams they contribute to the renormalization of mass and  coupling
constant in QCD. This is shown in Fig.30.

\begin{figure}
\centerline{\psfig{file= 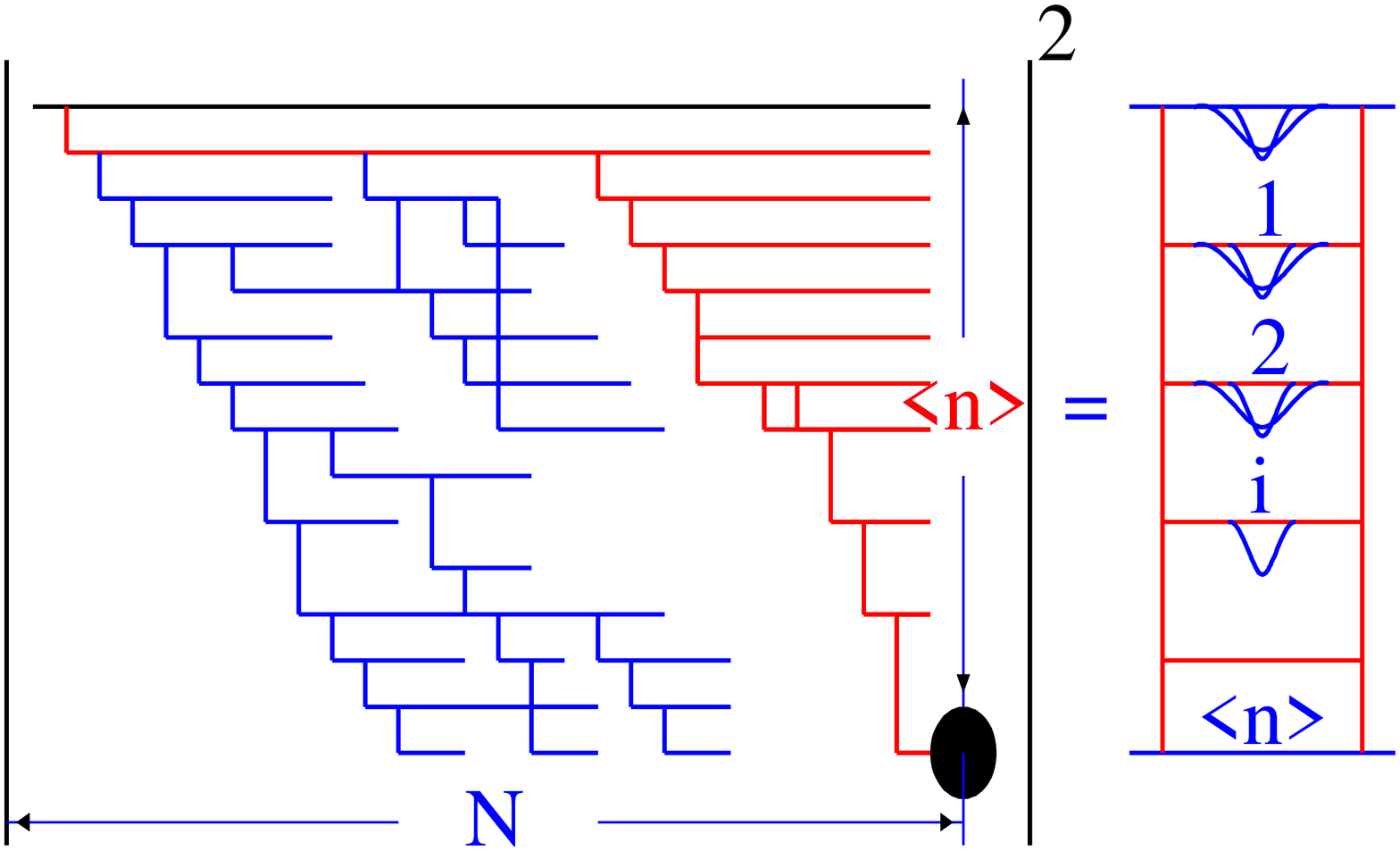, width=140mm,height=70mm}}
\vspace{2cm}
\caption{ \it Parton cascade and evolution equation ( Feyman diagrams).}
\label{fig31 }
\end{figure}

In Fig. 30 one can see that $< n >$ means the average number of cells in
our ``ladder" diagram or, in other words, in our evolution equations.
On the other had, $< n >$ is the average number of produced jets ( mini -
jets ) in the whole region of rapidity accessible in the experiment.
I found very instructive to plot what we know from HERA experiment
about the value of $< n >$ ( see Fig.31 )

\begin{figure}
\centerline{\psfig{file= 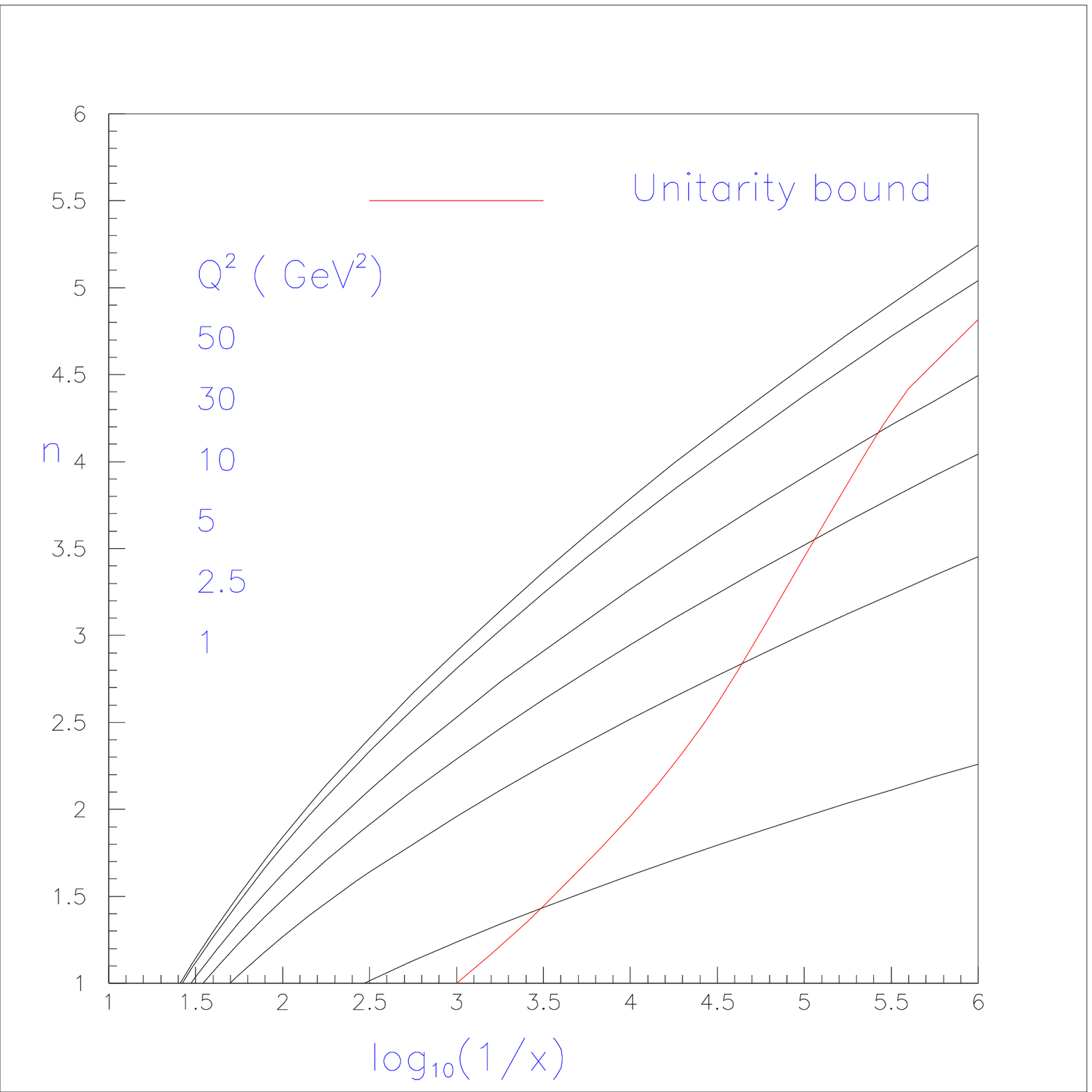, width=100mm}} 
\vspace{2cm} 
\caption{ \it $< n >$ versus $x$ at different values of $Q^2$ from HERA
experiment.}  
\label{fig32 }                                                            
\end{figure}  
In Fig. 32 $< n >$ is plotted down to $x\,=\,10^{-6}$ and looking at this
plot we can see, that
\begin{itemize}

\item  $<  n  >$  increases mostly due to cascading rather in
$Q^2$ than
in $\ln(1/x)$\,\,;

\item The maximum value of $< n > $ is about 5\,\,;

\item $ < n >$ is the total multiplicity of produced jets and/or minijets.
Therefore, the average density of produced jets ( minijets)
 is less than $d < n > /d y\,\,\approx\,\,0.5$. This number suggests that
we need to calculate only the Feynman diagrams of up to $\alpha^2_S$
or/and $\alpha^3_S$ order\,\,;

\item High energy resummation is only a way
to get $N$
without calculating $<   n  >$ \,\,.

\end{itemize}

The main conclusion from this discussion is very simple: we have to use
the evolution equations only because it is the only known way how to
calculate $N$ without evaluation of  $< n > $. However, if we will find a
way to calculate $< n >$ we will be able to predict our structure
functions with better accuracy restricting yourselves by calculation only
limited number of diagrams.

\section{SC FOR  ``HARD" POMERON}
 The main outcome of our previous discussion is the fact that
$N\,\,\gg\,\,1$ at high energies. It happens both for ``hard" ( see
\eq{1.15} and/or \eq{HP11}) and ``soft" processes. Therefore, in the
collisions at high energy the new interesting system of parton has been
created with large density of partons. For the ``hard" processes all these
partons are at the short distances where the coupling QCD constant is
small enough to be use as a small parameter. However, we cannot apply for
such a system the usual methods of perturbative QCD because the density of
partons is large. In essence, the theoretical problems here are non
perturbative, but the origin of the non perturbative effects does not lie
in long distances and large values of $\alpha_S$, which are typical for
the confinement region.
\subsection{Which parton density is large?}

 The quantative estimates of which  density is high, we can obtain from
the
$s$-channel unitarity \cite{GLR} which we can use in two different form:
\begin{itemize}
\item  $\sigma_{tot}( \gamma^* p ) \,\,\leq\,\alpha_{em}\,2\,\pi\,R^2$,
where $R$ is the size of the target, $\alpha_{em}$ is the fine
structure constant ( see  Ref. \cite{GRIB} for explanation why we 
have $\alpha_{em}$ ).   This constraint can be rewritten as
\cite{GLR} \cite{MUQI}\cite{MU90}
\beq \label{HSC1}
\kappa\,\,=\,\,\frac{3\,\pi\,\alpha_S x
G(x,Q^2)}{Q^2\,R^2}\,\,\leq\,\,1\,\,.
\eeq

\item $\sigma^{DD}( \gamma^* p )\,\,\leq\,\,\sigma_{tot}( \gamma^* p )$,
where $DD$ stands for diffractive dissociation. This inequality
leads also to  \eq{HSC1}.
\end{itemize}

Therefore,
\begin{enumerate}  
\item   If $\kappa$ is very small ($\kappa \ll 1$ ), we have a
low density QCD in
which
 the parton cascade can be perfectly described by the DGLAP evolution
equations \cite{DGLAP};

\item  If  $ \kappa\,\,\leq\,\,1$, we are in the
transition region between low and high density QCD . In this region we
can still use pQCD, but have to take into account   the interaction
between
partons inside  the partons cascade; 

\item If $\kappa\,\,\geq\,\,1 $, we
reach the region of high parton density QCD, where we need to use quite
different methods from pQCD
\end{enumerate}

First,  we want to make a remark on parton densities in a
nucleus. Taking into account that for nucleus $R_A\,\,=\,\,R_N
\,\times\,A^{\frac{1}{3}}$ and $xG_A(x,Q^2)\,\,=\,\,A\,xG_N(x,Q^2)$, we
can rewrite \eq{HSC1} as
\beq \label{HSC2}
\kappa_A\,\,=\,\,\frac{3\,\pi\,\alpha_S A\, x
G_N(x,Q^2)}{Q^2\,R^2_N\,\,A^{\frac{2}{3}}}\,\,=\,\,A^{\frac{1}{3}}\,\kappa_N
\,\,\leq\,\,1\,\,.
\eeq
Therefore, for the case of an interaction with nucleii, we can reach a
hdQCD region at smaller parton density than  in a nucleon .

Fig.32  gives
the kinematic plot $(x,Q^2)$ with the line $\kappa = 1 $, which shows that
the hdQCD effect should be  seen  at HERA.

\begin{figure}
\centerline{\psfig{file= 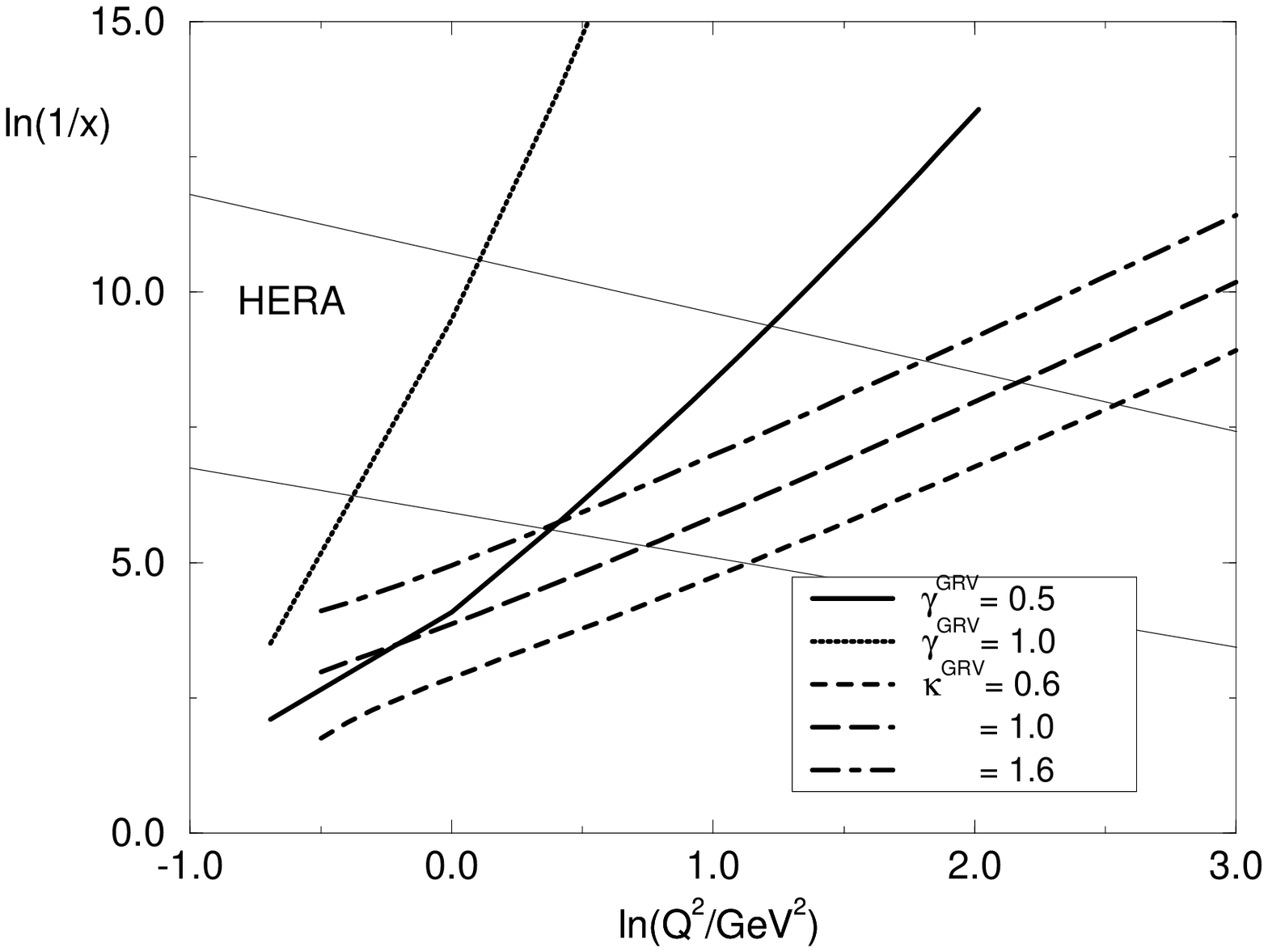,width=100mm}}
\caption{ \it Contours for $< \gamma >$  and $\kappa$   for the GRV'94
gluon density and HERA kinematic region.}
\label{fig32}
\end{figure}

\subsection{Two different theoretical approaches}
To understand these two approaches we have to look back at Fig.1 which
gives you a picture of high energy interaction in the parton approach.
In \eq{1.1} $N$ is the flux of partons. The main idea of SC is the fact
that this flux should be { \it renormalized } in the case of high parton
density. Indeed, if the number of ``wee" partons is so large that, let
say, two partons have the same kinematic variables : $x_i$ and
$\vec{k}_{ti}$, we do not need to take into account the interaction of
these two partons two times. Let me recall, 
that the total cross section is the number of interaction but not
the number of partons. It means that if the first of two partons has
interacted with the target the total cross section does not depend upon
the fact did the second interact or not. It is obvious that the
probability to have two partons which could interact with the target
semalteneously is equal to
\beq \label{HSC3}
P_2\,\,=\,\,\frac{\sigma_0\,\,N}{\pi R^2(x_i)}\,\,,
\eeq
where $\pi R^2$ is the area which is populated by partons with the
fraction of energy $x_i$. Therefore, we have to renormalize the flux
\beq \label{HSC4}
RENORMALIZED\,\,\,\,FLUX\,\,=\,\,N_{REN}\,\,=\,\,
N\,\times\,\{\,1\,\,-\,\,P_2\,\}\,\,\,.
\eeq
One can see that this renormalized flux gives the total cross section
in the form:
\beq \label{HSC5}
\sigma_{tot}\,\,=
\,\,\sigma^{HP}_{tot}\,\,-\,\,\frac{(\,\sigma^{HP}_{tot}\,)^2}{\pi
\,R^2}\,\,,
\eeq 
which is a Glauber formula for SC in the limit of small SC.

\parbox{5.5cm}{
\begin{flushleft}
If $ N\,\,\approx\,\,1$, we expect that the renormalization
of the flux
will be small, and we use an approach with the following typical
ingredients:

$ \bullet$  Parton Approach;\\
$ \bullet$ Shadowing Corrections;\\
$ \bullet$ Glauber Approach;\\
$ \bullet$ Reggeon-like Technique;\\
$ \bullet$ AGK cutting rules\,.\\
\end{flushleft}}
\parbox{7.5cm}{
\begin{flushleft}
However, when
$ N\,\,\gg\,\,1$,  we have to change our approach completely from  the
parton cascade to one based on semiclassical field approach, since due to
the uncertainty principle
 $\Delta N \Delta \phi \,\approx\,1$, we can consider the phase as a small
parameter.   
Therefore, in this kinematic region our magic words are:\\
$ \bullet$   Semi-classical gluon fields\,;\\
$ \bullet$  Wiezs$\ddot{a}$cker-Williams approximation;\\
$ \bullet$ Effective Lagrangian for hdQCD;\\
$ \bullet$  Renormalization Wilson group Approach.
\end{flushleft}}

It is clear, that for $N \,\approx\,\,1$ the most natural way is to
approach the hdQCD looking for corrections to the perturbative parton
cascade. In this approach the pQCD evolution has been naturally included,
and it aims to describe the transition region. The key problem
 is to penetrate into the hdQCD region where $\kappa $ is large.
Let us call this approach ``pQCD motivated approach ".

For $N \,\gg\,\,1$, the most natural way of doing is to use the effective
Lagrangian approach,  and remarkable progress has been achieved both in
writing of the explicit form  of this effective Lagrangian,  and in
understanding physics behind it  \cite{MCLER}. The key problem
for this approach was to find a correspondence  with pQCD. This problem
has been
solved \cite{KOV}.

Fig.33 shows the current situation on the frontier line in the offensive
on
hdQCD.

\begin{figure}
\centerline{\psfig{file= 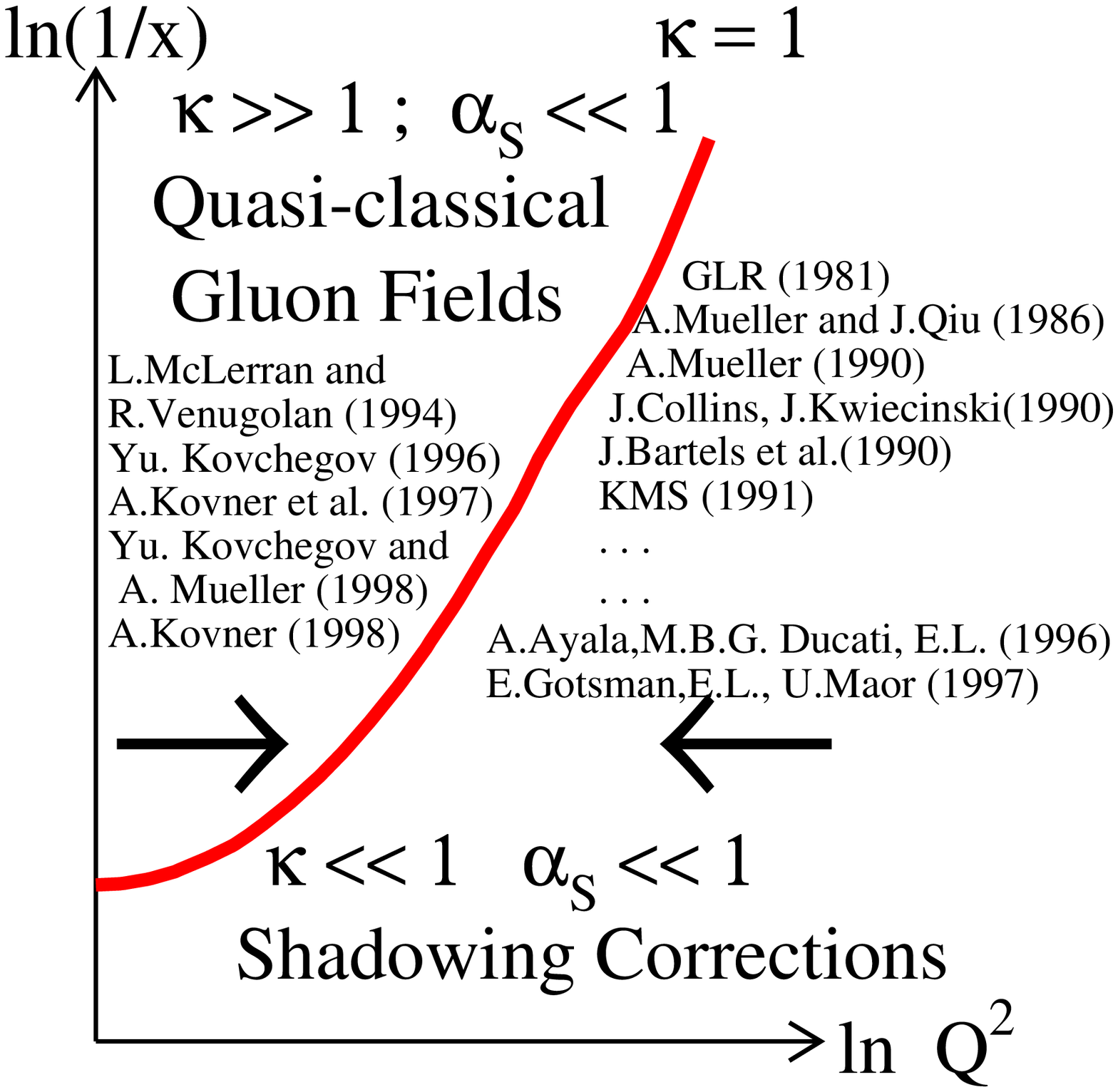,width=100mm}}
\caption{ \it The current situation in the battle for hdQCD.}
\label{fig33}
\end{figure}

\subsection{The picture of interaction}
To understand the picture of interaction 
  in the region of small $x$ it is better to by
examine  the
parton distribution in the transverse plane (see Fig.
those
partons with size $ r_p\,\sim \,\frac{1}{Q}$. At $x\,\
approx\,1 $ a few
parton are distributed in the hadron disc. If we choose $Q$ such that
$r^2_p\,\ll\,R^2_h$ then the distance between partons in the transverse
plane is much larger than their size, and we can neglect the interaction
between partons. The only essential process is the emission of
partons,which has been taken into account in QCD evolution. As $x$
decreases
for fixed $Q^2$ , the number of partons increases. and at value of
$x=x_{cr}$, partons start to populate the whole hadron disc densely.
For $x\,<\,x_{cr}$ the partons overlap spatially and begin  to interact
throughout the disc. For such small $x$ values, the processes of
recombination and annihilation of partons should be as essential as
their emission. However, neither process is incorporated into any
evolution equation. What happens in the kinematic region
$x\,<\,x_{cr}$ is anybody's guess. We suggested that parton density
saturates, i.e. the parton density is constant in this domain.

\subsection{The GLR equation}
The first attempt to take into account the parton - parton interaction in
the pQCD motivated approach was done  long ago \cite{GLR}. It was based 
on the simple idea that there are two processes in a parton cascade (see
 Fig.1) : (i) the probability of the emission of an extra gluon is
proportional to $\alpha_S \,\rho$ where $\rho $ is the density of gluon in
the transverse plane, namely
\beq \label{HSC6}
\rho\,\,=\,\,\frac{xG(x,Q^2)}{\pi\,R^2}\,\,;
\eeq
and (ii) the annihilation   of a gluon,  or in other words a process in
which  the  probability is proportional to $\rho^2\,\times\,
\sigma_{annihilation}$.
$ \sigma_{annihilation}$ can be estimated as $
\sigma_{annihilation}\,\propto \,\alpha_S\,r^2$, where $r$ is the size of
the parton produced in the annihilation process. For deep inelastic
scattering, $r^2\,\,\propto\,\,\frac{1}{Q^2}$.
Therefore, in the parton cascade we have
\begin{eqnarray}
&
Emission\,\,(\,\,1\,\,\rightarrow\,\,2\,\,)\,:
\,\,\,probability\,\,=\,\,P^{emission}\,\,\propto\,\,\alpha_S\,\,\rho\,\,;
& \label{HSC7}\\
&
Annihilation\,\,(\,\,2\,\,\rightarrow\,\,1\,\,)\,:
\,\,\,probability\,\,=\,\,P^{annihilation}\,\,\propto\,\,\alpha^2_S\,\,
\frac{1}{Q^2}\,\,\rho^2 \,\,.& \label{HSC8}
\end{eqnarray}

At $x\,\sim\,1$ only emission of new partons is essential,  because
$\rho\,\ll\,1$ and this emission is described by the DGLAP
evolution equations. However, at $x \,\rightarrow\,0$ the value of
$\rho$ becomes so large that the annihilation of partons becomes
important,  and so the value of $\rho$ is diminished.  The competition of
these  two processes we can write as an equation for the number of
partons in a phase space cell ( $\Delta y
\,=\,\Delta \,\ln(1/x)\,\Delta \,\ln (Q^2/Q^2_0) $ ):
\beq \label{HSC9}
\frac{\partial^2 \rho}{\partial \ln (1/x) \partial \ln (Q^2/Q^2_0)}\,\,=  
\,\,\frac{\alpha_S \,N_c}{\pi}\,\rho\,\,-\,\,\frac{\alpha^2_S\,\,\tilde
\gamma}{Q^2}\,\,\rho^2\,\,,
\eeq
or in terms of the gluon structure function $xG(x,Q^2)$
\beq \label{HSC10}
\frac{\partial^2 x G(x, Q^2)}{\partial \ln (1/x) \partial \ln
(Q^2/Q^2_0)}\,\,=
\,\,\frac{\alpha_S \,N_c}{\pi}\, x G(x,
Q^2\,\,-\,\,\frac{\alpha^2_S\,\,\tilde
\gamma}{\pi R^2\,\,Q^2}\,\,(\, x G(x, Q^2)\,)^2\,\,.
\eeq
This is the GLR equation which gave the first theoretical  basis
for the  consideration  of hdQCD.  This equation describes the
transition region at very large values of $Q^2$, but  a glance  at Fig.32
shows that we need a tool to  penetrate the kinematic
region of moderate and even small $Q^2$.
\subsection{Glauber - Mueller Approach}
We found that it is very instructive to start with the Glauber
approach to SC. The idea of how to write the Glauber formula in QCD was
originally formulated in two papers Ref.\cite{LR87} and Ref.
\cite{MU90}. However, the key paper for our problem is the second
paper of A. Mueller, who considered the Glauber approach for the
gluon structure function. The key observation is that the  fraction of
energy,  and the transverse coordinates of the fast  partons can be
considered as frozen,  during the high energy interaction with the
target \cite{LR87} \cite{MU90}. Therefore, the cross section of the
absorption of gluon($G^*$) with virtuality $Q^2$ and Bjorken $x$
can be written in the form:
\beq \label{HSC11}
\sigma^N_{tot}(\,G^*\,)\,\,=\,\,
\eeq
$$
\int^1_0 d z \,\,\int \,\frac{d^2 r_t}{2 \pi}\,\,
\int\frac{d^2 b_t}{2 \pi}
\Psi^{G^*}_{\perp} (Q^2, r_t,x,z)\,\,
 \sigma_N (x,r^2_t)\,\,\,\,[{\Psi^{G^*}_{\perp}} (Q^2, r_t,x,z)]^* \, ,
$$
where $z$ is the fraction of energy
which is carried by the gluon, $\Psi^{G^*}_{\perp}$ is the wave  
function of the transverse polarized gluon,  and $\sigma_N (x,r^2_t)$ is
the
cross section of the interaction of  the $GG$- pair with transverse
separation
$r_{t}$ with the nucleus.
Mueller showed that \eq{1.11} can be reduced to
\beq \label{MF}
x G^{MF}(x,Q^2) = \frac{4}{\pi^2} \int^{1}_{x} \frac{d x'}{x'}
\int_{\frac{1}{Q^2}}^{\infty} \frac{d^2 r_t}{\pi r_{t}^{4}}
\int_{0}^{\infty} \frac{d^2 b_t}{\pi}  2
\left\{ 1 - e^{- \frac{1}{2}\,\sigma_{N}^{GG} ( x^{\prime},r^2_t )
S(b^2_t) }
\right\}
\eeq
with
$$
\sigma_{N}^{GG}\,\,=\,\,\frac{4 \,\pi^2\,\alpha_S }{3}\,\,r^2_t\,x
G^{DGLAP}(x,\frac{1}{r^2_t})\,\,
$$
and profile function $S(b_t)$ chosen in the Gaussian form:
$
S(b^2_t)\,\,=\,\,\frac{1}{\pi R^2}\,\,e^{-\,\frac{b^2_t}{R^2}}\,\,.
$

Obviously, the Mueller formula has a defect, namely, only the fastest
partons ($GG$ pairs) interact with the target.
This assumption is an artifact of the Glauber approach, which looks
strange in
 the parton picture of the interaction. Indeed, in the parton model
we   expect that all partons not only the fastest ones should
interact
 with the target.  At first sight we can solve this problem by  iteration
\eq{MF}( see Ref. \cite{AGL} ). It means that the first iteration will
take into account that not only the fastest parton, but the next one will
interact with the target, and so on.

\subsection{ Why equation?}
We would like to suggest a new approach based on the  evolution
equation
 to sum all SC. However, we  want to argue first,  why an equation is
 better than any iteration procedure. To illustrate this point of view,
let
us
differentiate the Mueller formula with respect to $y\,=\,\ln(1/x)$ and $
\xi\,=\,\ln Q^2$. It is easy to see that this derivative is equal to
\beq \label{DER}
\frac{\partial^2 x G(x, Q^2)}{\partial y\, \partial \xi}\,\,=\,\,
\frac{4}{\pi^2}\,\int d b^2_t
\,\,\{\,\,1\,\,-\,\,e^{-\,\frac{1}{2}\,\sigma(x,
r^2_{\perp}\,=\,\frac{1}{Q^2})\,S(b^2_t)}\,\,\}\,\,.
\eeq
The advantages  of \eq{DER} are
\begin{itemize}
\item Everything enters at small
distances\,\,;

\item  Everything is under theoretical control\,\,;
 
\item  Everything that is not known  ( mostly  non perturbative )
is hidden in the initial and boundary condition.
\end{itemize}

Of course, we cannot get rid of our problems changing the procedure 
of solution. Indeed, the non-perturbative effects coming from the
large distances are still important,  but they are absorbed in the
boundary and initial conditions of  the equation. Therefore, an
equation is a good ( correct ) way to separate  what we know (
short distance contribution) from what we don't ( large distance
contribution).

\subsection{Equation}
We suggest the following way to take into account the interaction of all
partons in a parton cascade with the target.
Let us differentiate the $b_t$-integrated Mueller
formula of \eq{MF} in $y \, = \,\ln (1/x)$ and $ \xi = \ln(Q^2/Q^2_0)$.
 It gives
\beq \label{EQ}
\frac{\partial^2 x G ( y,\xi)}{\partial y \partial \xi}\,\,=\,\,
\frac{2 \,Q^2 R^2 }{ \pi^2}\,\,\,\left\{\,\, C \,\,+ \,\,\ln(\kappa_{G} (
x',
 Q^2 )) \,\,+\,\,
E_1 (\kappa_{G} ( x', Q^2 ))  \right\} \,\,\,,
\eeq
where $\kappa_G (x,Q^2)$ is given by
\beq \label{KAPPA}
\kappa ( x,Q^2) \,\,=\,\,
\frac{N_c \alpha_S \pi }{2 Q^2 R^2 }\,x G(x,Q^2)\,\,.
\eeq
Therefore, we consider $\kappa_G$ on the l.h.s. of \eq{EQ} as the
observable which is written through the solution of \eq{EQ}.

\eq{EQ} can be rewritten in the form ( for fixed $\alpha_S$
)
\beq \label{EQKA}
\frac{\partial^2 \kappa_G( y,\xi)}{\partial y \partial \xi}\,\,+\,\,
\frac{\partial \kappa_G(y, \xi)}{\partial y}\,\,=
\eeq
$$
\,\,
\frac{ N_c\, \alpha_S}{\pi}\,\, \left\{\,\, C \,\,+ \,\,\ln(\kappa_{G})
\,\,+\,\,
E_1 (\kappa_{G})  \right\}
\,\,\equiv\,\,F(\kappa_G)\,\,.
$$
This is the equation which we propose.

This equation has the following desirable properties:

\begin{enumerate}
\item   It  sums all contributions of the order $
(\,\alpha_S\,y\,\xi\,)^n$,
absorbing them in $x G (y,\xi)$, as well as all contributions of the order
of $\kappa^n$.   
Therefore, this equation solves the old problem, formulated in
Ref.\cite{GLR},
and
 for $N_c\,\rightarrow \,\infty $ \eq{EQKA} gives the complete
solution to our problem, summing all SC\,\,;
\item The solution of this equation matches with the solution of the DGLAP
 evolution equation in the DLA of perturbative QCD at $\kappa\,\rightarrow
\,0$\,\,;

\item   At small values of $\kappa$ ( $\kappa\,<\,1$ )
 \eq{EQKA} gives the GLR equation. Indeed,
for small $\kappa$ we can expand the r.h.s of \eq{EQKA} keeping
only the
 second term. Rewriting the equation through the gluon structure function
 we obtain
  the GLR equation \cite{GLR} with
 the coefficient in front of the second term as
 calculated by Mueller and Qiu \cite{MUQI}\,\,;

\item The first iteration of this equation gives the Mueller formula ( see
 Ref. \cite{MU90})\,\,;

\item In general, everything that we know about SC is included in
\eq{EQKA}\,\,;

\end{enumerate}

\subsection{ The size of SC}
\begin{figure}
\centerline{\psfig{file= 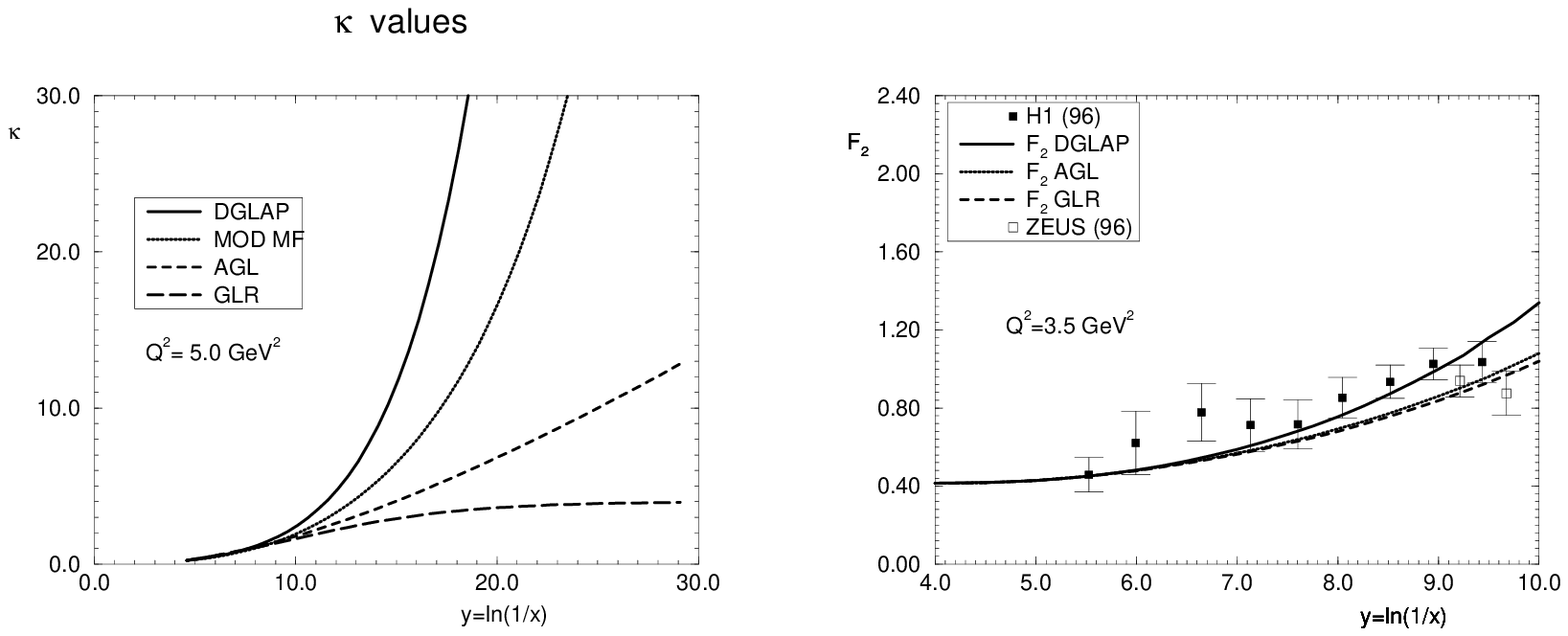,width=130mm}}
\caption{ \it The estimates of SC based on \eq{EQKA}.}
\label{scc}
\end{figure}   
In Fig.\ref{scc} we plot our calculation for the DGLAP evolution
equation, for the Glauber - Mueller approach, for the GLR equation
and for the new evolution equation. We can conclude, that:
\begin{enumerate}
\item SC even in the Glauber - Mueller approach are essential
for the gluon structure function\,\,;

\item The Glauber - Mueller approach considerably underestimates
 the value of SC\,\,;

\item The GLR equation leads to stronger SC than the solution to
\eq{EQKA}\,\,;

\item The new evolution equation does not reproduce the saturation of
 the gluon density
in the region of small $x$ which the GLR equation leads to\,\,.
\end{enumerate}

We  firmly believe that the new evolution equation gives  the
correct way of evaluation of the value of SC. However, the
difficult question  arises: why the SC have not been seen at HERA?
Our answer is:

1. The value of SC is rather large  but only in the gluon structure
function while their contribution to $F_2$ is rather small
\cite{AGL} (see Fig.\ref{scc} ),  and cannot be seen on the background of
the experimental
errors\,\,;

2. The theoretical  determination of the gluon
structure function is not very precise  and we evaluate the errors  as
50\%
\cite{LERIHC} and the SC in $xG$ is hidden in such  large
errors\,\,;

3. The statement that SC have not been seen is also not quite
correct. Our estimates \cite{GLMSLOPEF2} show that the contribution
of SC is rather large in the $F_2$ slope and incorporating  the SC one can
describe the recent experimental data \cite{DATACOLD} on  
$\frac{\partial F_2(x,Q^2)}{\partial \ln (Q^2/Q^2_0)}$ (see
Ref.\cite{GLMSLOPEF2} for detail ).

\subsection{Effective Lagrangian Approach.}
 Deeply in my heart,  I firmly
believe that there will be a bright  future for the effective Lagrangian
approach in  which one
combines the physics of hdQCD, with the formal methods of the
quantum field theory, and gives  a way to incorporate the lattice
calculation for the non-perturbative observables. However, the
parameter or better to say a new scale  $Q^2_0(x)$  which is the
solution of the equation $ \kappa(x,Q^2_0(x))\,\,=\,\,1 $ only can be
found  in the pQCD motivated approach, since the effective Lagrangian was 
 only derived  and justified assuming this new scale.

\begin{figure}
\centerline{\psfig{file= 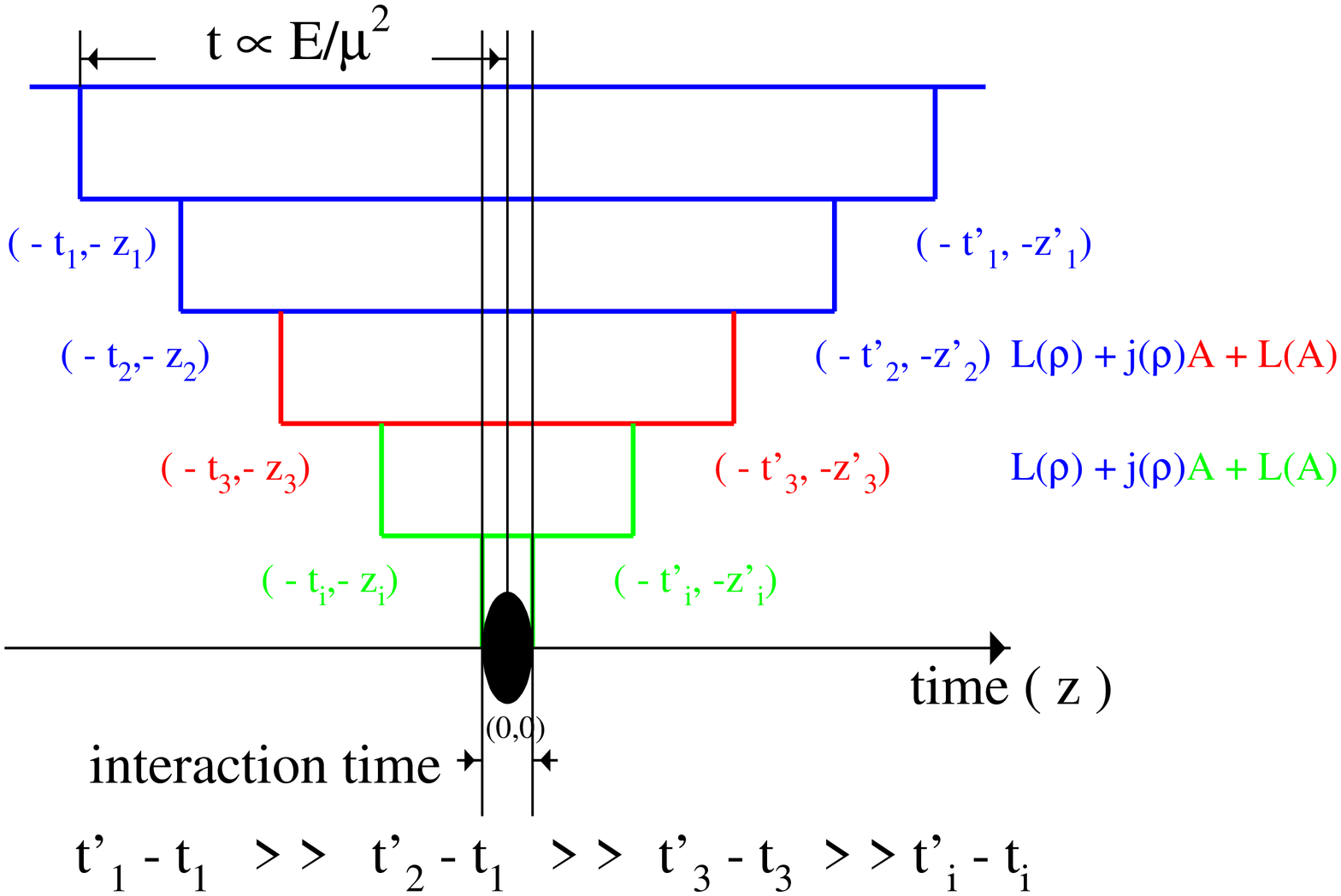,width=130mm, height=65mm}}
\caption{ \it The space - time structure of the parton cascade.}
\label{fig36}
\end{figure}

Here, I would like to show what kind of physics have been used to obtain
the effective Lagrangian. The space - time picture of the parton cascade
is given in Fig.35. The main and very important feature of this cascade is
that in the region of low $x$ ( in  so called leading log (1/x)
approximation of pQCD ) is the fact that a parton with higher energy in
the cascade lives much longer than a parton with smaller  energy.  
Let us look for the interaction of the parton emitted at time $t'_2$. This
parton lives much shorter time than all partons emitted before. Therefore,  
all these emitted before partons will have enough time form a current
which depends on their density. We assume that this density is large and
because of this fact our current actually is a normal classical current. 
Finally, the Lagrangian of the interaction of the parton emitted at time
$t'_2$ can be written as 
$$ L(\rho)\,\,+\,\,j_{\mu}(\rho) \cdot A_{\mu} \,\,+\,\,L(A)\,\,,$$ 
where $A$ is the field of a parton emitted at $t'_2$.
However, we can consider a parton emitted at $t= t'_3$ and include the
previous one in the system with density $\rho$. The form of Lagrangian
should be the same. This is a very strong condition on the form of
Lagrangian, so called Wilson renormalization group approach. In
Ref.\cite{MCLER}, this approach has been used to obtain the effective
Lagrangian.

\newpage
\centerline{\Huge \bf III T H E\,\,\, B F K L\,\,\, P O M E R O N}
\centerline{}
\section{The BFKL parton cascade and its evolution.}
The BFKL Pomeron \cite{BFKL} is an asymptotic of the scattering amplitude
in
perturbative QCD in the kinematic region where the log scale $L
\,\,=\,\,\alpha_S\,\ln(1/x) \,\,\gg\,\,1$ but the values of virtualities 
of incoming particles are moreless the same  $Q^2_0 \,\,\sim\,\,Q^2$.
Practically, this means that we calculate the same diagrams as before but
in the kinematic region where \eq{HP11} holds while there is no special
requirements for transverse momentum integration ( there is no \eq{HP12}
). Of course, the BFKL Pomeron should match the ``hard" Pomeron in the
situation when $Q^2\,\,gg\,\,Q^2_0$. 

First thing, that I would like to stress is the fact that the origin of
the $ln(1/x)$ contributions is very simple. As it has been mentioned the
factor $d x_i/x_i $ in \eq{HP1} comes from the phase space. Therefore, if
we write the optical theorem for total cross section of two particles
with virtualities $Q^2$ and $Q^2_0$ in the form
\beq \label{B1}
\sigma_{tot}(x,Q^2,Q^2_0)\,\,=\,\,\sum^{\infty}_{n =2}\,\int 
\prod_{i} \frac{d x_i}{x_i}  d^2 k_{t i} |M_{ 2 \,\rightarrow  n} (
x_i,k_i)|^2  
\eeq
on can see that to obtain $log(1/x)$
 contribution we can safely put all $x_i = 0$ in amplitudes $M_n$. It
gives
\beq \label{B2} 
\sigma_{tot}(x,Q^2,Q^2_0)\,\,=\,\,\sum^{\infty}_{n =0}\,\int
\prod_{i} \frac{d x_i}{x_i}\, \times\, 
 \int d^2 k_{t i} |M_{ 2\,\rightarrow \,2 +  n} (0 ,k_i)|^2\,\,=
\eeq
$$
\,\,\sum^{\infty}_{n =2}\,
\frac{\ln^n(1/x)}{n!} \int\,\prod^n_{i} d^2 k_{t i}\, |M_{ 2\,\rightarrow
\,2 + n}
(0
,k_i)|^2\,\,.
$$
Deriving \eq{B2} we used the ordering in $x_i$ of \eq{HP11}.

Assuming that the integral over $k_{ti}$ in \eq{B2} gives
$\int \prod^n_{i} d^2 k_{t i}\, |M_{ 2\,\rightarrow \,2 +
n}(0,k_i)|^2\,\,=\,\,\omega^n_L$ we can easily obtain that
\beq  \label{B4}
\sigma_{tot}(x,Q^2,Q^2_0)\,\,=\,\,\frac{\sigma_0}{\sqrt{Q^2\,Q^2_0}}
\,\,\,\,\,\,(\frac{1}{x}\,)^{\omega_L}\,\,,
\eeq
where $\omega_L\,\,\propto\,\alpha_S$. The result of \eq{B4} can be
rewritten in the form $\sigma_{tot}\,\,\propto\,\,e^{< n >}$ with $< n > =
\omega_L \,\ln(1/x)$. Note, that factor in front can be obtained just from
dimension of the total cross section and the fact that the scattering
amplitude in the region of $Q^2 \,\,\approx\,\,Q^2_0$ should be
symmetric with respect to $Q^2 \,\rightarrow\,\,Q^2_0$ and
$Q^2_0\,\rightarrow\,Q^2$.

 It is interesting to point out that the simple feature of QCD,namely, the
fact that we have a dimensionless coupling constant in QCD. In such
theory each emission leads to the change of the $\ln k_{t}$ of a parton by
some value ( $d$ ) :
\beq \label{B5}
\Delta \ln(k^2_{ti}/Q^2_0)\,\,=\,\,d \,\,.
\eeq
After $n$- emissions the total change of the transverse momentum will be
equal to;
\beq \label{B6}
< \ln^2\,\,(k^2_{ti}/Q^2_0)
>|_{after\,\,\,n\,\,\,\,emissions}\,\,=\,\,d\,n\,\,.
\eeq
It means that in average for the scattering amplitude we have
\beq \label{B7}
< \ln^2\,\,( Q^2/Q^2_0) >\,\,=\,\,d\,< n >\,\,=\,\,d\,\omega_L
\ln(1/x)\,\,=\,\,4\,D\,\ln(1/x)\,\,.
\eeq
Therefore, the transverse momentum dep[endence of the scattering amplitude
is a distribution  with $< ln( Q^2/Q^2_0) >$ given by \eq{B6}.

Collecting everything that has been discussed we have 
\beq \label{B8}
\sigma_{tot}(x,Q^2,Q^2_0)\,\,=\,\,\frac{\sigma_0}{\sqrt{Q^2\,Q^2_0}}   
\,\,\,\,e^{\omega_L\,Y}\,\,\frac{1}{\sqrt{4\,D \,\pi\,Y}}\,\,\,\,e^{-
\frac{\ln^2(Q^2/Q^2_0)}{4\,D\,Y}}\,\,.
\eeq 
\eq{B8} gives the answer. The values of three constants:
$\sigma_0\,\,=\,\,\alpha^4_S$\,\,, \,\,\,$\omega_L\,\,\propto\,\,\alpha_S$
\,\,and\,\,$D\,\,\propto\,\,\alpha_S$  have been found in the BFKL
approach to perturbative QCD\cite{BFKL}. It should be stressed that we
have used heavily a fact that out theory has no dimension scale. Actually,
it has because the QCD coupling depends on the transverse momentum.
Therefore, strictly speaking, \eq{B8} is valid only for fixed QCD
coupling.

\section{The BFKL equation.}
\subsection{$\mathbf \alpha_S(Q^2_0)$ fixed.}
This section will be a bit more formal than everything that was before.
 Nevertheless, I would like to share with you a beautiful result, that the
``soft" Pomeron might be the BFKL Pomeron with running QCD coupling +
confinement.

The natural function for which the BFKL equation can be written is
 the gluon  density $\phi(y,q^2)$ which is closely related to the gluon
structure function. 
\beq \label{BE1}
 xG(x,Q^2)\,\,=\,\,\int^{Q^2}\,\,\alpha_S(q^2)\,\phi(y,q^2)\,\,,
\eeq
where we  introduce from the beginning a new symmetric
variable $y = \ln \frac{s}{Q Q_0}$. In the leading order $y$ is equal to
$y = \ln(1/x)$ but in the next to leading order this variable turns out yo
be much more convenient.

The BFKL equation \cite{BFKL} has a form:
\beq \label{BE2}
\frac{d \phi( y,q^2)}{d y}\,\,
=
\eeq
$$
\,\,\delta^{(2)}((\,\vec{q}\,\,-\,\,\vec{q}_0\,)\,\,\,+\,\,\int\,\,d^2
\,q'\,K(\,\vec{q},\vec{q}'\,)\,\phi(\,y,\,\vec{q}',\vec{q}_0)\,)\,\,-\,\, 
2\,\alpha_G(q^2)\,\phi(\,y,\,\vec{q} ,\vec{q}_0)\,)\,\,,
$$ 
where all notations are clear from Fig.36.

\begin{figure}
\centerline{\psfig{file= 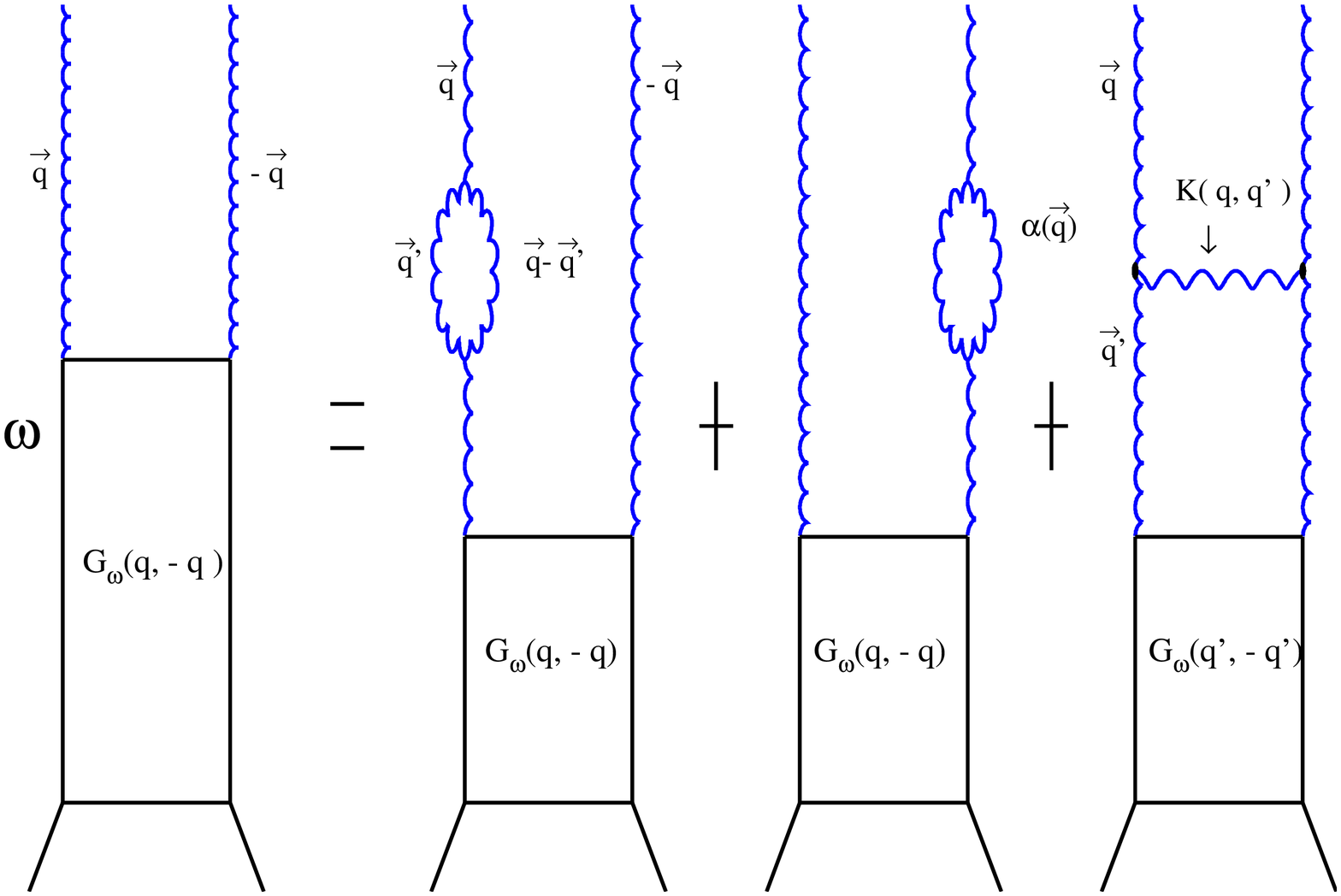,width=130mm, height=80mm}}
\caption{ \it The BFKL equation.}
\label{fig37}
\end{figure}

The BFKL kernel has a remarkable property that function
\beq \label{BE3}
\phi_f(q^2)\,\,=\,\,\frac{1}{\sqrt{q^2}}\,(\,q^2\,)^f\,\,
\eeq
is an eigenfunction, namely
\beq \label{BE4}
\int\,\,d^2
\,q'\,K(q,q')\,\phi_{\omega}(q')\,\,=\,\,\omega(f)\,\phi_f(q^2) \,\,,
\eeq
with
\begin{eqnarray} 
&
\omega(f)\,\,=\,\,\frac{N_c \alpha_S(Q^2_0)}{\pi}\,\{\,\Psi(f) + \Psi (1 -
f) - 2 \Psi(1)\,\}\,\,= &\label{BE5}\\
&
\,\, \omega_L
\,\,+\,\,D (f - \frac{1}{2} )^2\,\,+\,\,O((f - \frac{1}{2})^3)
\,\,, &\label{BE51}
\end{eqnarray}
where $\Psi(z)\,\,=\,\,\frac{d \ln \Gamma(z)}{d z}$  ($\Gamma(z)$ is the
Euler gamma-function.
For the Mellin transform of $\pi$
\beq \label{BE6}
\phi(y,q^2)\,\,=\,\,\int^{a + i\infty}_{a - i
\infty}\,\,\frac{d\,\omega}{2 \,\pi\,i}\,\,e^{\omega\,y} \,G_{\omega}(q^2)
\eeq
the BFKL equation reduces to the simple form:
\beq \label{BE7}
\omega \,G_{\omega}(f)\,\,=\,\,1\,\,+\,\,\,\omega(f)\,G_{\omega}(f)\,\,,
\eeq
where
\beq \label{BE8}
G_{\omega}(q^2)\,\,=\,\,\int^{a + i\infty}_{a - i
\infty}\,\,\frac{d\,f}{2 \,\pi\,i}\,\,e^{- f \,(r - r_0)}\,\,G_{\omega}(f)
\,\,,
\eeq
with $r - r_0 = ln(Q^2/Q^2_0)$.

Substituting  solution to \eq{BE7} $G_{\omega}(f)\,\,=\,\,\frac{1}{\omega
- \omega(f)}$ into \eq{BE6} and \eq{BE8}, we obtain after integration over
$\omega$ since we can close contour on the pole in $\omega$
\beq \label{BE9}
\phi(y,q^2)\,\,=\,\,\int^{a + i\infty}_{a - i
\infty}\,\,\frac{d\,f}{2 \,\pi\,i}\,\,e^{\omega(f)\,y\,\,- \,\,f \,(r -
r_0)}\,\,G_{\omega}(f)\,\,,
\eeq
where $G_{\omega}(f)$ should be found from initial nonperturbative parton
distributions.  Using the expansion of $\omega(f)$ at $f\,\rightarrow
\,\frac{1}{2}$, one can see that we are able to take the integral by
saddle point method and obtain the answer of \eq{B8} with
\begin{eqnarray} 
&
\omega_L\,\,=\,\,\frac{N_c \alpha_S(Q^2_0)}{\pi}\,4\,\ln
2\,\,;&\label{BE10}\\
&
D\,\,=\,\,\frac{N_c \alpha_S(Q^2_0)}{\pi}\,14\,\zeta(3)\,\,.&\label{BE11}
\end{eqnarray}

\subsection{Running QCD coupling.}
The previous subsection was written mostly because I would like to point
out that the running QCD coupling leads to a result which is very
important since it gives the only theoretical way that I know to obtain
the Pomeron - a Reggeon with intercept bigger than 1. Indeed, let us
consider the BFKL kernel to be proportional to the running
$\alpha_S(q^2)\,\,=\,\,\frac{\alpha_s(Q^2_0))r_0}{r}$ where $r_0\,=\,ln
Q^2_0/\Lambda^2)$ and $r\, =\,lnQ^2/\Lambda^2)$ . $\Lambda^2$ is the
position
of Landau pole in $\alpha_S$ where QCD coupling constant becomes very
large.
The generalized BFKL equation\cite{RUNBFKL} looks as follows:
\beq \label{BR1}
\frac{d \phi( y,q^2)}{d y}\,\,
=
\eeq
$$
\delta^{(2)}((\,\vec{q}\,\,-\,\,\vec{q}_0\,)\,\,\,+\,\,\frac{r_0}{r}\,
\,\int\,\,d^2
\,q'\,K(\,\vec{q},\vec{q}'\,)\,\phi(\,y,\,\vec{q}',\vec{q}_0)\,)\,\,.
$$
This equation can be easily rewritten as the differential equation in
double Mellin transform with respect to $y$ and $r$ ( see \eq{BE3} ).
\beq \label{BR2}
-\,\omega \frac{d g (\omega, f)}{d f}\,\,=\,\,r_0\,\omega_{conf} (f) \,g
(\omega,      
f)\,\,+\,\,r_0 \,e^{-f r_0}\,\,.
\eeq
The solution of homogeneous equation ( \eq{BR2} without the last term
) can be easily found and it has the  form ( see Refs. \cite{RUNBFKL}
for details ):
\beq \label{SOL}   
g(\omega, f)\,\,=\,\,\tilde g(\omega) \,e^{-
\,\frac{r_0}{\omega}\,\int^f_{f_0}\,\omega_{conf} (f') d f'}\,\,.
\eeq
Function $\tilde g (\omega)$ should be specified from initial or boundary
conditions.   The value of $f_0$ can be arbitrary since its redefinition
is included in function $\tilde g(\omega)$. Unless  it is  specially
stipulated  $f_0$ = 0.

Let us find a Green function of the BFKL equation with running QCD
coupling ( $G_r (y,r)$ ) which  satisfies the following boundary
condition:
\begin{eqnarray} \label{GREENR}
&
G_r (y,r):&  \\
&   
G_r(y, r = r_0)\,\,=\,\,\delta(y \,-\,y_0) &\nonumber
\end{eqnarray}
This Green function allows us to find us the solution of the BFKL equation
for any boundary input  distribution  $G_{in}(y,q^2=q^2_0)$
 at $q^2 = q^2_0$ ( $r = r_0$ ). Indeed, such a solution is equal to
\beq \label{SOLR}
G(y,r)\,\,=\,\,\int \,d y_0 \,G_r(y, r) \,G_{in}(y_0,q^2=q^2_0)
\,\,.
\eeq

Such a  Green function is very useful for study of the boundary
condition for the DGLAP evolution. Using $G_r(y,r)$ and \eq{SOLR},  we can
investigate the $y$-dependence at $q^2\,\cong\,q^2_0$. We can distinguish
two cases with different solutions:
\begin{enumerate}
\item  The integral over $y_0$ depends mostly on properties of input
function
$G_{in}$;

\item  The integral over $y_0$ is sensitive to the Green function. In this
case we can claim that the energy behaviour of our boundary condition is
defined by the BFKL dynamics.
\end{enumerate}
Therefore, this Green function ( $G_r (y,r)$ ) can provide us with  an
educated
guess  for the energy dependence of the boundary condition in the DGLAP
evolution equations \cite{DGLAP}.

Substituting in \eq{SOL} the expansion of $\omega_{conf}$ at
$f\,\rightarrow \,\frac{1}{2} $ ( see \eq{BE51} ) we obtain that integral
over $f$ in \eq{SOL} gives the Airy function $Ai\left( (\frac{\omega}{r_0
D})^{\frac{1}{3}}\,[\,r\,-\,\frac{\omega_L}{\omega}r_0\,]\,\right)$.
Therefore
 to satisfy the boundary condition of \eq{GREENR} we have to choose a
function $\tilde g(\omega) \,=\,Ai^{-1}\left( (\frac{\omega}{r_0
D})^{\frac{1}{3}}\,[\,r_0\,-\,\frac{\omega_L}{\omega}r_0\,]\,\right)$.

Finally \cite{RUNBFKL}, $G_r(y-y_0,r,r_0)$ is equal to
\beq \label{GRNRF}
G_r(y-y_0,r,r_0)\,\,=
\eeq
$$
\,\,\sqrt{\frac{r}{r_0}}\,\int^{a + i\infty}_{a -
i\infty}\,\frac{d \omega}{2\pi i}\,e^{\omega\,(y - y_0)}\,\,\frac{Ai\left(
(\frac{\omega}{r_0
D})^{\frac{1}{3}}\,[\,r\,-\,\frac{\omega_L}{\omega}r_0\,]\,\right)}{Ai\left(
(\frac{\omega}{r_0
D})^{\frac{1}{3}}\,[\,r_0\,-\,\frac{\omega_L}{\omega}r_0\,]\,\right)}\,\,.
$$

\begin{figure}
\centerline{\psfig{file= 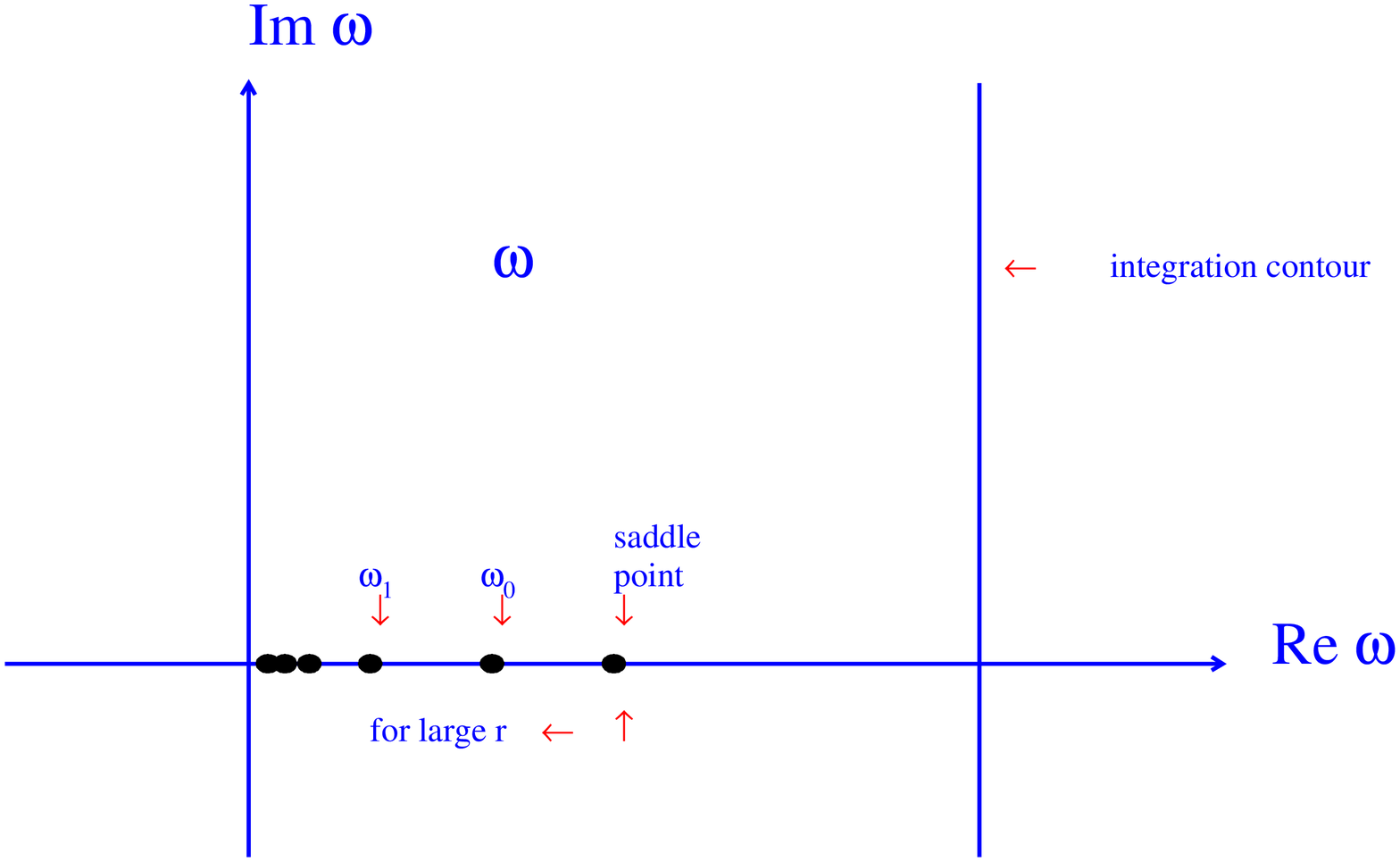,width=100mm}}
\caption{ \it Singularities in $\omega$-plane for the BFKL equation with 
running QCD coupling.}
\label{fig38}
\end{figure}
Airy function is an analytical function which has zeros and a singular
behaviour at large values of the argument $Ai(z)|_{z\,>\,0 ;
|z|\,\gg\,1}\,\,\rightarrow\,\,\frac{1}{2
z^{\frac{1}{4}}}
e^{-\frac{2}{3} \,z^{\frac{3}{2}}}$. Therefore, we have two possibility to
take the integral over $\omega$ (see Fig.37, which shows the structure of
singularities in $\omega$ - plane ):
\begin{enumerate}
\item To draw contour through the saddle point\,\,;  
\item To close contour on the zeros of the denominator\,\,.
\end{enumerate}
It turns out that the position of the saddle point
$\omega_{SP}\,=\,\omega_L(r_0)\,\frac{r_0}{r_0 + r }$ and at large values
of $q^2$ ( $r$ ) it moves to the left. Therefore, for large $r$ the
rightmost singularity is the $\omega_0(r_0)$ - the zero of the
denominator. Closing contour on pole at $\omega\,=\,\omega_0$ we obtain
Regge asymptotic $\sigma \,\,\propto\,\,e^{\omega_0\,y}$ and all $r$-
dependence enters only in the value of residue.

For not very large $r$ the saddle point contribution turns out to be
dominant and leads to the Green function:
\beq \label{BR6}
G_r(y - y_0,r,r_0)\,\,\propto\,\,\,e^{\omega_L\,(\, y\,-\,y_0\,)
\,\,-\,\,\frac{(\,r\,-\,r_0\,)^2}{4\,D\,(\,y\,-\,y_0\,)}\,\,+
\,\,\frac{D\,\omega^2_L}{12\,\,r^2_0}\,\,(\,y\,-\,y_0\,)^3}\,\,,
\eeq
where $\omega_L$ and $D$ the running coupling constant $\alpha_S (r_0)$
should be taken at different scale: $\frac{r + r_0}{2}$ instead of $r_0$
\cite{KM}. One can see that \eq{BR6} gives a Regge-BFKL asymptotic of
\eq{1.19} but only at restricted values of
$y\,-\,y_0\,\,\leq\,\,\frac{1}{\alpha^{\frac{5}{3}}_S}$. For larger
$y\,-\,y_0$ the last term in exponent appears which has clear non - Regge
behaviour \cite{KM}.

\subsection{Summary.}

The BFKL Pomeron in pQCD gives an example of a different asymptotic
behaviour of the scattering amplitude than the Regge pole (``soft" Pomeron
). It is very instructive to understand that the space - time picture of
the BFKL Pomeron is very similar to the ``soft" Pomeron and a difference
is
only in the particular properties of the QCD cascade.

For the running QCD coupling the BFKL asymptotics depend on the non
perturbative QCD initial condition which, in general, generates a ``soft"
Pomeron asymptotic behaviour at large values of virtuality. 

Certainly, the BFKL Pomeron is our window to non-perturbative regime
and we need more experience especially with the BFKL asymptotic in the
next-to-leading order approximation ( see Refs. \cite{NLO} ) to make a
reasonable educated guess about high energy asymptotic in non-perturbative
QCD.

\section{My several last words.}

In this talk,  I tried to explain what we have in mind when we are saying
Pomeron.  This is only an introduction in which I had to avoid many
important details. I feel pity that I did not discuss the open
theoretical questions and how the experimental measurements could help us
to solve them. However, this is an interesting subject but certainly
it cannot be considered in such an introductory talk.    
I also apologize that in the last section about the BFKL Pomeron, I was
using mathematics too much, but I did this only to show you the only one
example how  a ``soft" Pomeron can appear in QCD.

I am very grateful to A. Santoro who invited me to his conference in Rio
de Janeiro and without whom this talk will never be written. I thank all
my brasilian friends:Francisco Caruso, Helio del Motta, 
Maria - Elena Pol
 and Ronald Shellard, whose hospitality recall me once more that I
left a part of my heart in Rio. My special thanks go to E. Gotsman and U.
Maor, who started with me re-analyze the situation in ``soft" interaction
more than five years ago and who spoke  through my talk too, but with my
full responsibility for all inconsistency here. 

I would like to acknowledge the kind  hospitality of the Theory Group at
DESY where this paper was completed.

 \end{document}